\documentclass[9pt,a4paper,twosided]{article}

\usepackage{geometry}
\usepackage[utf8]{inputenc}
\usepackage[T1]{fontenc}

\usepackage[english]{babel}

\usepackage{microtype}

\ifxetex
  \usepackage{fontspec}
\fi

\usepackage{amsmath,amsthm,amssymb,stmaryrd,thmtools,upgreek}

\newtheorem{theorem}{Theorem}
\newtheorem{lemma}[theorem]{Lemma}
\newtheorem{corollary}[theorem]{Corollary}

\newtheorem{conjecture}[theorem]{Conjecture}

\theoremstyle{definition}
\newtheorem{definition}[theorem]{Definition}
\newtheorem{remark}[theorem]{Remark}
\newtheorem{example}[theorem]{Example}

\usepackage{graphicx}
\usepackage[obeyclassoptions,mode=tex]{standalone}
\usepackage{tikz}
\usetikzlibrary{backgrounds}
\usetikzlibrary{shapes.geometric}
\usetikzlibrary{decorations.markings}
\usetikzlibrary{positioning}
\usetikzlibrary{automata}
\usetikzlibrary{tikzmark}
\usetikzlibrary{patterns}
\usetikzlibrary{arrows}

\tikzset{every state/.style={minimum size=1pt}}
\usepackage{tikz-cd}

\usepackage{pgfornament}

\usepackage{hyperref}
\usepackage[capitalise,noabbrev,nameinlink]{cleveref}
\Crefname{assumption}{Assumption}{Assumptions}
\Crefname{conjecture}{Conjecture}{Conjectures}
\usepackage[electronic,hyperref,xcolor,cleveref]{knowledge}
\knowledgeconfigure{notion}

\usepackage{booktabs}
\usepackage{varwidth}

\usepackage{algorithm2e}
\Crefname{algocfline}{Algorithm}{Algorithms}
\crefname{algocfline}{Algorithm}{Algorithms}
\crefname{algocf}{Algorithm}{Algorithms}
\Crefname{algocf}{Algorithm}{Algorithms}

\usepackage{xparse}
\usepackage{xpatch}
\usepackage{tokcycle}
\usepackage{ifthen}

\usepackage{bussproofs}

\usepackage{ensps-colorscheme}
\knowledgestyle{intro notion}{color={A2}, emphasize}
\knowledgestyle{notion}{color={B1}}
\hypersetup{
    colorlinks=true,
    breaklinks=true,
    linkcolor=A4,
    citecolor=A4,
    urlcolor=A4,
    filecolor=A4,
}

\newcommand{\klogo}{%
\begin{tikzpicture}[scale=0.2,line/.style={draw, line width=0.2pt, line cap=round, line join=round}]
\coordinate (A00) at (0,0);
\coordinate (A01) at (0,1);
\coordinate (A10) at (1,0);
\coordinate (B10) at (1,0.2);
\coordinate (B01) at (0.2,1);

\coordinate (C01) at (0.4,0.7);
\coordinate (C10) at (0.7,0.4);
\coordinate (C12) at (0.4,1.2);
\coordinate (C21) at (1.2, 0.4);
\coordinate (C22) at (1.2, 1.2);

\coordinate (D00) at (C10);
\coordinate (D01) at (0.8,0.5);
\coordinate (D10) at (0.8,0.3);

\coordinate (E01) at (0.3,0.7);
\coordinate (E10) at (0.5,0.7);

\draw[line] (B01) -- (A01) -- (A00) -- (A10) -- (B10);
\draw[line] (C01) -- (C12) -- (C22) -- (C21) -- (C10);

\draw[line] (D01) -- (D00) -- (D10);
\draw[line] (E01) -- (E10);

\end{tikzpicture}%
}

\newcommand{\defined}{\mathrel{:=}}
\newcommand{\eqdef}{\stackrel{\mathsf{def}}{=}}
\newcommand{\seqof}[2]{(#1)_{#2}}
\newcommand{\setof}[2]{\mathopen{\{} #1 \mid #2 \mathclose{\}}}
\newcommand{\set}[1]{\mathopen{\{} #1 \mathclose{\}}}

\NewDocumentCommand{\card}{ O{} m }{ \mathopen{|} #2  \mathclose{|}_{#1}}

\newcommand{\Nat}{\mathbb{N}}
\newcommand{\Rel}{\mathbb{Z}}
\newcommand{\Rat}{\mathbb{Q}}

\newcommand{\MSO}{\ensuremath{\mathsf{\kl[\MSO]{MSO}}}}
\knowledge{\MSO}{notion}

\newcommand{\FO}{\ensuremath{\mathsf{\kl[\FO]{FO}}}}
\knowledge{\FO}{notion}

\newcommand{\ind}[1]{\mathbf{1}_{#1}}
\newcommand{\bigO}{\mathcal{O}}

\NewDocumentCommand{\Span}{m}{\mathop{\kl[\Span]{\mathsf{Span}_{#1}}}}
\knowledge{\Span}{notion}

\newcommand{\topartial}{\rightharpoonup}

\newcommand{\dom}{\operatorname{dom}}

\newcommand{\CoveredPoly}{\kl[\CoveredPoly]{\mathsf{PolyNNeg}}}
\newcommand{\CorrectPoly}{\kl[\CorrectPoly]{\mathsf{PolyStrNNeg}}}
\knowledge{\CoveredPoly}{notion}
\knowledge{\CorrectPoly}{notion}

\NewDocumentCommand{\Monomials}{o}{\mathop{\kl[\Monomials]{\IfNoValueTF{#1}{\mathsf{Mon}}{\mathsf{Mon}[#1]}}}}
\NewDocumentCommand{\MaximalMonomials}{}{\mathop{\kl[\MaximalMonomials]{\mathsf{maxmon}}}}
\knowledge{\Monomials}{notion}
\knowledge{\MaximalMonomials}{notion}

\newcommand{\restr}[2]{\withkl{\kl[\restr]}{\cmdkl{[}{#1}\cmdkl{]}_{#2}}}
\knowledge{\restr}{notion}

\NewDocumentCommand{\Poly}{O{}}{\kl[\Poly]{\mathsf{Poly}_{#1}}}
\NewDocumentCommand{\SF}{O{}}{\kl[\SF]{\mathsf{SF}_{#1}}}

\NewDocumentCommand{\Mealy}{O{}}{\kl[\Mealy]{\mathsf{Mealy}_{#1}}}
\NewDocumentCommand{\Sequential}{O{}}{\kl[\Sequential]{\mathsf{Seq}_{#1}}}
\NewDocumentCommand{\Rational}{O{}}{\kl[\Rational]{\mathsf{Rat}_{#1}}}
\NewDocumentCommand{\Regular}{O{}}{\kl[\Regular]{\mathsf{Reg}_{#1}}}
\knowledge{\Mealy}{notion}
\knowledge{\Sequential}{notion}
\knowledge{\Rational}{notion}
\knowledge{\Regular}{notion}

\NewDocumentCommand{\AMealy}{O{}}{\kl[\AMealy]{\mathsf{AMealy}_{#1}}}
\NewDocumentCommand{\ASequential}{O{}}{\kl[\ASequential]{\mathsf{ASeq}_{#1}}}
\NewDocumentCommand{\ARational}{O{}}{\kl[\ARational]{\mathsf{ARat}_{#1}}}
\NewDocumentCommand{\ARegular}{O{}}{\kl[\ARegular]{\mathsf{AReg}_{#1}}}
\knowledge{\AMealy}{notion}
\knowledge{\ASequential}{notion}
\knowledge{\ARational}{notion}
\knowledge{\ARegular}{notion}

\NewDocumentCommand{\NPoly}{O{}}{\kl[\NPoly]{\mathbb{N}\mathsf{Poly}_{#1}}}
\NewDocumentCommand{\ZPoly}{O{}}{\kl[\ZPoly]{\mathbb{Z}\mathsf{Poly}_{#1}}}
\NewDocumentCommand{\NSF}{O{}}{\kl[\NSF]{\mathbb{N}\mathsf{SF}_{#1}}}
\NewDocumentCommand{\ZSF}{O{}}{\kl[\ZSF]{\mathbb{Z}\mathsf{SF}_{#1}}}
\NewDocumentCommand{\ZRat}{}{\kl[\ZRat]{\mathbb{Z}\mathsf{Series}}}
\NewDocumentCommand{\NRat}{}{\kl[\NRat]{\mathbb{N}\mathsf{Series}}}
\NewDocumentCommand{\RatRat}{}{\kl[\RatRat]{\mathbb{Q}\mathsf{Series}}}
\NewDocumentCommand{\PRatRat}{}{\kl[\PRatRat]{\mathbb{Q}_+\mathsf{Series}}}
\NewDocumentCommand{\ZCommut}{}{\kl[\ZCommut]{\mathsf{Commut}}}

\knowledge{\Poly}{notion}
\knowledge{\SF}{notion}
\knowledge{\NPoly}{notion}
\knowledge{\ZPoly}{notion}
\knowledge{\NSF}{notion}
\knowledge{\ZSF}{notion}
\knowledge{\ZRat}{notion}
\knowledge{\NRat}{notion}
\knowledge{\RatRat}{notion}
\knowledge{\PRatRat}{notion}
\knowledge{\ZCommut}{notion}

\NewDocumentCommand{\commute}{O{\cdot}}{\{\!\{ #1 \}\!\}}

\NewDocumentCommand{\BadPoly}{}{\kl[\BadPoly]{\mathsf{P}_{\mathsf{bad}}}}
\knowledge{\BadPoly}{notion}

\NewDocumentCommand{\translate}{m}{\mathop{\kl[\translate]{\tau_{#1}}}}
\knowledge{\translate}{notion}

\NewDocumentCommand{\Diff}{m m}{ \mathop{\kl[\Diff]{\Delta_{#1}}}(#2) }
\knowledge{\Diff}{notion}

\NewDocumentCommand{\npolyleq}{ O{} }{\mathrel{\kl[\npolyleq]{\preceq_{\Nat #1}}}}
\knowledge{\npolyleq}{notion}
\NewDocumentCommand{\zpolyequiv}{ O{} }{\mathrel{\kl[\zpolyequiv]{\equiv_{\Rel #1}}}}
\knowledge{\zpolyequiv}{notion}

\NewDocumentCommand{\app}{ m m }{\mathop{{#2} \mathrel{\kl[\app]{\triangleright}} {#1}}}
\knowledge{\app}{notion}

\NewDocumentCommand{\Res}{}{ \mathop{\kl[\Res]{\mathsf{Res}}}}
\knowledge{\Res}{notion}

\NewDocumentCommand{\resleq}{ m m }{\mathrel{\kl[\resleq]{\leq_{{#1},{#2}}}}}
\knowledge{\resleq}{notion}
\NewDocumentCommand{\resleqsf}{ m m }{\mathrel{\kl[\resleqsf]{\leq_{{#1},{#2}}^{\mathsf{sf}}}}}
\knowledge{\resleqsf}{notion}

\NewDocumentCommand{\prefleq}{}{\mathrel{\kl[\prefleq]{\sqsubseteq_{\mathsf{pref}}}}}
\knowledge{\prefleq}{notion}
\NewDocumentCommand{\prefle}{}{\sqsubset_{\mathsf{pref}}}

\NewDocumentCommand{\aTransd}{}{\mathcal{A}}

\NewDocumentCommand{\ModuloTypes}{O{}}{\kl[\ModuloTypes]{{#1}\mathsf{Types}}}
\knowledge{\ModuloTypes}{notion}
\NewDocumentCommand{\moduloType}{O{}}{\kl[\moduloType]{{#1}\mathsf{type}}}
\knowledge{\moduloType}{notion}

\NewDocumentCommand{\floor}{m}{\lfloor #1 \rfloor}

\NewDocumentCommand{\BadExOk}{}{\mathsf{f}}
\NewDocumentCommand{\BadExKo}{}{\mathsf{g}}

\NewDocumentCommand{\pbinom}{ m m }{\withkl{\kl[\pbinom]}{\cmdkl{\bullet}\!\!\binom{#1}{#2}\!\!\cmdkl{\bullet}}}
\knowledge{\pbinom}{notion}

\NewDocumentCommand{\CutPref}{ m }{\text{pref}_{#1}}
\NewDocumentCommand{\CutSuff}{ m }{\text{suff}_{#1}}

\NewDocumentCommand{\Deriv}{ m m m }{\mathop{\withkl{\kl[\Deriv]}{\cmdkl{\Delta}_{#3}^{#2}\mathopen{\cmdkl{[}}#1\mathclose{\cmdkl{]}}}}}
\knowledge{\Deriv}{notion}

\NewDocumentCommand{\counting}{ m }{\mathop{\withkl{\kl[\counting]}{\cmdkl{\text{nbr}[}#1\cmdkl{]}}}}
\knowledge{\counting}{notion}

\NewDocumentCommand{\resequiv}{ m m }{\mathrel{\kl[\resequiv]{\equiv_{{#1},{#2}}}}}
\knowledge{\resequiv}{notion}

\knowledge{notion}
 | aperiodic monoid
 | aperiodic@monoid
 | aperiodic monoids

\knowledge{notion}
 | production function

\knowledge{notion}
 | aperiodic@language
 | aperiodic language
 | star-free language
 | star-free@language

\knowledge{notion}
 | growth rate
\knowledge{notion}
 | commutative
 | commutativity

\knowledge{notion}
 | $\Rel $-rational polynomial
 | $\Rel $-rational polynomials

\knowledge{notion}
 | $\Nat $-rational polynomial
 | $\Nat $-rational polynomials

\knowledge{notion}
 | Fatou extension
 | Fatou property
\knowledge{notion}
 | (noncommutative) $\mathbb {S}$-rational series
 | $\mathbb {S}$-rational series
 | weighted automata
 | $K$-rational series
 | $K$-weighted automata
 | $\mathbb {S}$-weighted automata
 | $\Nat $-weighted automaton
 | $\Rel $-rational series
 | $\Rel $-weighted automaton
 | $\Nat $-rational series
 | rational series 

\knowledge{notion}
 | regular function
 | regular functions

\knowledge{notion}
 | polyregular function
 | polyregular functions
 | Polyregular functions

\knowledge{notion}
 | star-free polyregular functions
 | star-free polyregular
 | star-free polyregular function
 | aperiodic polyregular function
 | star-free
 | aperiodic@polyregular
 | aperiodicity@polyregular
\knowledge{notion}
 | polyblind functions
 | polyblind
\knowledge{notion}
 | star-free polyblind

\knowledge{notion}
 | $\Rel $-polyregular
 | $\Rel $-polyregular function
 | $\Rel $-polyregular functions

\knowledge{notion}
 | $\Nat $-polyregular
 | $\Nat $-polyregular function
 | $\Nat $-polyregular functions

\knowledge{notion}
 | star-free $\Rel $-polyregular function
 | star-free $\Rel $-polyregular functions
 | star-free $\Rel $-polyregular 

\knowledge{notion}
 | star-free $\Nat $-polyregular function
 | star-free $\Nat $-polyregular functions
 | star-free $\Nat $-polyregular

\knowledge{notion}
 | polynomial growth
\knowledge{notion}
 | representing
 | represented
 | represents
\knowledge{notion}
 | ultimately polynomial
 | ultimate polynomiality
\knowledge{notion}
 | $(k,\Nat )$-combinatorial
 | $(1,\Nat )$-combinatorial
 | combinatorial

\knowledge{notion}
 | computed
 | computing
 | computes
 
\knowledge{notion}
 | downwards closed

\knowledge{notion}
 | divides
 | divided
 | strictly divides
 | divide
 | divisibility preordering

\knowledge{notion}
 | non-negative

\knowledge{notion}
 | non-negative coefficient
 | non-negative coefficients

\knowledge{notion}
 | monomial
 | monomials
\knowledge{notion}
 | maximal monomial
 | maximal monomials

\knowledge{notion}
 | discrete derivative
 | discrete differentiation
 | discrete derivatives
\knowledge{notion}
 | translations
 | translation function
 | translation operators
 | translation operator

\knowledge{notion}
 | noncommutative derivative
 | non-commutative derivatives

\knowledge{notion}
 | well-quasi-ordering
 | well-quasi-orderings
 | well-quasi-ordered
 | wqo
 | wqos

\knowledge{notion}
 | well@relation
 | well

\knowledge{notion}
 | good
 | good sequence
 | good sequences

\knowledge{notion}
 | bad
 | bad sequence

\knowledge{notion}
 | residuals
 | residual

\knowledge{notion}
 | residual transducer
 | $k$-residual transducer
 | $0$-residual transducer
 | $1$-residual transducer
 | $k$-residual transducers
 | residual transducers

\knowledge{notion}
 | Hoare ordering

\knowledge{notion}
 | counters
 | counter
 | counter-free

\knowledge{notion}
 | $\NPoly [0]$-transducers
 | $\NSF $-transducer
 | $\NPoly [0]$-transducer
 | $\mathcal {H}$-transducers
 | $\mathcal {H}$-transducer
 | $\NPoly [k-1]$-transducer
 | $\NPoly [1]$-transducer
 | $\NSF [0]$-transducer
 | $\NSF [k-1]$-transducer
 | $\reintro *\ZPoly [k-1]$-transducer
 | $\reintro *\NPoly [k-1]$-transducer
 | $\reintro *\NPoly [k]$-transducers
 | $\reintro *\NSF [k]$-transducers

\knowledge{notion}
 | types modulo $\omega $
 | $\omega $-type

\knowledge{notion}
 | prefix ordering

\knowledge{notion}
 | aperiodic ordering
 | aperiodic@ordering

\knowledge{notion}
 | pumping pattern
 | pumping patterns
 | Pumping patterns
\knowledge{notion}
 | binomial monomial
 | binomial monomials
\knowledge{notion}
 | simple binomial function
 | simple binomial functions
\knowledge{notion}
 | natural binomial functions
 | natural binomial function
\knowledge{notion}
 | natural binomial polynomial
 | natural binomial polynomials
\knowledge{notion}
 | integer binomial polynomials
 | integer binomial polynomial
\knowledge{notion}
 | strongly natural binomial polynomials
 | strongly natural binomial polynomial

\knowledge{notion}
 | integer-valued
 | integer-valued polynomials

\knowledge{notion}
  | rational function
  | rational functions

\makeatletter
\newcommand\mathgr[1]{\tokcycle
  {\addcytoks{##1}}
  {\processtoks{##1}}
  {\ifcsname up\expandafter\@gobble\string##1\endcsname
   \addcytoks[1]{\csname up\expandafter\@gobble\string##1\endcsname}%
    \else\addcytoks{##1}\fi}
  {\addcytoks{##1}}{#1}%
  \expandafter\mathrm\expandafter{\the\cytoks}%
}
\makeatother

\NewDocumentCommand{\rightmarginnote}{m}{%
    \checkoddpage%
    \ifoddpage%
        \marginpar{#1}%
    \else%
        \reversemarginpar%
        \marginpar{#1}%
    \fi
}

\NewDocumentEnvironment{proofof}{o}{%
    \IfValueTF{#1}{%
        \def\insideRestate{1}
        \ifcsname #1\endcsname
            \csname #1\endcsname*
        \fi
        \begin{proof}[Proof of \cref{#1} on page \pageref{#1}]
        \phantomsection\label{#1:proof}
    }{%
        \begin{proof}
    }
    \let\oldqedsymbol\qedsymbol
    \renewcommand\qedsymbol{\hyperref[#1]{\oldqedsymbol}}
}{%
    \end{proof}
    \renewcommand\qedsymbol{\oldqedsymbol}
}

\NewDocumentCommand{\proofref}{m}{%
    \ifdefined\insideRestate%
        \rightmarginnote{\vspace{0.6em}\ttfamily\small\hyperref[#1]{Go to \cref{#1} p.\pageref{#1:proof}}}%
    \else%
    \IfRefUndefinedExpandable{#1:proof}{}{%
        \rightmarginnote{\vspace{0.6em}\ttfamily\small\hyperref[#1:proof]{Go to proof of \cref{#1} p.\pageref{#1:proof}}}%
    }%
    \fi%
}

\newtheorem{faketheorem}[theorem]{Flawed Theorem}
\newtheorem{problem}{Problem}
\Crefname{problem}{Problem}{Problems}
\crefname{problem}{Problem}{Problems}

\usepackage{lowco2}

\title{Commutative \(\Nat\)-rational Series of Polynomial Growth}

\author{
Aliaume Lopez\thanks{University of Warsaw}
}

      \date{\today}
  
\newcommand{\repositoryUrl}{\url{https://github.com/AliaumeL/polynomial-n-rational-series}}

\knowledge{notion}
 | kl-usage

\begin{document}
\maketitle
\begin{abstract}
    This paper studies which functions computed by \(\Rel\)-weighted
    automata can be realised by \(\Nat\)-weighted automata, under two
    extra assumptions: commutativity (the order of letters in the input
    does not matter) and polynomial growth (the output of the function
    is bounded by a polynomial in the size of the input). We leverage
    this effective characterization to decide whether a function
    computed by a commutative \(\Nat\)-weighted automaton of polynomial
    growth is star-free, a notion borrowed from the theory of regular
    languages that has been the subject of many investigations in the
    context of string-to-string functions. Furthermore, we open the road
    to a generalization of our results to non-commutative functions, by
    formalizing a canonical computational model for \(\Nat\)-weighted
    automata of polynomial growth based on the notion of residual
    transducer.
\end{abstract}

\paragraph{ACM CSS 2012:}
Quantitative automata;
Formal languages and automata theory;
Computations on polynomials;
Transducers.
\paragraph{Repository:} \repositoryUrl
\paragraph{Related versions:}
\href{https://doi.org/10.4230/LIPIcs.STACS.2025.67}{STACS'25 (Conference
version)} \cite{lopez:LIPIcs.STACS.2025.67}

\paragraph{Acknowledgments.}
Aliaume Lopez was supported by the Polish National Science Centre (NCN)
grant ``Polynomial finite state computation'' (2022/46/A/ST6/00072). 

\bigskip

\klogo\ This document uses \href{https://ctan.org/pkg/knowledge}{knowledge}:
\kl[kl-usage]{notion} points to its \intro[kl-usage]{definition}. 

\lowcotwo~This is a low-co2 research paper:
\lowcotwourl[\lowcotwoversion]. This research was developed, written,
submitted and presented without the use of air travel.

\section{Introduction}\label{introduction:sec}

\AP\ Given a semiring $\mathbb{S}$, and a finite alphabet $\Sigma$, the class
of \intro{(noncommutative) $\mathbb{S}$-rational series} is defined as
functions from $\Sigma^*$ to $\mathbb{S}$ that are computed by
\kl{$\mathbb{S}$-weighted automata}~\cite[Chapter~1]{BERE10}. This computational model is
a generalization of the classical notion of non-deterministic finite automata
to the weighted setting, where transitions are labeled with elements of
$\mathbb{S}$, and the set of accepting states is replaced by a function from
states to weights. The semantics of \kl{$\mathbb{S}$-weighted automata} on a
given word $w$ is defined by the sum over all accepting runs reading $w$, of
the product of the weights of the transitions taken along this run. A 
concrete example of a \kl{$\Rel$-weighted automaton} in
\cref{weighted-automaton:example}.

When the semiring $\mathbb{S}$ is the Boolean semiring $\mathbb{B}$, one
recovers the classical class of regular languages via non-deterministic finite
automata. When the semiring $\mathbb{S}$ is a commutative field, for instance
in the case of the rational numbers $\Rat$, the class of
\kl{$\mathbb{S}$-rational series} is well understood, and enjoys many
decidability properties by virtue of its correspondence with \emph{linear
representations}~\cite[Theorem 7.1]{BERE10}. However, the situation is much less clear when
$\mathbb{S}$ is merely a commutative ring  (e.g., $\Rel$) or worse, a
commutative semiring (e.g. $\Nat$): the corresponding theories of \emph{modules}
or \emph{semimodules} are much more involved than the one for vector spaces,
and many questions are left open. 

\begin{example}
  \label{weighted-automaton:example}
  The function $f \colon \set{a}^* \to \Rel$, that maps a word $w$
  to the size of $w$ if it is odd, and minus the size of $w$ if it is even, is a
  \kl{$\Rel$-rational series}, computed by the following $\Rel$-weighted automaton:
  \begin{center}
    \begin{tikzpicture}[shorten >=1pt,node distance=2cm,on grid,auto]
      \node[state,initial] (q_0)  at (-1.5,0) {$q_0$};
      \node[state] (q_1) at (1,0) {$q_1$};
      \node[state] (q_2) at (3.5,0) {$q_2$};
      \node[state] (q_3) at (6,0) {$q_3$};

      \node[draw=none, anchor=north] (f0) at (-1.5,-1) {$0$};
      \node[draw=none, anchor=north] (f1) at (1,-1) {$-1$};
      \node[draw=none, anchor=north] (f2) at (3.5,-1) {$2$};
      \node[draw=none, anchor=north] (f3) at (6,-1) {$-2$};
      \draw[->] (q_2) -- (f2);
      \draw[->] (q_3) -- (f3);
      \draw[->] (q_0) -- (f0);
      \draw[->] (q_1) -- (f1);

      \path[->] (q_0) edge[bend left] node {a / 1} (q_1);
      \path[->] (q_1) edge[bend left] node {a / 1} (q_0);
      \path[->] (q_1) edge node {a / 1} (q_2);
      \path[->] (q_2) edge[bend right] node[below] {a / 1} (q_3);
      \path[->] (q_3) edge[bend right] node[above] {a / 1} (q_2);
    \end{tikzpicture}
  \end{center}
\end{example}

\AP\ A well-studied question on \kl{rational series} is the characterization of
$\mathbb{S}$-rational series among $\mathbb{T}$-rational series, when
$\mathbb{S}$ is a subsemiring of $\mathbb{T}$. For instance, any
$\Rat$-rational series taking values in $\Rel$ is a $\Rel$-rational
series~\cite[Chapter 7, Theorem 1.1]{BERE10}. One says that $\Rat$ is a
\intro{Fatou extension} of $\Rel$, and in general, any field is a \kl{Fatou
extension} of its subfields~\cite[Chapter 7, Theorem 2.1]{BERE10}. In this
paper, we will focus on three specific semirings $\mathbb{S}$, and we will be
talking about $\Nat$-rational series ($\intro*\NRat$), $\Rel$-rational series
($\intro*\ZRat$), and $\Rat$-rational series ($\intro*\RatRat$). It is clear
that $\NRat \subsetneq \ZRat \subsetneq \RatRat$, and we already stated that
$\Rat$ is a \kl{Fatou extension} of $\Rel$. On the other hand, $\Rat$ is not a
\kl{Fatou extension} of $\Nat$, and a longstanding open problem is to provide an
algorithm that decides whether a given $\ZRat$ is in $\NRat$~\cite{KARH77}. 

\begin{problem}\label{n-in-z-rat:problem}
    Input: A $\ZRat$ $f$. Output: Is $f$ in $\NRat$?
\end{problem}

\AP\ \Cref{n-in-z-rat:problem} recently received attention in the context of
\kl{polyregular functions} ($\intro*\Poly$), a computational model that aims to
generalize the theory of regular languages to the setting of string-to-string
functions~\cite{BOJA18}. In the case of regular languages, \emph{star-free
languages} form a robust subclass of regular languages described equivalently
in terms of first order logic \cite{MNPA71}, counter-free automata
\cite{MNPA71}, or aperiodic monoids \cite{SCHU65}. Intuitively, star-free
languages are those that do not exhibit non-trivial periodic behaviour: for
instance, the language $(aa)^*$ is not star-free, while the language $(ab)^*$
is. Analogously, there exists a \emph{star-free} fragment of \kl{polyregular
functions} called \reintro{star-free polyregular functions} ($\intro*\SF$)
\cite{BOJA18}. A typical example of a non-star-free \kl{polyregular function}
is a function that replaces every letter ``$a$'' in an odd position of the
input by a letter ``$b$''. One open question in this area is to decide whether
a given \kl{polyregular function} is \kl{star-free}.

\begin{problem}
    \label{sf-polyregular:problem}
    Input: A \kl{polyregular function} $f$. Output: Is $f$ \kl{star-free}?
\end{problem}

\AP In order to approach decision problems on \kl{polyregular functions},
restricting the output alphabet to a single letter has proven to be a fruitful
method~\cite{DOUE21,DOUE22}. Because words over a unary alphabet are
canonically identified with natural numbers, unary output \kl{polyregular
functions} are often called \kl{$\Nat$-polyregular functions} ($\NPoly$), and
their \emph{star-free} counterpart \kl{star-free $\Nat$-polyregular functions}
($\NSF$). Coincidentally, \kl{polyregular functions} with unary output forms a
subclass of \kl{$\Nat$-rational series}, namely the class of
\kl{$\Nat$-rational series} of \emph{polynomial growth}, i.e. the output of the
function is bounded by a polynomial in the size of the input \cite{SCHU62}.
However, even in this restricted setting, the decidability of
\cref{sf-polyregular:problem} remains open~\cite[Concluding remarks]{CDTL23}.

In \cite{CDTL23}, the authors introduced the class of \kl{$\Rel$-polyregular
functions} ($\ZPoly$) as a subclass of \kl{$\Rel$-rational series} that
generalizes \kl{$\Nat$-polyregular functions} by allowing negative outputs, and
showed that membership in the \emph{star-free} subclass $\ZSF$ inside $\ZPoly$
is decidable \cite[Theorem V.8]{CDTL23}. The paper also provides numerous
characterizations of $\ZSF$ inside $\ZRat$, hinting at the possibility of
defining a notion of \emph{star-free $\Rel$-rational series} that would
generalize this notion beyond polynomial growth functions.

Although these results on $\ZRat$ could not be immediately leveraged to decide
$\NSF$ inside $\NPoly$, it was conjectured that $\NPoly \cap \ZSF =
\NSF$~\cite[Conjecture 7.61]{DOUE23}: intuitively, the extra computing power
granted by negative outputs is \emph{orthogonal} to the extra computing power
granted by periodic behaviors. It was believed that understanding the
membership problem of $\NPoly$ inside $\ZPoly$, that is, a restricted version
of \cref{n-in-z-rat:problem}, would be a key step towards proving $\NPoly \cap
\ZSF = \NSF$, which itself would give hope in designing an algorithm for
\cref{sf-polyregular:problem}. We illustrate in
\cref{previously-known-inclusions:fig} the known inclusions and related open
problems between the discussed classes of functions.

\begin{figure}
    \centering
    \begin{tikzpicture}[
    xscale=1.5,
    cls/.style={minimum width=2cm, minimum height=1cm},
    axis/.style={
        ->,
    },
    decidable/.style={
        thick, Prune, double, ->,
        postaction={
            decorate,
            decoration={markings,
                        mark=at position 0.5 with {\node[midway,above=0.2cm,sloped]{decidable};}
            }
        }
    },
    inclusion/.style={
        dashed,
        thick,
        Prune,
        ->,
        postaction={
            decorate,
            decoration={markings,
                        mark=at position 0.5 with {\node[midway,above,sloped]{?};}
            }
        }
    },
    group/.style={
        draw, thick,
        rounded corners=3mm,
        opacity=0.05,
    },
    polygrowth/.style={
        fill=D4,
    },
    starfree/.style={
        fill=D5,
    },
    zoutput/.style={
        fill=D2,
    },
    noutput/.style={
        fill=D3,
    }
    ]
    \node[cls] (SF)  at (0,  -9) {$\SF$};
    \node[cls] (P)   at (4,  -9) {$\Poly$};
    \node[cls] (ZSF) at (0, -3) {$\ZSF$};
    \node[cls] (NSF) at (0, -6) {$\NSF$};
    \node[cls] (ZP)  at (4, -3) {$\ZPoly$};
    \node[cls] (NP)  at (4, -6) {$\NPoly$};
    \node[cls] (ZR)  at (8, -3) {$\ZRat$};
    \node[cls] (NR)  at (8, -6) {$\NRat$};

    \draw[decidable] (ZSF) to (ZP);
    \draw[decidable] (ZP) to (ZR);
    \draw[decidable] (NP) to (NR);
    \draw[decidable] (NP) to (P);
    \draw[decidable] (NSF) to (SF);

    \draw[inclusion] (NP) to 
        node[midway,right] {\cite[Open question 5.55]{DOUE23}}
    (ZP);
    \draw[inclusion] (NSF) to 
        (ZSF);
    \draw[inclusion] (NR) to 
        node[midway,right] {\cref{n-in-z-rat:problem}}
        (ZR);
    \draw[inclusion] (SF) to 
        node[midway,below] {\cref{sf-polyregular:problem}}
        (P);
    \draw[inclusion] (NSF) to 
        node[midway,below] {\cite[Conjecture 7.61]{DOUE23}}
    (NP);

\end{tikzpicture}
    \caption{
        Decidability and inclusions of classes of functions,
        arranged along two axes. The first one is the complexity
        of the output alphabet ($\Rel$, $\Nat$, $\Sigma$). The second
        one is the allowed computational power
        (\kl{star-free polyregular functions}, \kl{polyregular functions}, 
        \kl{rational series}).
        Arrows denote strict inclusions,
        and effectiveness (both in terms of decidability and of effective
        representation) is represented by thick double arrows. Inclusions that are
        suspected to be effective are represented using a dashed arrow together with a
        question mark.
    }
    \label{previously-known-inclusions:fig}
\end{figure}

\subsection{Contributions.}
\label{contributions:sec}

In this paper, we work under the extra
assumption of \emph{commutativity}, that is, assuming that the function is
invariant under the permutation of its input. In this setting, we prove that
$\NPoly \cap \ZSF = \NSF$ \cite[Conjecture 7.61]{DOUE23} and design an
algorithm that decides whether a function in $\ZPoly$ is in $\NPoly$ \cite[Open
question 5.55]{DOUE23}. As a consequence, the upper left square of
\cref{previously-known-inclusions:fig} has all of its arrows decidable and with
effective conversion procedures under this extra assumption. Because
\kl{$\Rel$-rational series} with \emph{polynomial growth} are exactly
\kl{$\Rel$-polyregular functions} \cite{CDTL23}, this can be seen as decision
procedure for \cref{n-in-z-rat:problem} under the extra assumption of
\emph{commutativity} and \emph{polynomial growth}. Similarly, our results
provide an algorithm for \cref{sf-polyregular:problem} under the extra
assumption of \emph{commutativity} and \emph{unary output alphabet}.

As an intermediate step, we provide a complete and decidable characterization
of polynomials in $\Rat[\vec{X}]$ that can be computed using $\NRat$ (resp.
$\ZRat)$. These characterizations uncover a fatal flaw in the proof of a former
characterization of such polynomials \cite[Theorem 3.3, page 4]{KARH77}. We
also prove that this previous results holds for polynomials with at most two
indeterminates (\cref{lem:correct-covered-2}), which may explain why it was not
detected earlier. 

Failure of this previous characterization is welcomed, as it implied that it
was not possible to decide if a polynomial belonged to $\NRat$
(\cref{undecidable-non-negative:lem}), therefore proving \cref{n-in-z-rat:problem}
to be undecidable.
On the contrary, our new (decidable)
characterizations provide effective descriptions of polynomials that can be
expressed in $\ZRat$ as those obtained using integer combinations of products
of \emph{binomial coefficients} (called \kl{integer binomial polynomials},
defined page \kpageref{integer binomial polynomial}) and similarly for $\NRat$
by introducing the notion of \kl{strongly natural binomial polynomials}
(defined page \kpageref{strongly natural binomial polynomial}), which we
believe has its own interest. In turn, these characterizations demonstrate that
polynomials expressible by $\ZRat$ (resp. $\NRat$) are exactly those
expressible by $\ZSF$ (resp. $\NSF$), that is, polynomials are inherently
\emph{star free} functions.

There are two limitations of our work: the assumption of \emph{polynomial
growth}, and the assumption of \emph{commutative input}. As a first step
towards removing the latter assumption, we show that non-commutative
polyregular functions in $\NPoly$ (resp. $\NSF$) have a canonical
representation in terms of their \emph{residuals}. While aperiodicity is not
clearly visible on the canonical representations we propose, they form a first
step towards a better understanding of general functions in $\NPoly$.

\subsection{Related problems}
\label{related-problems:sec}

Let us now provide more context on
\cref{n-in-z-rat:problem,sf-polyregular:problem}, and relate them to other
subjects of interest in the literature of polyregular functions and rational
series. This will also clarify the scope of our contributions.

\paragraph*{Fatou extensions and rational series.} \AP\ Let us complete the
picture by mentioning that the analogue of \cref{n-in-z-rat:problem} for $\Rat$
and $\Rel$ is decidable precisely because $\Rat$ is a \kl{Fatou extension} of
$\Rel$~\cite[Chapter 7, Theorem 1.1]{BERE10}: on the one hand, one can guess a word $w$ such that $f(w)
\notin \Rel$ to disprove membership in $\ZRat$, and on the other hand, one can
guess a $\ZRat$ representation of $f$ and check equality to prove membership
in $\ZRat$. Furthermore, it is known that $\Rat_+$ (non-negative rational
numbers) is a \kl{Fatou extension} of $\Nat$~\cite[Chapter 7, Theorem 2.2]{BERE10}, hence $\ZRat \cap
\mathbb{Q}_+\mathsf{Series} = \NRat$. These two results allow us to relate
\cref{n-in-z-rat:problem} to another open problem in the area of weighted
automata, namely, deciding whether a given $\Rat$-rational series is in fact a
$\Rat_+$-rational series, which is of importance in the area of probabilistic
automata. In the following \cref{q-qplus-iff-z-nat:remark}, we prove that this
problem is essentially the same as \cref{n-in-z-rat:problem}. However, the
reduction does not preserve \kl{polynomial growth}, hence it does not fit our
context of \kl{polyregular functions}.

\begin{remark}
  \label{q-qplus-iff-z-nat:remark}
  \Cref{n-in-z-rat:problem} is interreducible with 
  the problem of deciding whether a given $\Rat$-rational series is in
  $\Rat_+$-rational series. 
\end{remark}
\begin{proof}
  On the one hand, given $f \in \ZRat$, one can 
  ask whether $f$ is a $\Rat_+$-rational series. If the answer is yes, 
  then $f$ is in fact in $\NRat$ because $\ZRat \cap \Rat_+ = \NRat$. If the 
  answer is no, then in particular $f$ is not in $\NRat$.

  On the other hand, given $f \in \RatRat$, one can compute 
  a number $\alpha \in \Nat$ such that 
  $f' \defined w \mapsto \alpha^{|w|} f(w)$ is in $\ZRat$: 
  it is obtained by multiplying all the weights of a $\Rat$-weighted
  automaton computing $f$ by $\alpha$, and $\alpha$ can be chosen as the
  least common multiple of the denominators of the weights of this automaton.
  Now, if $f'$ is in $\NRat$, then $f$ is in $\Rat_+$ because one can simply 
  divide all the weights of a $\NRat$-weighted automaton computing $f'$ by
  $\alpha$. Conversely, if $f$ is a $\Rat_+$-rational series, then 
  $f'$ is also a $\Rat_+$-rational series, but $f'$ is in $\ZRat$, hence
  $f'$ is in $\NRat$.
\end{proof}

\paragraph*{Aperiodicity for rational series.} Beware that several
non-equivalent notions of aperiodicity coexist for rational series
\cite{REUT80,DRGA19,CDTL23}, and that we only refer to the one applying to
rational series of polynomial growth introduced in \cite{CDTL23}. Let us
briefly illustrate these difference and justify our choice to consider the one
introduced in \cite{CDTL23}. 

First, \cite[Proposition III.4.4, p. 477]{REUT80} introduces a notion of
aperiodicity for rational series based on a characterization of aperiodic
regular languages by Schützenberger \cite{SCHU65}: these are precisely those
languages that can be obtained from letters, $\Sigma^*$, and closures under
sum, concatenation, and complementation. Now, the analogue for rational series
was chosen to be subalgebra of $\RatRat$ generated by the geometric series
\cite[Paragraph 2.a]{REUT80} and characteristic functions of letters. In particular, the
function $f \colon \set{a}^* \to \Rel$ mapping $w$ to $(-1)^{|w|}$ is aperiodic
in this sense, and because one can consider sums and products of such functions, the
indicator function of the language $(aa)^*$ is also aperiodic in this sense
as $\ind{(aa)^*} = \frac{1}{2}(1 + f)$.
We
believe this to be problematic since $(aa)^*$ is not aperiodic as a language.

Then, \cite[Section 5]{DRGA19} introduces a notion of aperiodicity for rational
series based on weighted first-order logic. Similarly to the previous notion,
this notion of aperiodicity allows one to express $w \mapsto (-1)^{|w|}$, by
using the weighted first-order formula $\prod_x (-1)$. Because weighted
first-order logic is closed under sums, one can express $2 \times \ind{(aa)^*}$
using the formula $(\prod_x 1) + (\prod_x (-1)))$, which exhibits a similar
problem as before.

On the contrary, the notion of aperiodicity for rational series introduced in
\cite{CDTL23} is based on the notion of \emph{star-free polyregular functions},
and in particular the pre-image of any fixed output by an aperiodic rational
series is an aperiodic language \cite{BOJA18}.

\paragraph*{Aperiodicity for transducer models.} For string-to-string
transducer models, aperiodicity has a well established definition, that
generalizes the classical notion of aperiodic languages, and satisfies (in
particular) that an aperiodic function $f \colon \Sigma^* \to \Gamma^*$ has the
property that for every $c \in \Gamma^*$, the pre-image $f^{-1}(c)$ is an
aperiodic language, or more generally, that the pre-image of an aperiodic
language is aperiodic. Let us briefly survey some of these transducer models 
and their aperiodic counterparts.\footnote{This list is far from exhaustive.} 

\AP
We will only consider in this survey the following well-known models of
string-to-string functions: Mealy Machines ($\intro*\Mealy$) \cite{MEAL55},
sequential functions ($\intro*\Sequential$) \cite{SCHU77}, \intro{rational functions}
($\intro*\Rational$) \cite{EILE74}, \intro{regular functions}
($\intro*\Regular$) \cite{ENMA02}, and \reintro{polyregular functions}
($\reintro*\Poly$) \cite{BOKL19}. These models stem from different
characterizations of regular languages, respectively deterministic finite state
automata for $\Mealy$ and $\Sequential$, non-deterministic finite state
automata for $\Rational$, two-way deterministic finite state automata with
output for $\Regular$, and ``pebble automata'' for $\Poly$ \cite{BOKL19}. While
all these automata model define regular languages, they express different
classes of string-to-string functions, all strictly included one into the
other, as depicted in \cref{computational-models:fig}.

For all of these
models, an \emph{aperiodic} counterpart can be defined (respectively,
$\intro*\AMealy$, $\intro*\ASequential$, $\intro*\ARational$,
$\intro*\ARegular$, $\reintro*\SF$), lifting the correspondence between
counter-free automata, star-free languages and aperiodic monoids to the
functional setting~\cite{FKT14,BOJA14,CADA15,DJR16,BDK18,BOKL19,DGK21}. The
inclusions between these classes of functions are all known to be strict, and
we depicted in \cref{computational-models:fig} the status of the
\emph{membership problem} associated to these strict inclusions (that is, given
a function $f$, decide if $f$ can be computed by a function of a proper
subclass). Note that in \cref{computational-models:fig}, if $\mathsf{B}$ is a
proper subclass of $\mathsf{C}$, then the \emph{aperiodic} variant of
$\mathsf{B}$ is precisely the intersection of $\mathsf{B}$ with the
\emph{aperiodic} variant of the larger class $\mathsf{C}$
(\cref{aperiodic-intersection:remark}). As an example, to conclude that a
function can be computed by an \emph{aperiodic sequential function}
($\ASequential$) it is enough to prove that it is sequential ($\Sequential$),
and computed by an \kl{aperiodic polyregular function} ($\SF$).

\begin{remark}
    \label{aperiodic-intersection:remark}
    For all classes of functions $\mathsf{A}$, $\mathsf{B}$ appearing in 
    \cref{computational-models:fig},
    if $\mathsf{B}$ is a proper subclass of $\mathsf{C}$, then $\mathsf{Aperiodic B} =
    \mathsf{B} \cap \mathsf{Aperiodic C}$.
\end{remark}
\begin{proof}[Proof sketches.]
  Because we are still in the introduction, and we did not introduce formally 
  the computational models of \cref{computational-models:fig}, we only provide
  proof sketches that will illustrate how \emph{semantic characterizations} of
  the classes of functions can be used to prove such statements.

  First, $\AMealy = \Mealy \cap \ASequential$ is a folklore statement, that can
  be recovered from the fact that a Mealy machine is a sequential function that
  only outputs one letter at a time, or equivalently, a sequential function
  that is \emph{size-preserving}. 
  The proof that a sequential function that is size-preserving can be
  re-arranged to be computed by a Mealy machine 
  follows from standard pumping arguments, and only works by re-arranging 
  the outputs of the transitions, not by changing the underlying automaton.
  As a consequence, if the sequential function is aperiodic, then the
  resulting Mealy machine is also aperiodic.

  The equalities $\ASequential = \Sequential \cap \ARational$ and $\ARational =
  \Rational \cap \ARegular$ are consequences of the work of \cite[Section 4]{RESCH95}. The key property
  underlying these proofs is that aperiodic functions (in these two models) can be characterized
  by the following pumping property:
  \begin{equation}
    \label{eq:aperiodic-sequential-pumping}
    \exists N_0 \in \Nat,
    \forall u,v,w \in \Sigma^*,
    \exists z_1, z_2, z_3 \in \Sigma^*,
    \forall n \in \Nat,
    f(u v^{n + N_0} w) = z_1 z_2^n z_3
    \quad . 
  \end{equation}
  Since this property is model-independent, we obtain in particular 
  the desired equalities.

  Finally, $\ARegular = \Regular \cap \SF$ is proved in \cite{BOJA14},
  since it proves that $\ARegular$ is exactly the subclass of
  $\SF$ that has linear growth. Hence, if $f \in \Regular \cap \SF$
  then in particular $f$ has linear growth, hence $f \in \ARegular$.
\end{proof}

In light of \cref{computational-models:fig}, and
\cref{aperiodic-intersection:remark}, the conjecture that $\NPoly \cap \ZSF =
\NSF$ \cite[Conjecture 7.61]{DOUE23} can be thought of as yet another instance
of the general phenomenon that the \emph{aperiodicity} is orthogonal to
\emph{computing power}. For completeness, let us briefly mention that for some
subclass of \kl{polyregular functions} called \intro{polyblind functions}
\cite{LENP21}, one can find examples for which \cref{aperiodic-intersection:remark}
does not generalize: the function $f \colon \set{a}^* \to \set{a}^*$ mapping $a^n$ to $a^{
  n(n+1)/2 }$ is \kl{star-free polyregular} (concatenating all prefixes of the
input), and \kl{polyblind} (for every even letter, it outputs the whole input),
but not \intro{star-free polyblind}.\footnote{This example was communicated to us
by Nguyễn Lê Thành Dũng (Tito).}

\begin{figure}
    \centering
    \begin{tikzpicture}[
        xscale=1.2,
        decidable/.style={
            ->, thick, color=black
        },
        unknown/.style={
            ->, dashed, color=gray
        }
        ]
        \node (mealy)  at (0, 2) {$\Mealy$};
        \node (seq)    at (2, 2) {$\Sequential$};
        \node (rat)    at (4, 2) {$\Rational$};
        \node (reg)    at (6, 2) {$\Regular$};
        \node (poly)   at (8, 2) {$\Poly$};
        \node (amealy) at (0,0) {$\AMealy$};
        \node (aseq)   at (2,0) {$\ASequential$};
        \node (arat)   at (4,0) {$\ARational$};
        \node (areg)   at (6,0) {$\ARegular$};
        \node (apoly)  at (8,0) {$\SF$};
        \draw[decidable] (mealy) --
            node[midway, above] {folklore}
            (seq);
        \draw[decidable] (seq)   -- 
            node[midway, above] {\cite{CHOF03,RESCH95}}
            (rat);
        \draw[decidable] (rat)   -- 
            node[midway, above] {\cite{FGRS13,BGMP15,RESCH95}}
            (reg);
        \draw[decidable] (reg)   -- 
            node[midway, above] {\cite{BOKL19}}
            (poly);
        \draw[decidable] (amealy) --
            (aseq);
        \draw[decidable] (aseq) --
            (arat);
        \draw[decidable] (arat) --
            (areg);
        \draw[decidable] (areg) --
            (apoly);
        \draw[decidable] (amealy) --
            node[midway, left] {\cite{SCHU65,MNPA71}}
            (mealy);
        \draw[decidable] (aseq) -- 
            node[midway, left] {\cite{CHOF03,RESCH95}}
            (seq);
        \draw[decidable] (arat) -- 
            node[midway, left] {\cite{FGL16,FGLM18,RESCH95}}
            (rat);
        \draw[unknown] (areg) --
            node[midway, left] {$\simeq$\cite{BOJA14}}
            (reg);
        \draw[unknown] (apoly) --
            node[midway, left] {?}
            (poly);
    \end{tikzpicture}
    \caption{
        In this picture, arrows denote strict inclusions. Plain
        arrows signify that the corresponding membership problem is decidable,
        while dashed arrows signify that the corresponding membership problem
        is conjectured to be decidable. In the case of \kl{regular functions}
        ($\Regular$),
        the decidability has only been obtained under a different notion
        of equivalence of functions \cite{BOJA14}.
    }
    \label{computational-models:fig}
\end{figure}

\subparagraph*{Outline of the paper.} In \cref{preliminaries:sec}, we provide a
combinatorial definition of \kl{$\Nat$-polyregular functions} (resp.
\kl{$\Rel$-polyregular functions}), show that one can decide if a function $f
\in \ZPoly$ is \kl{commutative} (\cref{decidable-commutative-poly:lemma}). In
\cref{polynomials:sec}, we provide a counterexample to the flawed result of
\cite[Theorem 3.3, page 4]{KARH77} (\cref{thm:counter-example}), and correct it
by providing effective characterizations of polynomials computed by $\ZRat$
(\cref{integer-binomial-polynomials:cor}) and $\NRat$
(\cref{decide-rat-poly-npoly:cor}). Then, in \cref{beyond-polynomials:sec}, we
answer positively to \cite[Open question 5.55]{DOUE23}
(\cref{decidable-n-poly:thm}) and to \cite[Conjecture 7.61]{DOUE23}
(\cref{zsf-npoly-nsf:thm}), both under the extra assumption of
\emph{commutativity}. We open the way to removing the assumption of
\emph{commutativity} by providing canonical representations of functions in
\cref{canonical-models:sec}, where we provide a semantic characterization of
\kl{$\Nat$-polyregular functions} using their \kl{residuals}
(\cref{non-commutative-npoly:thm}).

\section{Preliminaries}
\label{preliminaries:sec}

\AP The capital letters $\Sigma,\Gamma$ denote
fixed alphabets, i.e. finite set of letters, and $\Sigma^*, \Gamma^*$ (resp.
$\Sigma^+, \Gamma^+$) are the set of words (resp. non-empty words) over
$\Sigma, \Gamma$. The empty word is written $\varepsilon \in \Sigma^*$. When $w
\in \Sigma^*$ and $a \in \Sigma$, we let $\card{w} \in \Nat$ be the length of
$w$, and $\card[a]{w}$ be the number of occurrences of $a$ in $w$. 

\AP We assume that the reader is familiar with the basics of automata theory,
in particular the notions of monoid morphisms, idempotents in monoids, monadic
second-order ($\intro*\MSO$) logic and first-order ($\intro*\FO$) logic over finite words
(see e.g. \cite{THOM97}). As aperiodicity will be a central notion of this
paper, let us recall that a monoid $M$ is \intro(monoid){aperiodic} whenever
for all $x \in M$, there exists $n \in \Nat$ such that $x^{n+1} = x^n$. If the
monoid $M$ is finite, this $n$ can be uniformly chosen for all elements in $M$.

\AP We use the notation $\commute \colon \Sigma^* \to \Nat^\Sigma$ for the map
that counts occurrences of every letter in the input word (that is, computes
the Parikh vector) namely: $ \commute[w] \defined \seqof{a \mapsto
\card[a]{w}}{a \in \Sigma}$. Given a set $X$, a function $f \colon \Sigma^* \to
X$ is \intro{commutative} whenever for all $u \in \Sigma^*$, for all
permutations $\sigma$ of $\set{1, \dots, \card{w}}$, $f(\sigma(u)) = f(u)$.
Equivalently, it is \reintro{commutative} whenever there exists a map $g \colon
\Nat^\Sigma \to X$ such that $g \circ \commute = f$.

\AP Let $k \in \Nat$, and let $\Sigma$ be a finite alphabet. Given a function
$\eta \colon \set{1, \dots, k} \to \Sigma$, we define the $\eta^\dagger \colon
\Nat^k \to \Sigma^*$ as $\eta^\dagger(\vec{x}) \defined \eta(1)^{x_1} \dots
\eta(k)^{x_k}$. A function $f \colon \Nat^k \to X$ is \intro{represented} by a
\kl{commutative} function $g \colon \Sigma^* \to X$ if there exists a map $\eta
\colon \set{1, \dots, k} \to \Sigma$ such that $g \circ \eta^\dagger = f$. This
notion will be useful to formally state that a polynomial ``is'' a
\kl{commutative} \kl{polyregular function}. For instance, the polynomial
function $P(X,Y) = X \times Y$ is \kl{represented} by the \kl{commutative}
function $g \colon \set{a,b}^* \to \Rel$ defined by $g(w) \defined \card[a]{w}
\times \card[b]{w}$.

\subsection{Polynomials} \AP A polynomial $P \in \Rel[X_1, \dots, X_k]$ is
\intro{non-negative} when for all non-negative integer inputs $n_1, \dots, n_k
\geq 0$, the output  $P(n_1, \dots, n_k)$ of the polynomial is non-negative. In
the case of at most three indeterminates, we use variables $X,Y,Z$ instead of
$X_1, X_2, X_3$ to lighten the notation. Beware that we do not consider
negative values as input, as the numbers $n_i$ will ultimately count the number
of occurrences of a letter in a word. As an example, the polynomial $(X - Y)^2$
is \kl{non-negative}, and so is the polynomial $X^3$, but the polynomial $X^2 -
2X$ is not.

\AP A \intro{monomial} is a product of indeterminates and integers. For
instance, $XY$ is a \kl{monomial}, $3 X$ is a \kl{monomial}, $-Y$ is a
\kl{monomial}, but $X + Y$ and $2X^2 + XY$ are not. Every polynomial $P \in
\Rel[X_1, \dots, X_n]$ decomposes uniquely into a sum of \kl{monomials}. A
\kl{monomial} $S$ \intro{divides} a \kl{monomial} $T$, when $S$ divides $T$
seen as polynomials in $\Rat$. For instance, $2X$ \kl{divides} $XY$, $-YZ$
\kl{divides} $X^2 Y Z^3$, and $Y$ does not \kl{divide} $X$. In the
decomposition of $P \in \Rel[X_1, \dots, X_k]$, a \kl{monomial} is a
\intro{maximal monomial} if it is a maximal element for the \kl{divisibility
preordering} of \kl{monomials}. In the polynomial $P(X,Y) \defined X^2 - 2XY +
Y^2 + X + Y$, the set of \reintro{maximal monomials} is $\set{X^2,  -2 XY,
Y^2}$. For instance, the \kl{non-negative} \kl{monomials} of $P(X,Y) \defined
(X - Y)^2$ are $X^2$ and $Y^2$.

\subsection{Polyregular Functions}
\label{polyregular:sec}

\AP Because the functions of interest in this paper have output in $\Nat$ or
$\Rel$, we will only provide the definition of \intro{polyregular functions}
for these two output semigroups, and we refer the reader to \cite{BOKL19} for
the general definition of \kl{polyregular functions} and their aperiodic
counterpart, the \intro{star-free polyregular functions}. We chose in this
paper to provide a combinatorial description of polyregular functions with
commutative outputs because it will play nicely with our analysis on
polynomials. This description is very similar in shape to the \emph{finite
counting automata} introduced by \cite{SCHU62}. 

\begin{definition}[$\Rel$-polyregular functions {\cite{CDTL23}}]
    \label{nat-rel-poly:def}
    Let $d \in \Nat$. The set $\intro*\ZPoly[d]$ of polyregular
    \intro{$\Rel$-polyregular functions} of degree at most $d$,
    is the set of functions $f \colon \Sigma^* \to \Rel$
    such that
    there exists a finite monoid $M$,
    a morphism $\mu \colon \Sigma^* \to M$,
    and a function $\pi \colon M^{d+1} \to \Rel$
    satisfying for all $w \in \Sigma^*$:
    \begin{equation*}
        f (w) = \pi^\dagger (w) \defined
        \sum_{w = u_1 \cdots u_{d+1}} \pi(\mu(u_1), \dots, \mu(u_{d+1}))
        \quad .
    \end{equation*}
    We call $\pi$ the \intro{production function} of $f$.
    If the function $\pi$ has codomain $\Nat$,
    then $f$ is \intro{$\Nat$-polyregular} of degree at most $d$,
    i.e., $f \in \intro*\NPoly[d]$.
    If the monoid $M$ is \kl(monoid){aperiodic}
    then
    the function $f$ is \intro{star-free $\Rel$-polyregular}
    ($\intro*\ZSF[d]$), resp. \intro{star-free $\Nat$-polyregular} ($\intro*\NSF[d]$).
\end{definition}

\AP We complete \cref{nat-rel-poly:def} by letting $\reintro*\NPoly \defined
\bigcup_{d \in \Nat} \NPoly[d]$, and similarly for $\ZPoly$, $\NSF$, and
$\ZSF$. In order to illustrate these definitions, let us provide an example of
an \kl{$\Nat$-polyregular function} computed using a finite monoid in
\cref{size-of-word-nsf:ex}. Let us also introduce in
\cref{q-polynomial-n-poly:ex} a function that serves as an example of
\emph{division} computed by a \kl{$\Nat$-polyregular function}.

\begin{example}
  \label{constant-function-nsf:ex}
  Let $c \in \Nat$, and 
  $f$ be the constant function $f \colon w \mapsto c$. 
  Then, $f \in \NSF[0]$.
\end{example}
\begin{proof}
  We consider the trivial monoid $M \defined (\set{1}, \times)$ (which is \kl(monoid){aperiodic})
  the morphism $\mu \colon \Sigma^* \to M$ defined by $\mu(w) \defined 1$,
  and the \kl{production function} $\pi \colon M \to \Nat$
  defined by $\pi(1) \defined c$.
\end{proof}

\begin{example}
    \label{size-of-word-nsf:ex}
    The map $f \colon w \mapsto \card{w} + 1$
    belongs to $\NSF[1]$.
\end{example}
\begin{proof}
  Similarly to \cref{constant-function-nsf:ex},
  let us define $M \defined (\set{1}, \times)$ which is 
    a finite \kl{aperiodic monoid}, $\mu \colon \Sigma^* \to M$
    defined by $\mu(w) \defined 1$, and
    $\pi \colon M^2 \to \Nat$
    that is the constant function equal to $1$.
    We check that for all $w \in \Sigma^*$:
    $
        \pi^\dagger(w)
        =
        \sum_{uv = w} 1
        =
        \card{w} + 1
        = f(w)
        $.
\end{proof}

\begin{example}
    \label{q-polynomial-n-poly:ex}
    Let $f \colon \Sigma^* \to \Nat$ be the function that maps a word $w$
    to the number of distinct pairs of positions in $w$,
    i.e., $f(w) = \binom{\card{w}}{2} = \card{w}(\card{w} - 1) / 2$.
    Then, $f \in \NSF[2]$.
\end{example}
\begin{proof}
    Let us remark that the set $P_w$ of distinct pairs of positions
    $i < j$ in a word $w$ is in bijection with the set $D_w$ of
    decompositions of the form $w = xyz$,
    where $x$ and $y$ are non-empty, 
    via the map $(i,j) \mapsto (w_{1,i}, w_{i+1,j}, w_{j+1,\card{w}})$.
    Let us write $M \defined (\set{0,1}, \max)$
    which is a finite aperiodic monoid, and $\mu \colon \Sigma^* \to M$
    that maps the empty word $\varepsilon$ to $0$ and the other words to $1$.
    Then, let us define $\pi \colon M^3 \to \Nat$
    via $\pi(x,y,z) = x \times y$.
    We conclude because:
    \begin{align*}
        \pi^\dagger(w) 
        &\defined
        \sum_{xyz = w} \pi(\mu(x), \mu(y), \mu(z)) 
        = 
        \sum_{xyz = w \wedge x \neq \varepsilon \wedge y \neq \varepsilon} 1
        = 
        \card{D_w}
        =
        \card{P_w}
        = f(w) \quad .
        \qedhere
    \end{align*}
\end{proof}

\AP One of the appeals of $\NPoly$ and $\ZPoly$ are the numerous
characterizations of these classes in terms of logic, weighted automata, and
the larger class of \kl{polyregular functions} \cite{CDTL23,DOUE23}. In this
paper, the main focus will be the connection to \kl{weighted automata}, which is
based on the notion of \emph{growth rate}. The \intro{growth rate} of a
function $f \colon \Sigma^* \to \Rel$ is defined as the minimal $d$ such that
$\card{f(w)} = \bigO\left(\card{w}^d\right)$. If such a $d$ exists, we say that
the function $f$ has \intro{polynomial growth}. It turns out that for all $k
\in \Nat$, $\ZPoly[d]$ (resp. $\NPoly[d]$) are precisely functions in $\ZRat$
(resp. in $\NRat)$ that have growth rate at most $d$.

\begin{lemma}[{\cite[Theorem 5.22]{DOUE23}}]
    \label{polyregular-polynomial-growth:lemma}
    Let $f \in \ZRat$. The following are equivalent:
    \begin{enumerate}
        \item $f \in \ZPoly[d]$.
        \item $f$ has \kl{polynomial growth} of degree at most $d$.
    \end{enumerate}
    And similarly for $\NPoly$ and $\NRat$.
\end{lemma}

Let us introduce some compositional properties of
\kl{$\Rel$-polyregular functions} that will be used in this paper
to construct \kl{$\Rel$-polyregular functions}.
\begin{lemma}[{\cite[Theorem II.20]{CDTL23}}]
    \label{stability-polyregular:lemma}
    Let $d \geq 1$,
    $f,g \in \NPoly[d]$ (resp. $\ZPoly[d]$, $\NSF[d]$, $\ZSF[d]$),
    $L$ be a star-free language over $\Sigma^*$,
    and $h \colon \Sigma^* \to \Gamma^*$ be a \kl{polyregular function}
    (resp. a \kl{star-free polyregular function}).
    Then, the following 
    are also in $\NPoly[d]$ (resp. $\ZPoly[d]$,
    $\NSF[d]$, $\ZSF[d]$):
    $f \circ h$,
    $f + g \defined w \mapsto f(w) + g(w)$,
    $f \times g \defined w \mapsto f(w) \times g(w)$,
    $\ind{L} \times f$.
    Furthermore, the above constructions preserve \kl{commutativity}.
\end{lemma}

Let us briefly state that \kl{commutativity} is a decidable property of
\kl{$\Rel$-rational series}, hence of \kl{$\Rel$-polyregular functions}. As a
consequence, we are working inside a relatively robust and decidable subclass of
\kl{$\Rel$-rational series}.

\begin{lemma}
    \label{decidable-commutative-poly:lemma}
    \label{decidable-commutative-rat:lemma}
    Let $f \in \ZRat$. One can decide if 
    $f$
    is \kl{commutative}.
\end{lemma}
\begin{proof}
    Remark that the group of permutations of $\set{1, \dots, n}$ is generated by
    the cycle $c \defined (n,1, \dots, n-1)$ and the transposition $t \defined (1, 2)$.
    As a consequence, a function $f$ is commutative if and only if
    $f \circ c = f = f \circ t$.
    When $f$ is a \kl{rational series},
    $f \circ c$ and $f \circ t$ are both \kl{rational series} that can be
    effectively computed from $f$,\footnote{
        This can be done by guessing the second (resp. last) letter of the input word, 
        remembering the first letter in a state, and 
        then running the original automaton for $f$ on the modified input, checking at the second position 
        (resp. the end of the word)
        if the guess was correct.
    } and since equivalence
    of rational series is decidable 
    \cite[Corollary 3.6]{BERE10},
    we have obtained a decision procedure.
\end{proof}

\section{$\Nat$-rational Polynomials}
\label{polynomials:sec}

\AP In this section, we will completely characterize which polynomials in
$\Rat[\vec{X}]$ are \kl{represented} by \kl{$\Nat$-rational series} (resp.
\kl{$\Rel$-rational series}). To that end, we start by characterizing these
classes for polynomials in $\Rel[\vec{X}]$. We say that a polynomial $P \in
\Rel[X_1, \dots, X_n]$ is an \intro{$\Nat$-rational polynomial} if it is
\kl{represented} by a \kl{$\Nat$-rational series}. It is an easy check that
polynomials with coefficients in $\Nat$ are \kl{$\Nat$-rational polynomials}
(\cref{n-poly-n-poly:example}). However, \cref{negative-but-npoly:ex} provides
a polynomial with negative coefficients that is an \kl{$\Nat$-rational
polynomial}. The problem of characterizing \kl{$\Nat$-rational polynomials} was
claimed to be solved in \cite{KARH77}, using the \cref{karh:def} to
characterize \kl{$\Nat$-rational polynomials}, as restated in \cref{karh:thm}.

\begin{lemma}[restate=n-poly-n-poly:example,label=n-poly-n-poly:example]
    Let $P \in \Nat[\vec{X}]$. Then, $P$
    is an \kl{$\Nat$-rational polynomial}.
\end{lemma}
\begin{proof}
    The polynomials of $\Nat[\vec{X}]$
    are obtained from basic functions (constant in $\Nat$,
    $X_i$ for some $1 \leq i \leq n$)
    by products and sums (\cref{stability-polyregular:lemma}). Because the basic functions are
    \kl{represented} by \kl{star-free $\Nat$-polyregular} functions (see
    \cref{size-of-word-nsf:ex}),
    that are closed under these operations, we conclude.
\end{proof}

\begin{example}[restate=negative-but-npoly:ex,label=negative-but-npoly:ex]
    The polynomials $X$, $X^2 + 3$,
    and $X^2 - 2X + 2$
    are \kl{$\Nat$-rational polynomials},
    but $- X$ is 
    not an \kl{$\Nat$-rational polynomial}.
\end{example}
\begin{proof}
    The function $w \mapsto |w|$ is a \kl{$\Nat$-polyregular function}.
    Thus, 
    $P(X) \defined X$ is
    a \kl{$\Nat$-rational polynomial}. Similarly,
    $w \mapsto |w|^2 + 3$ is a \kl{$\Nat$-polyregular function},
    showing that $Q(X) \defined X^2 + 3$
    is a \kl{$\Nat$-rational polynomial}.

    Remark that $P = (X-1)^2 + 1$. Let $f  \colon \Sigma^* \to \Nat$
    that maps $aw \mapsto |w|^2$, and $\varepsilon \mapsto 0$.
    It is clear that $f \in \NPoly$, and that 
    $P$ is \kl{represented} by
    $\ind{|w| \geq 1} \times f + \ind{|w| = 0} + 1$.

    Finally, 
    $T(X) \defined - X$ cannot be 
    a \kl{$\Nat$-rational polynomial} as \kl{$\Nat$-polyregular functions}
    are non-negative.
\end{proof}

\begin{definition}[{\cite[Section 3, page 3]{KARH77}}]
    \label{karh:def}
    The class $\intro*\CoveredPoly[\vec{X}]$
    is the class of polynomials $P \in \Rel[\vec{X}]$
    that are \kl{non-negative}
    and such that every \kl{maximal monomial} is \kl{non-negative}.
    When the indeterminates are clear from the context, we write
    this class $\reintro*\CoveredPoly$.
\end{definition}

\begin{faketheorem}[{\cite[Theorem 3.3, page 4]{KARH77}}] 
    \label{karh:thm}
    Let $P \in \Rel[\vec{X}]$ be a polynomial. Then,
    $P$ is an \kl{$\Nat$-rational polynomial}
    if and only if 
    $P \in \CoveredPoly$.
\end{faketheorem}

Before giving a counterexample to the above statement, let us first exhibit in
\cref{non-neg-not-nrat:ex} some
\kl{non-negative} polynomial that is not an \kl{$\Nat$-rational polynomial}.
While the example will not be in $\CoveredPoly$, it illustrates the key
difference between \kl{non-negative} polynomials and \kl{$\Nat$-rational
polynomials}. In order to derive this example, we will need the following
fundamental result about the pre-image of regular languages by \kl{polyregular
functions}.\footnote{ In this particular case, one could have considered more
generally \kl{$\Nat$-rational series}, and replaced regular languages over a
unary alphabet by semi-linear sets. } Before that, let us remark that if a
polynomial $P$ is \kl{represented} by a \kl{$\Nat$-rational series}, then it is
in fact \kl{represented} by a \kl{$\Nat$-polyregular function} thanks to
\cref{polyregular-polynomial-growth:lemma}.

\begin{theorem}[{\cite[Theorem 1.7]{BOJA18}}]
    \label{pre-image-regular:fact}
    The pre-image of a regular language by a (string-to-string) \kl{polyregular function}
    is a regular language.
\end{theorem}

\begin{example}
    \label{non-neg-not-nrat:ex}
    Let $P(X, Y) \defined (X - Y)^2$.
    Then $P$ is \kl{non-negative}, but is
    not an \kl{$\Nat$-rational polynomial}.
    Indeed, assume by contradiction that
    $f \in \NPoly$ \kl{represents} $P$ over the alphabet $\Sigma \defined \set{a,b}$.
    Then, $f^{-1}(\set{0})$ is a regular language
    (\cref{pre-image-regular:fact}),
    but $f^{-1}(\set{0}) = \setof{ w \in \Sigma }{ \card[a]{w} = \card[b]{w} }$
    is not.
\end{example}

Please note that the same argument cannot be leveraged for proving that $P$ is
not represented by a \kl{$\Rel$-rational series}: \cref{pre-image-regular:fact}
only holds for
\emph{string-to-string} functions, and is applied to the specific case where
the output alphabet is $\set{1}$, i.e., where the output of the function
belongs to $\set{1}^*$ which is isomorphic to $\Nat$.

\AP
Let us now design a counterexample to \cref{karh:thm} by suitably tweaking
\cref{non-neg-not-nrat:ex} to ensure that the polynomial not only is
\kl{non-negative}, but also belongs to $\CoveredPoly$.
\label{def:bad-polynomial}
We define $\intro*\BadPoly(X,Y,Z) \defined Z (X + Y)^2 + 2 (X - Y)^2$.

\begin{lemma}
    \label{thm:counter-example}
    The polynomial $\BadPoly$ belongs to $\CoveredPoly$,
    but is not an \kl{$\Nat$-rational polynomial}.
    As a corollary, \cite[Theorem 3.3]{KARH77}, restated
    in \cref{karh:thm}, is false
    when allowing at least $3$ indeterminates.
\end{lemma}
\begin{proof}
    It is clear that $\BadPoly$ is \kl{non-negative}. We can expand
    the expression of $\BadPoly$ to 
    obtain
    $\BadPoly = ZX^2 + ZY^2 + 2ZXY + 2X^2 -4XY + 2Y^2$.
    The \kl{maximal monomials} of $P$
    are $ZX^2$, $ZY^2$, and $2ZXY$, all of which are
    \kl{non-negative}.

    Assume by contradiction that $\BadPoly$ is an \kl{$\Nat$-rational polynomial}.
    Let $\Sigma \defined \set{a,b,c}$ be a finite alphabet.
    There exists a \kl{commutative}
    \kl{$\Nat$-polyregular function} $f \colon \Sigma^* \to \Nat$
    such that for all $w \in \Sigma^*$,
    $\BadPoly(\card[a]{w}, \card[b]{w}, \card[c]{w}) = f(w)$.
    Remark that for all $x,y,z \geq 0$, $\BadPoly(x,y,z) = 0$
    if and only if $z(x+y)^2 = -2 (x-y)^2$. Hence,
    $\BadPoly(x,y,z)=0$ if and only if $z = 0$ and $x = y$, or 
    $z \neq 0$, and $x = y = 0$.
    Now, let us consider the language $L \defined \setof{w}{ f(w) = 0}$. By the
    above computation, we conclude that $L = \setof{ w \in \set{a,b}^* }{
    \card[a]{w} = \card[b]{w} } \cup \set{ c }^*$.
    Because $L \cap \set{a,b}^*$ is not a regular language,
    we
    conclude that $L$ is not a regular language.
    However, $L = f^{-1}(\set{0})$ is a regular language
    (\cref{pre-image-regular:fact}). 
\end{proof}

\AP We will discuss at the end of \cref{sec:poly-to-n-poly} why
\cref{thm:counter-example} is minimal in the number of indeterminates, which
first requires us to provide a \emph{correct} analogue of \cref{karh:thm}. Our
counterexample relies on the fact that $\CoveredPoly$ is not stable under
fixing indeterminates, while \kl{$\Nat$-rational polynomials} are. Indeed, the
polynomial $\BadPoly$ satisfies
$\BadPoly(X,Y,1) = 3X^2 + 3Y^2 - 2XY$, which has a negative coefficient for a
\kl{maximal monomial}. Let us now prove that closing
$\CoveredPoly$ under variable assignments is enough to recover from
\cref{karh:thm}. We use the following notation to fix the value of some
indeterminate, if $P(X,Y)$ is a polynomial in $\Rel[X,Y]$, then
$\intro*\restr{P(X,Y)}{X = 1}$ is the polynomial $P(1,Y) \in \Rel[Y]$. More
generally, if $\nu$ is a partial function from $\vec{X}$ to $\Nat$, written
$\nu \colon \vec{X} \topartial \Nat$, the restriction $\restr{P(\vec{X})}{\nu}$
is the polynomial with indeterminates $\vec{Y} \defined \vec{X} - \dom(\nu)$
obtained by fixing the variables of the domain of $\nu$.

\begin{definition}
	The class $\intro*\CorrectPoly[\vec{X}]$ is the collection of
	polynomials $P \in \Rel[\vec{X}]$ such that,
	for every partial function $\nu \colon \vec{X} \topartial \Nat$,
	every \kl{maximal monomial} of
	$\restr{P}{\nu}$ is \kl{non-negative}.
\end{definition}

First, let us remark that $\CorrectPoly \subseteq \CoveredPoly$, because
polynomials in $\CorrectPoly$ are \kl{non-negative}. We also remarked at the
beginning of this section that our counterexample $\BadPoly$ provided in
\cref{thm:counter-example} is not in $\CorrectPoly$. The rest of the
section is mainly concerned with proving the following corrected version of
\cref{karh:thm}.

\begin{theorem}[label=corrected-version:thm,]
	Let $P \in \Rel[\vec{X}]$.
	The following are equivalent:
	\begin{enumerate}
		\item \label{corrected-1:item} $P \in \CorrectPoly$,
		\item \label{corrected-2:item} $P$ is \kl{represented} by a \kl{$\Nat$-rational series},
		\item \label{corrected-3:item} $P$ is \kl{represented} by a \kl{$\Nat$-polyregular function},
		\item \label{corrected-4:item} $P$ is \kl{represented} by a \kl{star-free $\Nat$-polyregular function},
	\end{enumerate}
	Furthermore, the properties are decidable, and conversions effective.
\end{theorem}

\Cref{corrected-version:thm} is surprising given the
fact that it is not possible to decide whether a polynomial $P \in \Rel[\vec{X}]$
is \kl{non-negative} or if a polynomial $P$ belongs to $\CoveredPoly$
(\cref{undecidable-non-negative:lem}), by reduction to the undecidability of
Hilbert's Tenth Problem \cite{HILB1902,MATI1970}. That is, $\CorrectPoly$ is a
decidable class that strictly contains $\Nat[\vec{X}]$, and is contained in the
undecidable classes $\CoveredPoly$ and the class of \kl{non-negative}
polynomials.

\begin{remark}[restate=undecidable-non-negative:lem,label=undecidable-non-negative:lem]
    Checking whether a polynomial
    $P \in \Rel[\vec{X}]$ is \kl{non-negative} is undecidable.
    Similarly, checking whether a polynomial $P \in \Rel[\vec{X}]$
    belongs to $\CoveredPoly$ is undecidable.
\end{remark}
\begin{proof}
    Recall that by the undecidability of the Hilbert's Tenth Problem \cite[Problem 10 page 22]{HILB1902}, it is
    undecidable whether a given polynomial $P \in \Rel[X_1, \dots, X_k]$ has a solution in
    $\Rel^k$ \cite{MATI1970,DAVIS1973}.
    The result follows by observing that
    checking whether a polynomial $P \in \Rel[\vec{X}]$ has no solutions in $\Rel^k$ if and only 
    if $P^2 - 1$ is \kl{non-negative}. 

    Let us prove that checking whether $P \in \CoveredPoly$ is undecidable. Let
    $P \in \Rel[\vec{X}]$ be a polynomial, and let us write $P = P_+ - P_-$
    where both $P_+$ and $P_-$ belong to $\Nat[\vec{X}]$. Let $Z$ be a fresh
    indeterminate. We define $Q \defined P_+ + (Z - 1) P_-$. By construction, $Q$
    has \kl{non-negative} \kl{maximal monomials}. Furthermore, $Q$ itself is
    \kl{non-negative} if and only if $\restr{Q}{Z = 0} = P$ is \kl{non-negative},
    since for every $n > 0$, $\restr{Q}{Z = n} \in \Nat[\vec{X}]$ is
    \kl{non-negative}. We conclude that 
    $P$ is \kl{non-negative} if and only if $Q$ is \kl{non-negative},
    if and only if $Q \in \CoveredPoly$.
\end{proof}

The proof of \cref{corrected-version:thm} is divided into two parts. First, we
provide in \cref{sec:n-poly-to-poly} a fine combinatorial understanding of what
functions can be computed in $\NPoly$ and $\ZPoly$. This allows us to prove
that \kl{$\Nat$-rational polynomials} are in $\CorrectPoly$
(\cref{n-rat-correct:lem}). Then, in \cref{sec:poly-to-n-poly} we will show how
to compute polynomials in $\CorrectPoly$ using $\NPoly$
(\cref{lem:correct-to-n-rat}). Finally, we will next generalize
\cref{corrected-version:thm} to polynomials in $\Rat[\vec{X}]$ in
\cref{sec:rel-to-rat}.

\subsection{From $\Nat$-polyregular functions to polynomials}
\label{sec:n-poly-to-poly}

\AP Let us prove that \kl{$\Nat$-rational polynomials} are in $\CorrectPoly$.
This fact follows from the correct implication in the statement of
\cref{karh:thm}, but
we provide a self-contained proof using a refinement of the classical
combinatorial \emph{pumping arguments} for $\ZPoly$ \cite[Lemma 4.16]{CDTL23} and
$\NPoly$ \cite[Lemma 5.37]{DOUE23}. We take extra care to reprove in our
upcoming \cref{n-poly-combinatorics:lem} a strong statement that has
two main goals. Our first goal is to highlight the role of \kl{commutative}
\kl{polyregular functions} in the broader study of \kl{polyregular functions},
which is done by reformulating the traditional pumping argument as a
composition property involving said functions, which will be reused in the
upcoming \cref{k-combinatorial:def,ultimately-polynomial:def}
of
\cref{beyond-polynomials:sec}. Our second goal
is to give a precise shape of the functions that arise from such \emph{pumping
arguments}, which was lacking in former similar statements.

\AP To address our first goal, let us define that a function $q$ is a
\intro{pumping pattern} from $\Nat^p$ to $\Sigma^*$
whenever there exists words $\alpha_0, \dots, \alpha_p \in \Sigma^*$, and words
$u_1, \dots, u_p \in \Sigma^*$, such that $q(X_1, \dots, X_p) = \alpha_0
\prod_{i = 1}^p u_i^{X_i} \alpha_i$. That is, $q$ is syntactically defined by a
non-commutative monomial over the monoid $\Sigma^*$. \kl{Pumping patterns}
are \kl{commutative} \kl{polyregular functions}.

\AP Our second goal is achieved by understanding that \kl{$\Nat$-polyregular
functions} essentially compute binomial coefficients, as illustrated by the
polynomial $X(X-1)/2 = \binom{X}{2}$
of \cref{q-polynomial-n-poly:ex}. A \intro{simple binomial function} is a
function of the form $\binom{X - \ell}{k}$, where $\ell$ and $k$ are natural
numbers. We extend this to \intro{natural binomial functions} that are obtained
by considering $\Nat$-linear combinations of products of \kl{simple binomial
functions}, that is, we consider functions that have the following shape:
$f(x_1, \dots, x_k) = \sum_{i = 1}^n n_i \prod_{j = 1}^k \binom{x_j -
p_{i,j}}{k_{i,j}}$. Beware that $\binom{X - \ell}{k}$ is defined to be $0$ when
$X \leq \ell$, and is therefore not a polynomial. Let us immediately
prove that \kl{simple binomial functions}  can be \kl{represented} in $\NSF$,
generalizing \cref{q-polynomial-n-poly:ex}.
Conversely, we prove in \cref{n-poly-combinatorics:lem} that,
    when suitably pumping a \kl{$\Nat$-polyregular function}, one
always obtains \kl{natural binomial functions}.

\begin{lemma}
	\label{binomial-function-star-free:lem}
	Let $F$ be a \kl{simple binomial function} from $\Nat^k$ to $\Nat$.
	Then it is \kl{represented} by a \kl{star-free polyregular function}.
\end{lemma}
\begin{proof}
	Because of the stability properties of
	$\NSF$ (\cref{stability-polyregular:lemma}), we only need to
	check that given $r,s \in \Nat$,
	the function $x \mapsto \binom{x - r}{s}$ is \kl{represented}
	by a function $f_{r,s} \in \NSF$.
	Let us prove it when $r = 0$, since the other functions
	can be obtained by translating $f_{r,s}$.
    By definition, $\binom{x}{s} = \card{P_{x,s}}$,
    where $P_{x,s} \defined \setof{ (x_1, \dots, x_s) \in \Nat^s }{ 1 \leq x_1 < \cdots < x_s \leq x }$.
    Let us proceed as in \cref{q-polynomial-n-poly:ex} and define 
    $D_{w,s} \defined \setof{ (u_1, \dots, u_s, u_{s+1}) \in (\Sigma^+)^s \times \Sigma^* }{ w = u_1 u_2 \cdots u_s u_{s+1} }$.
    It is clear that $D_{a^x,s}$ is in bijection with $P_{x,s}$ for all $x \in \Nat$
    using the map $(x_1, \dots, x_s) \mapsto (a^{x_1}, \dots, a^{x_s})$.
    Now, using the monoid $M \defined (\set{0,1}, \max)$
    and the morphism $\mu(\varepsilon) \defined 0$ and $\mu(a) \defined 1$,
    one can compute $\card{D_{w,s}}$ 
    as $\pi^\dagger (w)$ where $\pi \colon M^{s+1} \to \Nat$ is defined by
    $\pi(m_1, \dots, m_s, m_{s+1}) \defined m_1 \times \cdots \times m_s$.
    We conclude that $f_{0,s} \in \NSF[s]$ is a \kl{star-free polyregular function}.
\end{proof}

\begin{lemma}[restate=n-poly-combinatorics:lem,label=n-poly-combinatorics:lem]
    \proofref{n-poly-combinatorics:lem}
	Let $f$ be an \kl{$\Nat$-polyregular} function.
	There exists a computable $\omega \in \Nat_{\geq 1}$
	such that for all \kl{pumping patterns}
	$q \colon \Nat^p \to \Sigma^*$,
	there exists a computable \kl{natural binomial function} $F$
	such that:
	\begin{equation*}
		f \circ q(\omega X_1, \dots, \omega X_p)
		=
		F
		\quad
		\text{ over } (\Nat_{\geq 1})^p
		\quad .
	\end{equation*}
\end{lemma}

The multiplicative factor $\omega$ is necessary in
\cref{n-poly-combinatorics:lem}. Indeed, the function $f \colon \set{a}^* \to
\Nat$ defined as $0$ when the input is of odd length and $1$ when the input is
of even length is \kl{$\Nat$-polyregular}, but $f(a^X)$ is not a polynomial. We
can trade off this multiplicative factor for a constant term addition under the
extra assumption that the function is \kl{star-free polyregular}, as described
in the following \cref{n-sf-combinatorics:lem}. This lemma is not immediately
of use, but is crucial for the upcoming characterization of $\Nat$-rational
polynomials in \cref{decide-rat-poly-npoly:cor}, which in turn is a key
ingredient of our main \cref{zsf-npoly-nsf:thm}.

\begin{lemma}[restate=n-sf-combinatorics:lem,label=n-sf-combinatorics:lem]
    \proofref{n-sf-combinatorics:lem}
	Let $f$ be a \kl{star-free $\Nat$-polyregular} function.
	There exists a computable $s \in \Nat_{\geq 1}$
	such that for all \kl{pumping patterns}
	$q \colon \Nat^p \to \Sigma^*$,
	there exists a computable \kl{natural binomial function} $F$
	such that:
	\begin{equation*}
		f \circ q(X_1 + s, \dots, X_p + s)
		=
		F
		\quad
		\text{ over } \Nat^p
		\quad .
	\end{equation*}
\end{lemma}

Because \kl{natural binomial functions} behave as polynomials with
\kl{non-negative} \kl{maximal monomials} on large enough inputs, we can
conclude from \cref{n-poly-combinatorics:lem}
that \kl{$\Nat$-rational polynomials} are in $\CorrectPoly$.

\begin{corollary}
	\label{n-rat-correct:lem}
	Let $P \in \Rel[X_1, \dots, X_p]$ be an \kl{$\Nat$-rational polynomial}.
	Then,
	$P \in \CorrectPoly$.
\end{corollary}
\begin{proof}
    Let $f$ be a \kl{commutative} \kl{$\Nat$-rational series}
	with domain defined as $\Sigma \defined \set{a_1, \dots, a_p}$
	that \kl{represents} $P$.
    Because $f$ has \kl{polynomial growth},
    $f \in \NPoly$ 
    (\cref{polyregular-polynomial-growth:lemma}).
	Using \cref{n-poly-combinatorics:lem},
	there exists a number $\omega \in \Nat_{\geq 1}$
	and \kl{natural binomial function} $Q$
	such that
	for all $n_1, \dots, n_p \geq 1$:
	\begin{equation*}
		f\left(
		\prod_{i = 1}^p a_i^{\omega n_i}
		\right)
		= Q(n_1, \dots, n_p)
		= P(\omega n_1, \dots, \omega n_p)
		\quad .
	\end{equation*}
    For large enough values of $X$, the \kl{simple binomial function}
	$\binom{X - p}{k}$ coincides with a polynomial whose leading coefficient
	is $1/k!$ which is \kl{non-negative}.
	We conclude that
	the \kl{maximal monomials} of
	$P(\omega X_1, \dots, \omega X_p)$ are \kl{non-negative},
	and since $\omega \geq 1$, we conclude that
	the \kl{maximal monomials} of $P$ have \kl{non-negative} coefficients.

	For every partial valuation $\nu \colon \vec{X} \topartial \Nat$,
	the polynomial $\restr{P}{\nu}$ continues to be represented
	by a \kl{$\Nat$-polyregular function}, namely
	$f_u \colon w \mapsto f(uw)$ 
    where $w$ belongs to a restricted alphabet.
    As a consequence,
	the \kl{maximal monomials} of
	$\restr{P}{\nu}$ are also \kl{non-negative},
	and
	we have proven that $P \in \CorrectPoly$.
\end{proof}

\subsection{From polynomials to $\Nat$-polyregular functions}
\label{sec:poly-to-n-poly}

\AP This section is devoted to proving that polynomials in $\CorrectPoly$ can
be \kl{represented} by \kl{star-free $\Nat$-polyregular functions}. The key
lemma of this section is \cref{lem:correct-to-n-rat}, which is proved by
induction on the number of indeterminates of a given polynomial $P$. In order
to prove that result, we use the combinatorial
\cref{derivation-stabilises-correct:lem} that allows us to transform a
polynomial $P \in \CorrectPoly$ into a polynomial in $\Nat[\vec{X}]$ through a
well-chosen translation of the indeterminates. This argument is based on the
notion of \intro{discrete derivative} which is built by translating the domain
of the polynomial. To that end, let us write $\intro*\translate{K}$ for the
\intro{translation function} that maps a polynomial $P \in \Rel[X_1, \dots,
X_k]$ to the polynomial $P(X_1 + K, \dots, X_k + K)$.

\begin{definition}
	\label{discrete-derivative:def}
	Let
	$K \in \Nat$,
	and
	$P \in \Rel[\vec{X}]$ be a polynomial,
	then
	$
		\intro*\Diff{K}{P} \defined
		\translate{K}(P) - P
	$.
\end{definition}

\begin{lemma}
	\label{all-positive-derivative:lem}
	Let $P \in \Nat[\vec{X}]$ that is non-constant, and $K \in \Nat$,
	then $\Diff{K}{P} \in \Nat[\vec{X}]$ and all of its
	coefficients are (positive) multiples of $K$.
	Furthermore, every monomial that strictly divides some monomial of $P$
	appears in $\Diff{K}{P}$.
\end{lemma}
\begin{proof}
	We prove the result for monomials, as it extends
	to $\Nat$-linear combinations by linearity.
	Let $P = \prod_{i = 1}^k X_i^{\alpha_i}$ be a monomial.
	Notice that $\translate{K}(P) = \prod_{i = 1}^k (X_i + K)^{\alpha_i}$,
	and using a binomial expansion
	we list all the possible divisors of $P$,
	all of which with coefficients that are positive integers and multiples of $K$ except the coefficient
	of the maximal monomial (equal to $P$ itself) which is $1$.
	As a consequence, $\translate{K}(P) - P$ is simply
	obtained by removing this maximal monomial, which concludes the proof.
\end{proof}

\begin{lemma}
	\label{derivation-stabilises-correct:lem}
	Let $P \in \CorrectPoly$,
	$P_1$ be the sum of \kl{maximal monomials} of $P$,
	and $P_2 \defined P - P_1$ be the sum of
	non-maximal monomials of $P$.
	There exists a computable number $K \in \Nat$,
	such that
	$Q \defined (\Diff{K}{P_1} + \translate{K}(P_2)) \in \Nat[\vec{X}]$.
\end{lemma}
\begin{proof}
	Let us first tackle the specific case where $P$ is a constant polynomial.
	In this case, $P_1 = P$ and $P_2 = 0$.
	Furthermore, $\Diff{K}{P_1} = 0$ for all $K \in \Nat$.
	We conclude that $\Diff{K}{P_1} + \translate{K}(P_2) = 0$
	for all $K \in \Nat$, hence belongs to $\Nat[\vec{X}]$. Selecting $K = 0$
	we conclude.
	Assume now that $P$ is not a constant polynomial. We will use
	\cref{all-positive-derivative:lem} on a well-selected value of $K$. Let us
	write $\alpha$ to be the maximal absolute value of a coefficient in $P$.
	Let $D$ be the number of unitary monomials that divide some monomial
	appearing in $P$. We can now define $K \defined D \times \alpha$,
	and let
	$Q \defined (\Diff{K}{P_1} + \translate{K}(P_2))$.
	Remark that $\Diff{K}{P_1}$ is already in $\Nat[\vec{X}]$,
	and the constant coefficient of $\translate{K}(P_2)$ is also
	in $\Nat$.
	For any other monomial of $P_2$, by the maximality of $P_1$,
	it strictly divides some monomial of $P_1$, and
	equals some monomial of $\Diff{K}{P_1}$ up to a multiplication by a factor in $\Rat$. Because every monomial
	of $\Diff{K}{P_1}$ has a coefficient that is a multiple of $K = \alpha \times D$, we can
	compensate every monomial of $P_2$ by a monomial of $\Diff{K}{P_1}$.
	Therefore,
	$Q \in \Nat[\vec{X}]$.
\end{proof}

\begin{lemma}
	\label{lem:correct-to-n-rat}
	Let $P \in \Rel[\vec{X}]$.
	If $P \in \CorrectPoly$, then $P$ is \kl{represented}
	by a \kl{star-free $\Nat$-polyregular function},
	which is computable given $P$.
\end{lemma}
\begin{proof}
	We prove the result by induction on the number of indeterminates of $P$.
	In the proof, we write $\vec{X}$ for the indeterminates appearing in $P$,
	that is, we assume without loss of generality that all indeterminates are used.

	\textbf{Base case:} If the (unique) \kl{maximal monomial} of $P$ is a
	constant term. Since $P \in \CorrectPoly$, $P = n \in \Nat$, and therefore
	$P$ is \kl{represented} by a constant \kl{star-free $\Nat$-polyregular
		function}.

	\textbf{Induction:} Assume that $P$ is not a constant polynomial, and let
	us write $P = P_1 + P_2$ where $P_1$ is the sum of the \kl{maximal
		monomials} of $P$. We compute a bound $K$ such that $Q \defined
		(\Diff{K}{P_1} + \translate{K}(P_2)) \in \Nat[\vec{X}]$ 
        (\cref{derivation-stabilises-correct:lem}). In particular, $Q$ is
	\kl{represented} by a \kl{star-free $\Nat$-polyregular function} using
	\cref{n-poly-n-poly:example}, the latter being effectively computable from
	$Q$. Let us now remark that $P_1 \in \Nat[\vec{X}]$, and is therefore
	(effectively) \kl{represented} by a \kl{star-free $\Nat$-polyregular
		function} (using again \cref{n-poly-n-poly:example}). As a consequence,
	$\translate{K}(P) = P_1 + Q$ is (effectively) \kl{represented} by a
	function $f_\Delta$.

	For all partial valuations $\nu \colon \vec{X} \topartial \set{0, \dots,
			K}$ fixing at least one indeterminate, one can use the induction hypothesis
	to compute a \kl{star-free $\Nat$-polyregular function} $f_\nu$ that
	\kl{represents} $\restr{P}{\nu}$. This is possible because we assumed that
	all indeterminates in $\vec{X}$ are used in $P$.

	Let us assume that the alphabet over which the (\kl{commutative}) functions
	$f_\Delta$ and $f_\nu$ are defined is $\set{a_1, \dots, a_k}$, with $a_i$
	representing the indeterminate $X_i$ of the polynomials. Now, let us define
	by case analysis the following \kl{commutative} \kl{star-free
		$\Nat$-polyregular function}, defined on words $w$ of the form $w \defined
		a_1^{x_1} \cdots a_k^{x_k}$, with $x_1, \dots, x_k \geq 0$.

	\begin{equation*}
		f(w) \defined
		\begin{cases}
			f_{[X_i \mapsto x_i]}(w)                     & \text{ if } \exists i \in \set{1, \dots, k}, x_i \leq K \\
			f_\Delta(a_1^{x_1 - K} \cdots a_k^{x_k - K}) & \text{ otherwise }
		\end{cases}
		\quad .
	\end{equation*}
	Remark that
	$f$ is a \kl{commutative} \kl{star-free $\Nat$-polyregular function}
	that
	\kl{represents} $P$.
\end{proof}

While \cref{lem:correct-to-n-rat} provides an effective conversion procedure,
it does not explicitly state that the membership is decidable to keep the proof
clearer. A similar proof scheme can be followed to conclude that membership is
decidable, and even show that elements in $\CorrectPoly$ are, up to suitable
translations, polynomials in $\Nat[\vec{X}]$
(\cref{derivation-translation:lem}). Beware that partial applications are still
needed in this characterization, as \cref{bad-poly-translate:ex} illustrates. 

\begin{lemma}[restate=derivation-translation:lem,label=derivation-translation:lem]
	\proofref{derivation-translation:lem}
	Let $P \in \Rel[\vec{X}]$.
	There exists a computable number $K \in \Nat$
	such that the following are equivalent:
	\begin{enumerate}
		\item \label{d-t-correct:item} $P \in \CorrectPoly$,
		\item \label{d-t-transl:item}
		      for
		      all partial functions $\nu \colon \vec{X} \topartial \Nat$,
		      $\translate{K}(\restr{P}{\nu}) \in \Nat[\vec{X}]$,
		\item \label{d-t-transl-fin:item}
		      for all partial functions
		      $\nu \colon \vec{X} \topartial \set{0, \dots, K}$,
		      $\translate{K}(\restr{P}{\nu}) \in \Nat[\vec{X}]$.
	\end{enumerate}
	In particular, the above properties are decidable.
\end{lemma}

\begin{example}
	\label{bad-poly-translate:ex}
	The polynomial $\BadPoly$ is not a
	\kl{$\Nat$-rational polynomial},
	but is \kl{non-negative} and satisfies
	$\translate{10}(\BadPoly) \in \Nat[\vec{X}]$.
\end{example}

We now have all the tools to prove the corrected version of \cref{karh:thm}.

\begin{proofof}[corrected-version:thm]
	The implications
	\cref{corrected-4:item} $\implies$
	\cref{corrected-3:item} $\implies$
	\cref{corrected-2:item} are obvious.
	\cref{lem:correct-to-n-rat} proves
	\cref{corrected-1:item} $\implies$ \cref{corrected-4:item},
	while \cref{n-rat-correct:lem}
	proves
	\cref{corrected-2:item} $\implies$ \cref{corrected-1:item}.
	Note that the lemmas provide effective conversion procedures,
	and that
	\cref{derivation-translation:lem}
	also provides a decision
	procedure.
\end{proofof}

For completeness, let us remark that the counterexample of
\cref{thm:counter-example} uses three indeterminates, and this is not a
coincidence: in the particular cases of one or two indeterminates,  the classes
$\CorrectPoly$ and $\CoveredPoly$ coincide. In particular, the examples
appearing in \cite{KARH77} are not invalidated, as they all use at most two
indeterminates. Note that the equivalence is clear for the univariate case,
where being non-negative and having non-negative maximal coefficient clearly
imply being an \kl{$\Nat$-rational polynomial}.

\begin{lemma}
	\label{lem:correct-covered-2}
	$\CorrectPoly[X,Y] = \CoveredPoly[X,Y]$.
\end{lemma}
\begin{proof}
	It is clear that $\CorrectPoly[X,Y] \subseteq \CoveredPoly[X,Y]$,
	by considering the empty valuation $\nu \colon \set{X,Y} \topartial \Nat$.
	For the converse inclusion, let us consider $P(X,Y)$
	that is \kl{non-negative}, such that the \kl{maximal monomials}
	are  \kl{non-negative}.

	If we fix none of the variables, then the \kl{maximal monomials}
	are \kl{non-negative} by assumption. If we fix one of the
	variables, we can assume without loss of generality that we
	fix $X = k$ for some $k \in \Nat$.
	Then $P(k,Y)$ is a \kl{non-negative} \emph{univariate} polynomial,
	and therefore must either have a positive leading coefficient
	(which is the unique \kl{maximal monomial} in this case)
	or be constant equal to 0. In both cases, the \kl{maximal monomials}
	have positive coefficient.
	The same reasoning applies \emph{a fortiori} in the case where
	we fix the two indeterminates, leading to a constant polynomial.
\end{proof}

\subsection{From $\Rel$ to $\Rat$}
\label{sec:rel-to-rat}

\AP Let us complete our analysis of polynomials \kl{represented} by $\NRat$ or
$\ZRat$ by considering polynomials with coefficients in $\Rat$, and justify
that all the combinatorial work has already happened in $\Rel$ and $\Nat$.

From \cref{n-sf-combinatorics:lem}, we know
that the polynomials that can be computed by \kl{star-free $\Nat$-polyregular
functions} are going to coincide (on large enough inputs) with \kl{natural binomial
functions}. For that reason, we introduce the following ``polynomial
counterpart" of a binomial coefficient: given two numbers $\ell,k \in \Nat$,
$\intro*\pbinom{X - \ell}{k}$ defined as $(X - \ell) \cdots (X - \ell - k) /
k!$,\footnote{ In particular, $\pbinom{X - \ell}{k}$ is defined to be $1$ when
$k = 0$, and $X - \ell$ when $k = 1$. } that we call a \intro{binomial
monomial}, and we introduce \intro{natural binomial polynomials} as $\Nat$-linear
combinations of products of \kl{binomial monomials}, i.e., of the shape:
$P(X_1, \dots, X_k) = \sum_{i = 1}^n n_i \prod_{j = 1}^k \pbinom{X_j -
p_{i,j}}{k_{i,j}}$. Similarly, we introduce the class of \intro{integer
binomial polynomials}, which are obtained by $\Rel$-linear combinations of
products of \kl{binomial monomials}.

\AP
These definitions are justified by the classical result of Pólya that
characterizes polynomials $P$ in $\Rat[X]$ that are \intro{integer-valued} (i.e., are such
that $P(\Rel) \subseteq \Rel$) as \kl{integer binomial polynomials}
\cite{POLYA1915,CACHA1996}. Note
that this result straightforwardly extends to multiple indeterminates as we
prove in \cref{integer-binomial-polynomial:lem}.

\begin{lemma}[restate=integer-binomial-polynomial:lem,label=integer-binomial-polynomial:lem]
	\proofref{integer-binomial-polynomial:lem}
	Let $P \in \Rat[X_1, \dots, X_k]$ be a polynomial.
	Then, $P$ is an \kl{integer binomial polynomial} if and only if
	$P(\Rel^k) \subseteq \Rel$, if and only if $P(\Nat^k) \subseteq \Rel$.
\end{lemma}

As an immediate corollary,  we completely characterize the class of polynomials
in $\Rat[\vec{X}]$ that are \kl{represented} by $\ZPoly$ as the \kl{integer
binomial polynomials}.

\begin{theorem}
	\label{integer-binomial-polynomials:cor}
	Let $P \in \Rat[\vec{X}]$. Then, the following properties are equivalent:
	\begin{enumerate}
        \item \label{int-bin-3:item} $P$ is \kl{integer-valued},
        \item \label{int-bin-2s:item} $P$ is \kl{represented} by a \kl{$\Rel$-rational series},
		\item \label{int-bin-2:item} $P$ is \kl{represented} by a \kl{$\Rel$-polyregular function},
		\item \label{int-bin-1:item} $P$ is \kl{represented} by a \kl{star-free $\Rel$-polyregular function},
		\item \label{int-bin-0:item} $P$ is an \kl{integer binomial polynomial}.
	\end{enumerate}
	These properties are furthermore decidable.
\end{theorem}
\begin{proof}
	The implications \cref{int-bin-1:item}
	$\implies$ \cref{int-bin-2:item} $\implies$ 
    \cref{int-bin-2s:item}
    $\implies$ \cref{int-bin-3:item} are obvious.
	Now, \cref{int-bin-3:item} $\implies$ \cref{int-bin-0:item} is a direct
	consequence of \cref{integer-binomial-polynomial:lem}.
	Finally, \cref{int-bin-0:item} $\implies$ \cref{int-bin-1:item} follows from the fact
	that $\pbinom{X - p}{k}$ is \kl{represented} by a \kl{star-free $\Rel$-polyregular function}
	defined by hardcoding the output values (in $\Rel$) when $0 \leq X \leq p$, and
	using a \kl{star-free $\Nat$-polyregular function} when $X > p$
	(\cref{binomial-function-star-free:lem}).
	Because
	$\ZSF$ is closed under products and $\Rel$-linear combinations, we conclude.
\end{proof}

\AP Obtaining an analogue of \cref{integer-binomial-polynomials:cor} for
\kl{$\Nat$-polyregular functions} requires a bit more work, as polynomials in
$\Rat[\vec{X}]$ that are \kl{represented} by $\NPoly$ are not exactly
\kl{natural binomial polynomials}
(see \cref{natural-binomial-polynomial-positive-bad:ex}). To address the issues
raised by the former example, we introduce the notion of \intro{strongly
natural binomial polynomials}, as the polynomials $P \in \Rat[X]$ such that for
all partial valuation $\nu \colon \topartial \Nat$, $\restr{P}{\nu}$ is a
\kl{natural binomial polynomial}, which characterizes the class of polynomials
that are \kl{represented} by $\NPoly$ (\cref{decide-rat-poly-npoly:cor}).

\begin{example}[restate=natural-binomial-polynomial-positive-bad:ex,label=natural-binomial-polynomial-positive-bad:ex]
	The polynomial
	$Q(X,Y,Z) \defined \pbinom{X - 4}{1} \pbinom{Y}{1} \pbinom{Z}{1}
		+ 8 \pbinom{Y}{2} + 8 \pbinom{Z}{2} + 4$
    is a \kl{non-negative} \kl{natural binomial polynomial} in $\Rel[X,Y,Z]$,
	but cannot be computed by a \kl{star-free $\Nat$-polyregular function}.
    Indeed, 
    $Q(0,Y,Z)$ has a negative maximal monomial, hence $Q \not \in \CorrectPoly$,
    and we conclude using 
    \cref{corrected-version:thm}.
\end{example}

\begin{lemma}[restate=polyrec-integer-strong:lem,label=polyrec-integer-strong:lem]
    \proofref{polyrec-integer-strong:lem}
    Let $P \in \Rat[\vec{X}]$ be an \kl{integer-valued} polynomial, 
    and $n \in \Nat_{\geq 1}$ be
    such that $n P \in \CorrectPoly$.
    Then, $P$ is a \kl{strongly natural binomial polynomial}.
\end{lemma}

\begin{theorem}
	\label{decide-rat-poly-npoly:cor}
	Let $P \in \Rat[\vec{X}]$ be a polynomial with \emph{rational}
	coefficients and let $\alpha$ be the smallest number in $\Nat_{\geq 1}$
	such that $\alpha P \in \Rel[\vec{X}]$. Then, 
    the following are equivalent:
	\begin{enumerate}
        \item \label{rat-npoly-4:item} $\alpha P \in \CorrectPoly$ 
            and $P$ is \kl{integer-valued},
        \item \label{rat-npoly-0:item} $P$ is \kl{represented} by a \kl{$\Nat$-rational series},
		\item \label{rat-npoly-1:item} $P$ is \kl{represented} by a \kl{$\Nat$-polyregular function},
		\item \label{rat-npoly-2:item} $P$ is \kl{represented} by a \kl{star-free $\Nat$-polyregular function},
		\item \label{rat-npoly-3:item}
		      $P$ is a \kl{strongly natural binomial polynomial}.
	\end{enumerate}
	In particular, the properties are decidable.
\end{theorem}
\begin{proof}
    Let us first remark that $\NPoly \subseteq \NRat$,
    and that if $P$ is \kl{represented} by a function $f \in \NRat$,
    then said function has \kl{polynomial growth}, and in particular
    $f \in \NPoly$ thanks to \cref{polyregular-polynomial-growth:lemma}.
    As a consequence, \cref{rat-npoly-0:item} $\iff$ \cref{rat-npoly-1:item}.
    For the implication \cref{rat-npoly-1:item} $\implies$
    \cref{rat-npoly-4:item}, we obtain $\alpha P \in \CorrectPoly$ via
    \cref{corrected-version:thm} by remarking that \kl{$\Nat$-polyregular functions} have output in $\Nat$ and are closed under
    multiplication by a constant $\alpha \in \Nat$. The fact that $P$ is
    \kl{integer-valued} follows from
    \cref{integer-binomial-polynomials:cor} and the fact that
    $\NSF \subseteq \ZPoly$.
    The implication \cref{rat-npoly-4:item} $\implies$
    \cref{rat-npoly-3:item} is obtained thanks to \cref{polyrec-integer-strong:lem}.

    Let us now prove by induction on the number of indeterminates that
    \cref{rat-npoly-3:item} $\implies$
    \cref{rat-npoly-2:item}. Note that by construction, there exists a
    number $K \in \Nat$ such that when the input values of $P$ are all greater
    than $K$, $P$ coincides with a \kl{natural binomial function}, which is
    itself \kl{represented} by a \kl{star-free $\Nat$-polyregular function}. If
    some input value $X_i$ is set to a number $x_i \leq K$, then one can
    leverage the fact that $\restr{P}{X_i = x_i}$ remains a \kl{strongly
    natural binomial polynomial} to conclude by induction that $\restr{P}{X_i =
    x_i}$ is \kl{represented} by a \kl{star-free $\Nat$-polyregular function}.
    Combining these, we obtain a \kl{star-free $\Nat$-polyregular function}
    \kl{representing} $P$.

    Finally, the implication  \cref{rat-npoly-2:item} $\implies$ \cref{rat-npoly-1:item} is
    immediate as $\NSF \subseteq \NPoly$.
\end{proof}

Let us remark that \cref{decide-rat-poly-npoly:cor} shows that the
class of polynomials \kl{represented} by $\NPoly$ is the same as the class of
polynomials \kl{represented} by $\NSF$, which is a non-trivial statement that
will be reused in the study of more general \kl{commutative} functions in
$\ZPoly$.
\section{Beyond Polynomials}
\label{beyond-polynomials:sec}
\label{star-free:sec}

In this section, we leverage the decidability results of \cref{polynomials:sec}
to decide membership in $\NPoly$
inside $\ZPoly$ and membership in $\NSF$ inside $\NPoly$, both under the extra
assumption of \kl{commutativity}. 
To characterize $\NPoly$ inside $\ZPoly$ we introduce the notion of
\kl{$(k,\Nat)$-combinatorial} function
(\cref{k-combinatorial:def}), following the spirit of previous
characterizations of subclasses of $\ZPoly$ in terms of \emph{polynomial
pumping arguments} as in \cref{aperiodic-intersection:remark} and \cite{DOUE21,DOUE22,CDTL23}.

\begin{definition}
    \label{k-combinatorial:def}
    Let $k \in \Nat$, and $f \colon \Sigma^* \to \Rel$
    be a \kl{$\Rel$-polyregular function}. The function $f$ is 
    \intro{$(k,\Nat)$-combinatorial} if there exists $\omega \in \Nat$,
    such that
    for all
    \kl{pumping patterns} $q \colon \Nat^k \to \Sigma^*$,
    there exists a \kl{strongly natural binomial polynomial} $P$
    satisfying:
    \begin{equation*}
        f \circ q(\omega X_1,\dots, \omega X_k)
        = 
        P
        \quad 
        \text{ over } (\Nat_{\geq 1})^k
        \quad .
    \end{equation*}
\end{definition}

\AP Let us now introduce a decomposition of \kl{commutative}
\kl{$\Rel$-polyregular functions} into \kl{integer binomial polynomials}. Given
a number $\omega \in \Nat$, let us write $\intro*\ModuloTypes[\omega]^k$ for
the collection of pairs $(S, \vec{r})$ where $S \subseteq \set{1, \dots, k}$
and $r \in \set{0, \dots, \omega - 1}^k$. To a tuple $\vec{x} \in \Nat^k$, one
can associate its \intro{$\omega$-type}, written
$\intro*\moduloType[\omega](\vec{x})$, which is the pair $(S, \vec{r})$ where
$S = \set{i \in \set{1, \dots, k} \mid x_i \geq \omega}$ and $\vec{r} = (x_i
\mod \omega)_{i \in \set{1, \dots, k}}$.

\begin{lemma}[restate=decompose-polynomial:lem,label=decompose-polynomial:lem]
    Let $f \colon \Sigma^* \to \Rel$ be a \kl{commutative}
    \kl{$\Rel$-polyregular function},
    where we fix the alphabet $\Sigma = \set{a_1, \dots, a_k}$.
    There exists a computable
    $\omega \in \Nat_{\geq 1}$,
    and computable 
    \kl{integer binomial polynomials} 
    $P_{(S,\vec{r})} \in \Rat[\seqof{X_i}{i \in S}]$ for $(S,\vec{r}) \in \ModuloTypes[\omega]^k$,
    such that for all $\vec{x} \in \Nat^k$,
    \begin{equation*}
        f\left(
            \prod_{i = 1}^k a_i^{x_i}
        \right)
        = P_{(S, \vec{r})}
        \left(
            \seqof{\floor{x_i / \omega}}{i \in S}
        \right)
        \text{ where } (S, \vec{r}) = \moduloType[\omega](\vec{x})
        \quad .
    \end{equation*}
\end{lemma}
\begin{proof}
    Let us first remark that one only needs to consider the case where $f$ is a
    function in $\NPoly$, because every function $f \in \ZPoly$ can be decomposed
    as $f = f_+ - f_-$ where $f_+$ and $f_-$ 
    are \kl{$\Nat$-polyregular functions}, and because \kl{integer binomial
    polynomials} are closed under subtraction.
    Now, let us prove this result by induction on the number of letters in the alphabet.
    When the alphabet is empty, the result is trivial because $f$ can only be applied to $\varepsilon$,
    hence can be represented using a constant polynomial.

    Assume that the result holds for all functions with an alphabet of size at
    most $k-1$. We first apply \cref{n-poly-combinatorics:lem} to obtain a number
    $\omega$ such that for all \kl{pumping patterns} $q \colon \Nat^\ell \to
    \Sigma^*$, there exists a (computable) \kl{natural binomial function} $F_q$
    such that $f \circ q (\omega X_1, \dots, \omega X_\ell) = F_q$ over
    $(\Nat_{\geq 1})^\ell$.
    Given a tuple $\vec{r} \in \set{0, \dots, \omega - 1}^k$, and a 
    set $S \subseteq \set{1, \dots, k}$,
    let us define
    the \kl{pumping pattern} $q_{(S,\vec{r})} \colon \Nat^{\card{S}} \to \Sigma^*$ as
    $q_{(S, \vec{r})}(\seqof{X_i}{i \in S}) \defined \prod_{i = 1}^k a_i^{r_i} \cdot \prod_{i \in S} a_i^{X_i}$.
    Let us write $F_{(S,\vec{r})} \colon \Nat^{\card{S}} \to \Nat$ 
    for the associated
    \kl{natural binomial function} satisfying:
    \begin{equation*}
        f \left(
            \prod_{i = 1}^k a_i^{r_i}
            \cdot
            \prod_{i \in S} a_i^{\omega x_i}
        \right)
        = 
        F_{(S,\vec{r})}(\seqof{x_i}{i \in S})
        \quad 
        \forall \vec{x} \in (\Nat_{\geq 1})^{S}
        \quad .
    \end{equation*}

    Such a function $F_{(S,\vec{r})}$ coincides with a \kl{natural binomial
    polynomial} $P_{(S,\vec{r})} \in \Rat[\seqof{X_i}{i \in S}]$ on large
    enough inputs. Let us write $N_0 \geq 1$ be a number such that
    $F_{(S,\vec{r})}(\seqof{x_i}{i \in S}) = P_{(S,\vec{r})}(\seqof{x_i}{i \in
    S})$ for all $(S, \vec{r}) \in \ModuloTypes[\omega]^k$, for all $\vec{x}
    \in (\Nat_{\geq N_0})^{S}$. This $N_0$ is computable because one can
    collect upper bounds from the syntax of the \kl{natural binomial functions}
    $F_{(S,\vec{r})}$.
    We conclude that:
    \begin{equation*}
        f\left(
            \prod_{i = 1}^k a_i^{x_i}
        \right)
        = 
        P_{(S, \vec{r})}(\seqof{\floor{x_i / \omega}}{i \in S})
        \quad 
        \text{ where } (S, \vec{r}) = \moduloType[\omega](\vec{x})
        \text{ and }
        \vec{x} \in (\Nat_{\geq N_0 \omega})^k
        \quad .
    \end{equation*}

    This provides a representation for $f$ on large enough inputs, and we can
    apply the induction hypothesis on the functions $f_{a^n} \colon (\Sigma
    \setminus \set{a})^* \to \Nat$ in $\NPoly$ defined by $f_{a^n} (w) \defined
    f(a^n \cdot w)$, for every $a \in \Sigma$ and $n \leq N_0\omega$. Remark
    that \kl{integer binomial polynomials} are closed multiplying the input by
    a constant, hence that we can always move from a decomposition using
    $\omega$ to a decomposition using a multiple of $\omega$.
    Let us then consider as our new $\omega'$ the least common multiple of
    $\omega$, $N_0$, and all the $\omega_{a^n}$'s obtained for each $f_{a^n}$.

    We can now, given an entry $\vec{x} \in \Nat^k$ with
    $\moduloType[\omega'](x) = (S, \vec{r})$ recognize if a some letter
    $a$ appears less than $N_0 \omega$ times in the input, and apply the
    corresponding polynomials from the induction hypothesis if it is the case.
    Whenever all letters appear at least $N_0 \omega$ times, we can apply the
    polynomial $P_{(S,\vec{r})}$.
\end{proof}

\begin{theorem}
    \label{decidable-n-poly:thm}
    Let $k,d \in \Nat$, and $f \in \ZPoly[d]$ be \kl{commutative}
    over an alphabet of size $k$.
    Then, the following are equivalent:
    \begin{enumerate}
        \item \label{f-combinatorial:item} $f$ is \kl{$(k,\Nat)$-combinatorial},
        \item \label{f-npoly-combi:item} $f \in \NPoly[d]$,
    \end{enumerate}
    Furthermore, the properties are decidable,
    and conversions effective.
\end{theorem}
\begin{proof}
    Let $f \in \ZPoly[d]$ be \kl{commutative} over an alphabet of size $k$.
    We apply \cref{decompose-polynomial:lem} to compute an $\omega \in \Nat$ and
    \kl{integer binomial polynomials} $\seqof{P_{(S,\vec{r})}}{(S,\vec{r}) \in
    \ModuloTypes[\omega]^k}$ such that for all $\vec{x} \in \Nat^k$,
    $f\left(\prod_{i = 1}^k a_i^{x_i}\right) = P_{(S, \vec{r})}(\seqof{\floor{x_i / \omega}}{i \in S})$,
    where $(S, \vec{r}) = \moduloType[\omega](\vec{x})$.
    We are first going to prove that $f \in \NPoly[d]$ if and only if $P_{(S,
    \vec{r})}$ is a \kl{strongly natural binomial polynomial} for all $(S,
    \vec{r}) \in \ModuloTypes[\omega]^k$. This will also provide decidability
    of \cref{f-npoly-combi:item}, since one can decide whether a
    polynomial is \kl{strongly natural binomial polynomial} using
    \cref{decide-rat-poly-npoly:cor}.

    Assume that $f \in \NPoly$, then by definition, the polynomials $P_{(S,
    \vec{r})}$ are \kl{represented} by an \kl{$\Nat$-polyregular function},
    hence are \kl{strongly natural binomial polynomials}
    (\cref{decide-rat-poly-npoly:cor}). Conversely, if
    $P_{(S, \vec{r})}$ is a \kl{strongly natural binomial polynomial} for all
    $(S, \vec{r}) \in \ModuloTypes[\omega]^k$, then $f \in \NPoly$ because one
    can compute the \kl{$\omega$-type} of the input using a \kl{polyregular
    function}, and then compute the suitable \kl{strongly natural binomial
    polynomial} $P_{(S, \vec{r})}$ which is possible in $\NPoly$ thanks to
    \cref{decide-rat-poly-npoly:cor}. The resulting composition
    belongs to $\NPoly$ thanks to
    \cref{stability-polyregular:lemma},
    and we conclude that $f \in \NPoly[d]$ because it has \kl{growth rate} at
    most $d$
    (\cref{polyregular-polynomial-growth:lemma}).

    Note that the same proof scheme can be used to conclude that
    \cref{f-npoly-combi:item} implies
    \cref{f-combinatorial:item}. For the converse implication, we are
    going to introduce $\omega_2$ associated to the fact that $f$ is
    \kl{$(k,\Nat)$-combinatorial}. Because polynomials \kl{represented} by
    \kl{$\Nat$-polyregular functions} and \kl{integer binomial polynomials} are
    both closed under multiplication of their input by a constant factor, we can assume that
    $\omega = \omega_2$ in the decomposition of $f$.
    Now, consider $(S, \vec{r}) \in \ModuloTypes[\omega]^k$. 
    Notice that for all vectors $\vec{x} \in (\Nat_{\geq 1})^k$, 
    the vector $(x_1 \omega \ind{S}(1) + r_1, \dots, x_k \omega \ind{S}(k) + r_k)$ has
    \kl{$\omega$-type} $(S, \vec{r})$.
    In particular, the following
    equality holds:
    \begin{equation*}
        f\left(\prod_{i = 1}^k 
            a_i^{ x_i \omega \ind{S}(i) + r_i}
        \right)
        =
        P_{(S, \vec{r})}(\seqof{x_i}{i \in S})
        \quad  \forall \vec{x} \in (\Nat_{\geq 1})^k
        \quad .
    \end{equation*}

    Let us therefore consider the \kl{pumping pattern} $q \colon \Nat^k \to
    \Sigma^*$ that is simply defined as $q(X_1, \dots, X_k) \defined \prod_{i =
    1}^k a_i^{ X_i \ind{S}(i) + r_i}$. 
    Because $f$ is \kl{$(k,\Nat)$-combinatorial} with parameter $\omega$,
    there exists a \kl{strongly natural binomial polynomial} $P \in \Rat[X_1,
    \dots, X_k]$ such that $f \circ q (\omega X_1, \dots, \omega X_k) = P (X_1, \dots, X_k)$ 
    over $(\Nat_{\geq 1})^k$.
    This proves that
    $P_{(S, \vec{r})}(\seqof{X_i}{i \in S})$
    equals
    $P(X_1, \dots, X_k)$ as polynomials,  hence,
    that $P_{(S, \vec{r})}$ is a \kl{strongly natural binomial polynomial}
    for all $(S, \vec{r}) \in \ModuloTypes[\omega]^k$. We have proven that
    $f \in \NPoly[d]$.
\end{proof}

It was already known that \kl{$\Rel$-polyregular functions} with unary input
that are non-negative are \kl{$\Nat$-polyregular} \cite[Proposition 2.1 p
137]{BERE10}. Let us derive this fact from our \cref{decidable-n-poly:thm}.

\begin{corollary}
    Let $f \colon \set{a}^* \to \Rel$ be a non-negative \kl{$\Rel$-polyregular function},
    then $f \in \NPoly$.
\end{corollary}
\begin{proof}
    Since $f$ has unary input, it is \kl{commutative}. Furthermore,
    $f$ is \kl{$(1,\Nat)$-combinatorial} because for all $q \colon \Nat \to
    \set{a}$ and all $\omega \geq 1$, $f(q(\omega X))$ is non-negative.
    When it is a polynomial function, it therefore belongs to $\CorrectPoly$,
    hence is a \kl{strongly natural binomial polynomial}.
    We conclude using \cref{decidable-n-poly:thm}.
\end{proof}

Let us now prove that the above characterizations of \kl{commutative}
\kl{$\Nat$-polyregular functions} can be combined with the recent advances in
the study of \kl{$\Rel$-polyregular functions} \cite{CDTL23} allowing to decide
the membership of $\ZSF$ inside $\ZPoly$. The key ingredient of this study is
the use of a semantic characterization of \kl{star-free $\Rel$-polyregular
functions} among \kl{$\Rel$-rational series} that generalizes the notion of
aperiodicity for languages to functions.

\begin{definition}[Ultimately polynomial]
    \label{ultimately-polynomial:def}
    Let $\Sigma$ be a finite alphabet. 
    A function $f \colon \Sigma^* \to \Rel$
    is \intro{ultimately polynomial}
    when there exists $N_0 \in \Nat$ such that
    for all $k \in \Nat$,
    for all \kl{pumping pattern} $q \colon \Nat^k \to \Sigma^*$,
    there exists a polynomial $P \in \Rat[X_1, \dots, X_k]$
    such that:
    \begin{equation*}
        f \circ q = P
        \quad 
        \text{ over } (\Nat_{\geq N_0})^k
        \quad .
    \end{equation*}
\end{definition}

It was observed in \cite[Claim V.6]{CDTL23}, and in \cite[Claim 7.45, Lemma
7.53]{DOUE23} that a regular language $L$ is \emph{star-free} if and only if its
indicator function $\ind{L}$ is \kl{ultimately polynomial}. We can now answer
\cite[Conjecture 7.61]{DOUE23} positively, by proving that $\NPoly \cap \ZSF =
\NSF$.

\begin{theorem}
    \label{zsf-npoly-nsf:thm}
    Let $\Sigma$ be a finite alphabet, 
    and $f \colon \Sigma^* \to \Rel$ be a \kl{commutative}
    \kl{$\Nat$-polyregular function}.
    Then, the following are equivalent:
    \begin{enumerate}
        \item \label{zsf-npoly-nsf:thm:up:item} $f$ is \kl{ultimately polynomial},
        \item \label{zsf-npoly-nsf:thm:zsf:item} $f \in \ZSF$,
        \item \label{zsf-npoly-nsf:thm:nsf:item} $f \in \NSF$.
    \end{enumerate}
    Furthermore, membership is decidable and conversions are effective.
    \proofref{zsf-npoly-nsf:thm}
\end{theorem}
\begin{proof}
    The implication \cref{zsf-npoly-nsf:thm:nsf:item} $\Rightarrow$
    \cref{zsf-npoly-nsf:thm:zsf:item} is immediate since $\NSF \subseteq \ZSF$.
    Furthermore,
    \cref{zsf-npoly-nsf:thm:zsf:item} implies \cref{zsf-npoly-nsf:thm:up:item}
    following previous results for \kl{star-free $\Rel$-polyregular functions}
    \cite[Theorem V.13]{CDTL23}.

    For the implication \cref{zsf-npoly-nsf:thm:up:item} $\Rightarrow$
    \cref{zsf-npoly-nsf:thm:nsf:item}, let us assume that $f$ is \kl{ultimately
    polynomial}. We prove the result by induction on the size of the alphabet
    $\Sigma$. By definition, there exists $N_0 \in \Nat$, and $P \in
    \Rat[\seqof{X_a}{a \in \Sigma}]$ such that:

    \begin{equation*} f\left(
        \prod_{a \in \Sigma} a^{x_a} \right)
        = 
        P(\seqof{x_a}{a \in \Sigma})
        \quad \forall \vec{x} \in (\Nat_{\geq N_0})^{\Sigma} \quad .
    \end{equation*}

    It is clear that $\translate{N_0}(P)$ is \kl{represented} by an
    \kl{$\Nat$-polyregular function}, namely, $f_u \colon w \mapsto f(uw)$
    where $u \defined \prod_{a \in \Sigma} a^{N_0}$, and is therefore represented by a
    \kl{star-free $\Nat$-polyregular function} thanks to
    \cref{decide-rat-poly-npoly:cor}.
    For every letter $a \in \Sigma$ and number $0 \leq n \leq N_0$, there exists, by induction hypothesis,
    a \kl{star-free $\Nat$-polyregular function} $g_{a^n}$ that \kl{represents} the
    function $f_{a^n} \colon (\Sigma \setminus \set{a})^* \to \Rel$
    that maps $w \in (\Sigma \setminus \set{a})^*$ to $f(a^n w)$.

    Let us conclude by computing $f$ 
    using the following \kl{star-free $\Nat$-polyregular function} $g \colon \Sigma^* \to \Rel$:
    \begin{equation*}
        g \colon w \mapsto \begin{cases}
            g_{a^n} (w) & \text{ if } \card[a]{w} = n \text{ for some } a \in \Sigma \text{ and } n \leq N_0 \\
            \translate{N_0}(P)(\seqof{\card[a]{w} - N_0}{a \in \Sigma}) & \text{ otherwise }
        \end{cases}
        \qedhere
    \end{equation*}
\end{proof}
\section{Canonical Models for $\NPoly$}
\label{canonical-models:sec}

In this section, we present a more open-ended research direction that aims at
providing canonical models for \kl{$\Nat$-polyregular functions}. The hope with
such models is to be able to decide membership, equivalence, and aperiodicity
problems for \kl{$\Nat$-polyregular functions}, which are currently open. The
key ingredient of our approach is the notion of \emph{noncommutative
integration}, which allows to re-interpret \kl{$\Nat$-polyregular functions} as
iterated integrals of the constant function equal to zero.

\subsection{Noncommutative Integration}
\label{noncommutative-integration:sec}

In this section, we re-introduce the notion of transducers with oracle of
\cite{CDTL23}, and re-interpret them as \emph{integration operators} on
functions. To illustrate the construction on a simple example, let us consider
a function $f \colon \set{1}^* \to \set{1}^*$, that is a function $f \colon
\Nat \to \Nat$ with numbers represented in unary.
Let us write $\Delta[f] (n) = f(n+1) - f(n)$ for the \emph{discrete derivative}
of $f$ at $n$, and let us remark that for all $n \in \Nat$,
\begin{equation}
    \label{discrete-derivative:eq}
    f(n) = f(0) + \sum_{i=0}^{n-1} \Delta[f](i)
    \quad .
\end{equation}
A presentation of \cref{discrete-derivative:eq} in terms of transducers
would lead to the following picture:
\begin{center}
    \begin{tikzpicture}
        \node[state, initial] (q0) {$q_0$};
        \node[right=1.2em of q0] (q1) {$f(0)$};
        \draw (q0) edge[loop above] node{$1/\Delta[f]$} (q0);
        \draw[->] (q0) edge (q1);

    \end{tikzpicture}
\end{center}
Where the semantics of such a model is to
start in node $q_0$, and when reading a $1$, output the value of $\Delta[f]$
on the rest of the input, and loop back to $q_0$. When the input is empty, the
automaton outputs $f(0)$.
It is a straightforward exercise to prove that a function $f \colon
\set{1}^* \to \Rel$ corresponds to a polynomial of degree $k$
if and only if its $k$th discrete derivative $(\Delta \circ \cdots \circ \Delta)[f]$ is constant.

To capture \kl{$\Nat$-polyregular functions} using this construction, we
generalize the integration operator in two ways: first, we take into account
that the input alphabet may contain more than one letter (picking one function
per letter in the automaton above), and second, we allow our integration
operators to have some finite inner state (adding states to the automaton
presented above). These two yield the following definition of
\kl{$\mathcal{H}$-transducers}.

\begin{definition}
    \label{oracle-transducer:def}
    Let $\mathcal{H} \subseteq \Rel^{\Sigma^*}$ be a set of
    functions. An \intro{$\mathcal{H}$-transducer}
    is a tuple $\aTransd \defined (Q, q_0, \delta, \lambda, F)$
    where
    \begin{itemize}
        \item $Q$ is a finite set of states,
        \item $q_0 \in Q$ is called the initial state,
        \item $\delta \colon Q \times \Sigma \to Q$ is called the transition function,
        \item $\lambda \colon Q \times \Sigma \to \mathcal{H}$ is called the integrated function,
        \item $F \colon Q \to \Rel$ is called the final condition.
    \end{itemize}
    We allow ourselves to write $\delta^* \colon Q \times \Sigma^* \to Q$ for
    the repeated application of the transition function $\delta$, that is
    $\delta^*(q,\varepsilon) \defined q$, and $\delta^*(q, aw) \defined \delta(\delta^*(q,a), w)$.
\end{definition}

\AP The semantics of an \kl{$\mathcal{H}$-transducer} is defined by induction.
We say that a transducer $\aTransd$ \intro{computes} a function $f \colon
\Sigma^* \to \Rel$ if for all $q \in Q$, $a \in \Sigma$, and $w \in \Sigma^*$,
$\aTransd(q, aw) = \aTransd(\delta(q,a), w) + \lambda(q, a)(w)$ and
$\aTransd(q, \varepsilon) = F(q)$.
That is, given a word $w$:
\begin{equation}
    \label{transducer-semantics:eq}
    \aTransd(q_0, w) 
    = \sum_{i=0}^{|w|-1} \lambda(\delta^*(q_0, w_{\leq i}), w_{i+1})(w_{> i+1}) + F(\delta^*(q_0, w))
    \quad 
    .
\end{equation}

\AP In that sense, the $\mathcal{H}$-transducer is really computing a
\emph{noncommutative integral} of the function $\lambda$ with (final) condition
$F$. The analogy between \cref{transducer-semantics:eq} and
\cref{discrete-derivative:eq}, does not stop here, as the notion of
\emph{polynomials} can be extended to \kl{polyregular functions} using the
following ``integration theorem for polyregular functions.'' In order to
formally state \cref{H-transducers:thm}, we need to introduce the
automata-theoretic counterpart of \kl{aperiodic monoids}, which is based on the
notion of \kl{counter}: a \intro{counter} in a finite state automaton is a pair
$(q,u)$ where $q$ is a state and $u$ is a word such that $\delta(q,u) \neq q$
and $\delta(q,u^n) = q$ for some $n \geq 2$. An automaton is
\reintro{counter-free} when it contains no \kl{counters}. This connection
between \kl{counter-free} automata and \kl{star-free} functions (with boolean
output) is well-known in automata theory \cite{MNPA71}.

\begin{theorem}[\cite{DOUE23}]
    \label{H-transducers:thm}
    Let $f \colon \Sigma^* \to \Rel$ (resp. $\Nat$) be a function and $k \geq 1$.
    Then,
    $f \in \NPoly[k]$
    if and only if $f$ is computed by an \kl{$\NPoly[k-1]$-transducer} 
    Furthermore, $f \in \NSF[k]$
    if and only if
    $f$ is computed by a \kl{counter-free} $\NSF[k-1]$-transducer.
    The same holds for the classes $\ZPoly[k]$ and $\ZSF[k]$.
\end{theorem}

However, \cref{H-transducers:thm} does not provide a \emph{canonical} model for
computing a function $f$, and therefore cannot help in deciding membership,
equivalence, or optimization problems. This can already be seen in the case of
unary input functions, as illustrated by the two distinct minimal (in the
number of states) \kl{$\NPoly[0]$-transducers} of
\cref{non-canonical-transd:fig} computing the function $\BadExOk$ of
\cref{non-canonical-transd:ex}.

\begin{figure}
    \centering
    \begin{tikzpicture}
        \begin{scope}[xshift=-3cm]
            \node[state, initial] (q0) {$q_0$};
            \node[state, below=of q0] (q1) {$q_1$};
            \node[right=1.5em of q0] (o0) {$1$};
            \node[right=1.5em of q1] (o1) {$0$};
            \draw[->] (q0) edge (o0);
            \draw[->] (q1) edge (o1);
            \draw[->] (q1) edge[loop left] node{$a/1$} (q1);
            \draw[->] (q0) edge node[midway, left] {$a/0$} (q1);
        \end{scope}
        \begin{scope}[xshift=3cm]
            \node[state, initial] (p0) {$q_0$};
            \node[state, below=of p0] (p1) {$q_1$};
            \node[right=1.5em of p0] (o0) {$1$};
            \node[right=1.5em of p1] (o1) {$0$};
            \draw[->] (p0) edge (o0);
            \draw[->] (p1) edge (o1);
            \draw[->] (p0) edge[bend right=30] node[midway, left] {$a/0$} (p1);
            \draw[->] (p1) edge[bend right=30] node[midway, right] {$a/ \lambda u. 2 \times \ind{u \neq \varepsilon}$} (p0);
        \end{scope}
    \end{tikzpicture}
    \caption{Two distinct $\NPoly[0]$-transducers computing the function $\BadExOk$ of \cref
    {non-canonical-transd:ex}.
}
    \label{non-canonical-transd:fig}
\end{figure}

\begin{example}
    \label{non-canonical-transd:ex}
    Let $\BadExOk \colon \set{a}^* \to \Nat$ that maps 
    $\varepsilon$ to $1$, and $aw$ to $\card{w}$.
    Then $\BadExOk$ is computed by the two transducers of 
    \cref{non-canonical-transd:fig}, and no \kl{$\NPoly[0]$-transducer} with fewer 
    states can compute $\BadExOk$.
\end{example}
\begin{proof}
    Assume by contradiction that there exists an \kl{$\NPoly[0]$-transducer}
    with one state that computes $\BadExOk$.
    Then, the transducer must output $1$ on the empty word,
    and has a self-loop $\delta(q_0) = q_0$.
    Assuming that it computes $\BadExOk$, we must have for all $u \in \Sigma^*$,
    $f(au) = \aTransd(q_0, au) =
    \aTransd(q_0, u) + \lambda(q_0, a)(u) = 
    f(u) + \lambda(q_0, a)(u)$.
    Therefore,
    $\lambda(q_0, a)(u) = f(au) - f(u)$ takes a negative
    value for $u = \varepsilon$, which is absurd because
    $\lambda(q_0, a) \in \NPoly[0]$.

    Let us prove that the left transducer of
    \cref{non-canonical-transd:fig}
    computes $f$.
    We prove by induction on $w$
    that $A(q_0, w) = f(w)$,
    and that $A(q_1, w) = f(aw)$.
    To that end, let us first remark that $f(\varepsilon) = 1$,
    and $f(au) = \card{u}$.
    When $w = \varepsilon$, $A(q_0, \varepsilon) = F(\varepsilon) = 1 = f(\varepsilon)$.
    Similarly, $A(q_1, \varepsilon) = F(q_1) = 0 = f(a)$.

    Assume that $w = au$. Then:
    \begin{align*}
        A(q_0, w) = A(q_0, au) &= \lambda(q_0, a)(u) + A(q_1, u) \\ 
                               &= 0 + A(q_1,u) \\
                               &= f(au) & \text{by induction hypothesis} \\
                               &= f(w) 
    \end{align*}
    Similarly,
    \begin{align*}
        A(q_1, w) = A(q_1, au) &= \lambda(q_1, a)(u) + A(q_1, u) \\ 
                               &= 1 + A(q_1,u) \\
                               &= 1 + f(au) & \text{by induction hypothesis} \\
                               &= 1 + \card{u} \\
                               &= \card{au} \\
                               &= f(aau) \\
    \end{align*}

    The proof for the other automaton is similar. The
    key ingredient in the induction hypothesis is that if 
    $\card{u} \geq 1$, then:
    $f(aau) - f(u) = 2$ and otherwise $f(aau) - f(u) = 0$.
    Hence, $f(aau) - f(u) = 2 \times \ind{u \neq \varepsilon} 
    = \lambda(q_1, a)(u)$.
\end{proof}

\subsection{Residual Transducers}
\label{residual-transducer:sec}

\AP In order to obtain a \emph{canonical object} based on
\kl{$\mathcal{H}$-transducers}, a key ingredient is to consider the so-called
\intro{residuals} of a function $f \colon \Sigma^* \to \Rel$,  defined by
$\intro*\app{f}{u} \defined w \mapsto f(uw)$. The collection of \kl{residuals}
of a function $f$ is denoted $\intro*\Res(f)$ and is defined as the set of
$\app{f}{u}$ where $u$ ranges over words in $\Sigma^*$. Given a function $f
\colon \Sigma^* \to \Nat$ and two words $u,v \in \Sigma^*$, we also define the
\intro{noncommutative derivative} $\intro*\Deriv{f}{u}{v} \defined \app{f}{u} -
\app{f}{v}$.

\AP To compute a function $f$ by ``integrating'' simpler functions, we
will try to detect when the \kl{noncommutative derivative} of $f$ is simpler
than $f$ itself. To that end, given $k \in \Nat$, we define the following
equivalence relation on $\Sigma^*$: $u \intro*\resequiv{f}{k} v$ if and only if
$\Deriv{f}{u}{v} \in \ZPoly[k-1]$. Note that $\resequiv{f}{k}$ is an
equivalence relation because $\ZPoly[k-1]$ is closed under $\Rel$-linear
combinations and multiplication by $(-1)$. The key construction of Colcombet,
Douéneau-Tabot, and Lopez in \cite{CDTL23} is to remark that this equivalence
relation characterizes $\ZPoly[k]$ in the following sense: given $f \colon
\Sigma^* \to \Rel$, $f \in \ZPoly[k]$ if and only if $\resequiv{f}{k}$ has
finite index.

\AP Let us briefly pause here to remark that the characterization of
$\ZPoly[k]$ using $\resequiv{f}{k}$ is reminiscent of the Myhill-Nerode theorem
for \kl{$\Rel$-rational series}: a function $f \colon \Sigma^* \to \Rel$ is
a \kl{$\Rel$-rational series} if and only if the set $\Res(f)$ of its \kl{residuals} is
finitely generated as a $\Rel$-module, using only elements in $\Res(f)$
\cite[Chapter 1, Corollary 5.4]{BERE10}. The latter statement exactly says
that there are finitely many words $u_1, \dots, u_n$ such that for all $a \in \Sigma$,
and $1 \leq i \leq n$, there exist $\alpha_{i,j} \in \Rel$ such that
$\app{f}{u_i a} = \sum_{j=1}^n \alpha_{i,j} (\app{f}{u_j})$, providing a 
recurrence relation to compute the value of $f$ on any input word.

\AP The analogue notion for $\NPoly[k-1]$ is defined by $v \intro*\resleq{f}{k}
u$ if and only if $\Deriv{f}{u}{v} \in \NPoly[k-1]$, this is only a partial
ordering relation. Given a function $f$, our
goal is to leverage $\resleq{f}{k}$ to build a canonical
\kl{$\NPoly[k-1]$-transducer} that \kl{computes} $f$. The idea is to consider
as states the minimal elements of $\Sigma^*$ for $\resleq{f}{k}$, and define
transitions by letting $\delta(u, a)$ be some state $v$ such that $v
\resleq{f}{k} ua$. To produce a canonical model, this has to be done carefully,
as illustrated by the two distinct \kl{$\NPoly[0]$-transducers} of
\cref{non-canonical-transd:fig}
computing
the function $\BadExOk$ of \cref{non-canonical-transd:ex}, and
having as states minimal elements for $\resleq{\BadExOk}{0}$. To ensure
unicity, we ask that the set of states is a \intro{downwards closed} subset of
$\Sigma^*$ for the \intro{prefix ordering} $(\intro*\prefleq)$, i.e. that every
prefix of a state is also a state. 

\begin{lemma}
    \label{good-residual-ordering:fact}
    Let $k \in \Nat$, and let $f \colon \Sigma^* \to \Nat$. Then,
    $(\resleq{f}{k})$ is a quasi-ordering, satisfying the following
    extra properties:
    \begin{enumerate}
        \item For all $u,v,w \in \Sigma^*$, $u \resleq{f}{k} v$
            implies $uw \resleq{f}{k} vw$,
        \item If $u \resleq{f}{k} v$ and $v \resleq{f}{k} u$,
            then $\Deriv{f}{u}{v} = 0$,
    \end{enumerate}
\end{lemma}
\begin{proof}
    Assume that $u \resleq{f}{k} v$ and $v \resleq{f}{k} w$.
    Then, $\Deriv{f}{u}{v} \in \NPoly[k-1]$ and $\Deriv{f}{v}{w} \in \NPoly[k-1]$,
    we conclude that $\Deriv{f}{u}{w} = \Deriv{f}{u}{v} + \Deriv{f}{v}{w} \in \NPoly[k-1]$.
    We have proven that $(\resleq{f}{k})$ is a quasi-ordering.

    Let $u,v,w \in \Sigma^*$ be such that $u \resleq{f}{k} v$.
    Then, $\Deriv{f}{u}{v} \in \NPoly[k-1]$.
    We have $\Deriv{f}{uw}{vw} = \app{\Deriv{f}{u}{v}}{w} \in \NPoly[k-1]$.

    Finally, let $u,v \in \Sigma^*$ be such that $u \resleq{f}{k} v$ and $v \resleq{f}{k} u$.
    Then, $\Deriv{f}{u}{v} \in \NPoly[k-1]$ and $\Deriv{f}{v}{u} \in \NPoly[k-1]$,
    as a consequence $\app{f}{u} - \app{f}{v}$ is non-negative,
    and $\app{f}{v} - \app{f}{u}$ is non-negative, leading to 
    $\app{f}{u} = \app{f}{v}$, i.e.,
    $\Deriv{f}{u}{v}  = \Deriv{f}{v}{u} = 0$.
\end{proof}

\begin{definition}
    \label{residual-transducer:def}
    Let $f \colon \Sigma^* \to \Nat$ and $k \in \Nat$.
    A transducer $\aTransd \defined (Q, q_0, \delta, \lambda, F)$
    is a \intro{$k$-residual transducer}
    of $f$ 
    when
    it is a \kl{$\NPoly[k-1]$-transducer}
    satisfying the following properties:
    \begin{enumerate}
        \item $\aTransd$ \kl{computes} $f$;
        \item $Q \subseteq \Sigma^*$ is a \kl{downwards closed}
            for $\prefleq$;
        \item $q_0 = \varepsilon$;
        \item every state $q \in Q$ is accessible from $q_0$;
        \item For all $u, a \in Q$,
            $\delta(u,a)$ is the $\prefleq$-maximal $v \in Q$
            such that $v \prefleq ua$, and $v \resleq{f}{k} ua$.
        \item For all $u,a \in Q$,
            $\lambda(u,a) = 
            \Deriv{f}{ua}{\delta(u,a)} \in \NPoly[k-1]$.
    \end{enumerate}
\end{definition}

Let us immediately prove that \kl{$k$-residual transducers} are unique due to
the way their state space  is defined and transitions are computed using
maximal prefixes.

\begin{lemma}
    \label{unique-res-transducer:fact}
    Let $f \colon \Sigma^* \to \Nat$ and $k \in \Nat$.
    Then $f$ has at most one \kl{$k$-residual transducer}.
\end{lemma}
\begin{proof}
    Let $\aTransd_1$ and $\aTransd_2$ be two
    \kl{$k$-residual transducers} for $f$.
    The two initial states must be $\varepsilon$.
    Let us prove by induction on $u \in \Sigma^*$ that
    $\delta_1(\varepsilon, u) = \delta_2(\varepsilon, u)$
    and that $Q_1$ equals $Q_2$ over prefixes of $u$.
    This will prove that 
    $Q_1 = Q_2$, hence that $\aTransd_1 = \aTransd_2$.

    Let $u \in \Sigma^* \cap Q_1 \cap Q_2$ and $a \in \Sigma$, $v_1 \in Q_1$ be
    defined as $v \defined \delta_1(u,a)$, and $v_2 \defined \delta_2(u,a)$.
    Remark that by induction hypothesis, for all $v \prefle u$, $v \in Q_1 \cap
    Q_2$. If $\delta_1(u,a) = Q_1$, it means that for all $v \in Q_2$ such that
    $v \prefle ua$, we have $\neg( v \resleq{f}{k} ua )$. The only possible
    transition in $\aTransd_2$ is therefore $\delta_2(u,a) = ua$, and $ua \in
    Q_2$. Similarly, if $\delta_1(u,a) \prefleq u$, then $\delta_2(u,a) =
    \delta_1(u,a)$ by definition of $\delta_2$ as a maximum.
\end{proof}

Let us now introduce \cref{residual:algo} to compute the \kl{$k$-residual
transducer} given a function $f$. Notice that this algorithm requires the
ability to test if a function belongs to $\NPoly[k]$, which is only known to be
feasible for \kl{commutative} \kl{polyregular functions}
(\cref{decidable-n-poly:thm}).
However, the
termination of this algorithm will prove the existence of the \kl{$k$-residual
transducer}. The key argument proving the termination of \cref{residual:algo}
is based on the fact that for a function $f \in \NPoly[k]$, the quasi-ordering
$(\resleq{f}{k})$ is a \kl{well-quasi-ordering}.

\AP A quasi-ordered set $(X, \leq)$ is a \intro{well-quasi-ordering} if every
infinite sequence of elements $\seqof{x_n}{n \in \Nat}$ of $X$ contains two
indices $i < j$ such that $x_i \leq x_j$. We say that such a sequence is a
\intro{good sequence}, on the contrary, a sequence is \intro{bad} if it
contains no such pair of indices.

\begin{algorithm}[t]
    \caption{Computing a \kl{$k$-residual transducer} of a function $f$.}
    \label{residual:algo}
    $Q \defined \{ \varepsilon \}$;

    $O \defined \setof{ a }{ a \in \Sigma}$;

    $\delta \defined \emptyset$;

    $\lambda \defined \emptyset$;

    $F \defined \emptyset$;

    \While{$O \neq \emptyset$}{
        choose $ua \in O$;

        $O \defined O \setminus \set{ ua }$;

        \eIf{$\exists v \in Q, v = \max_{\prefleq} \setof{w \in Q}{w \prefleq ua \wedge w \resleq{f}{k} ua}$}{
            $\delta(u, a) \defined v$;

            $\lambda(u, a) \defined \Deriv{f}{ua}{v}$;
        }{
            $Q \defined Q \uplus \set{ ua }$;

            $\delta(u,a) \defined ua$;

            $\lambda(u,a) \defined 0$;

            $O \defined O \cup \setof{ uab }{b \in \Sigma}$;
        }
    }
    \For{$u \in Q$}{
        $F(u) \defined f(u)$;
    }
    \Return{$(Q, \varepsilon, \delta, \lambda, F)$};

\end{algorithm}

\paragraph{Correctness of the Algorithm.} Let us now prove that
\cref{residual:algo}
computes
the \kl{$k$-residual transducer} of a function $f$. We start by collecting some
invariants about its execution in \cref{q-o-prefix-cool:fact},
allowing us to derive correctness in \cref{correct-residual:lemma}.
The termination of the algorithm is then obtained in
\cref{wqo-implies-termination:lemma}, under the assumption that
$\resleq{f}{k}$ is a \kl{well-quasi-ordering}.

\begin{lemma}
    \label{q-o-prefix-cool:fact}
    Let $f \colon \Sigma^* \to \Nat$ and $k \in \Nat$.
    At each step of the \texttt{while loop}
    of \cref{residual:algo}, the sets
    $Q$ and $O$ are such that
    \begin{enumerate}
        \item $Q \cup O$ is a \kl{downwards closed} subset of 
            $\Sigma^*$ for $\prefleq$;
        \item elements in $O$ are pairwise incomparable
            for $\prefleq$, and are maximal
            for $\prefleq$ inside $Q \cup O$.
    \end{enumerate}
\end{lemma}
\begin{proof}
    Let us write $Q_i$ and $O_i$ for the value of the variables
    $Q$ and $O$ at step $i$ of the \texttt{while loop}.
    We prove the desired property by induction on $i$.

    For $i=0$, the property is true because
    $Q_0 = \set{\varepsilon}$ and $O_0 = \setof{a}{a \in \Sigma}$.

    For $i+1$. Either the \texttt{if} branch was taken, in which case $Q_{i+1}
    \cup O_{i+1} = (Q_i \cup O_i) \setminus \set{u}$ for some $u \in O_i$. This
    set remains \kl{downwards closed}, and elements in $O_{i+1}$ remain maximal
    elements. 

    If the \texttt{else} branch was taken, then there exists $u \in O_i$ such
    that $Q_{i+1} = Q_i \cup \set{u}$ and $O_{i+1} = O_i \setminus \set{ u }
    \cup \setof{ ua }{ a \in \Sigma}$. We conclude that $Q_{i+1} \cup O_{i+1} =
    Q_i \cup O_i \cup \setof{ ua }{a \in \Sigma}$ continues to be \kl{downwards
    closed} for $\prefleq$. Let $v \in Q_{i+1} \cup O_{i+1}$ be such that $ua
    \prefleq v$ for some $a \in \Sigma$. Then $u \prefleq v$, and $u = v$ since
    $u$ was a maximal element. As a consequence, $ua$ is a maximal element for
    all $a \in \Sigma$. Assume by contradiction that $ua$ is comparable with
    some $v \in O_{i+1}$ with $ua \neq v$, it cannot be that $ua \prefleq v$ by
    the above argument, and if $v \prefleq ua$ with $v \neq ua$, then $v
    \prefleq u$ and $u = v$, which is absurd since $v \not \in O_{i+1}$.
    We have concluded that $O_{i+1}$ continues to have pairwise incomparable
    elements.
\end{proof}

\begin{lemma}
    \label{correct-residual:lemma}
    If \cref{residual:algo} terminates on 
    an input $f \colon \Sigma^* \to \Nat$, $k \in \Nat$,
    then it computes the \kl{$k$-residual transducer} of $f$.
\end{lemma}
\begin{proof}
    Because of \cref{q-o-prefix-cool:fact},
    we already know that $q_0 = \varepsilon$,
    $Q$ is a \kl{downwards closed} subset of $\Sigma^*$
    for $\prefleq$, 
    that every state of $Q$ is accessible from $q_0$.
    Notice that at every step,
    $\lambda(u,a)$ is defined as
    $\Deriv{f}{ua}{\delta(u,a)} = \app{f}{ua} - \app{f}{\delta(u,a)}$.
    Finally, since $Q \cup O$ is a \kl{downwards closed} subset of $\Sigma^*$
    at every step,
    we have that at step $i$,
    for all $ua \in O_i$,
    $\setof{w \in Q}{w \prefleq ua} = \setof{w \in Q_i}{ w \prefleq ua}$,
    which proves that the maximum considered in the algorithm
    is indeed computing correctly.

    Let us now prove that $\aTransd$ \kl{computes} $f$.
    To that end, let us prove by induction on $w \in \Sigma^*$ 
    that for all $q \in Q$, $\aTransd(q, w) = f(qw)$.
    The base case is trivial, as $\aTransd(q, \varepsilon) = F(q) \defined f(q)$.
    For the induction step, 
    let $w = au$ with $a \in \Sigma$ and $u \in \Sigma^*$.
    Then $\aTransd(q, w) = \aTransd(\delta(q,a), u) + \lambda(q,a)(u)$.
    By induction hypothesis, $\aTransd(\delta(q,a), u) = f(\delta(q,a)u)$.
    Furthermore, $\lambda(q,a)(u) = \Deriv{f}{qa}{\delta(q,a)} = f(qau) - f(\delta(q,a)u)$.
    As a consequence, $\aTransd(q, w) = f(qw)$ which concludes the proof.
\end{proof}

\begin{lemma}
    \label{wqo-implies-termination:lemma}
    Let $f \colon \Sigma^* \to \Nat$, and $k \in \Nat$ be such that
    every infinite, $\prefleq$-increasing sequence is \kl{good}
    in $(\Sigma^*, \resleq{f}{k})$.
    Then, \cref{residual:algo} terminates on the input $(f,k)$.
\end{lemma}
\begin{proof}
    Assume towards a contradiction that
    \cref{residual:algo} does not terminate.
    Then, the \texttt{else} branch in the \texttt{while loop}
    must be taken infinitely often.
    This means that the set $Q$ of states grows arbitrarily large.

    Let us write $\seqof{Q_i}[i \in \Nat]$ for the set of states $Q$ at step
    $i$ of the execution of \cref{residual:algo}. Applying
    \cref{q-o-prefix-cool:fact}, we know that for all $i \in \Nat$, $Q_i$ is
    \kl{downwards closed} for $\prefleq$. Let us write $Q_\infty \defined
    \bigcup_{i \in \Nat} Q_i$. The set $Q_\infty$ is infinite, and is
    \kl{downwards closed} for $\prefleq$. As a consequence, it is an infinite
    tree with a finite branching (at most $\card{\Sigma}$), and has an infinite
    branch $\seqof{u_j}{j \in \Nat}$ thanks to König's lemma.

    Let us prove that this infinite branch is a \kl{bad sequence} for the
    ordering $\resleq{f}{k}$.
    Let $j < p$, and assume by contradiction that $u_j \resleq{f}{k} u_p$. We
    know that $u_j \in Q_j$ and $u_p \in Q_p$. Then, at step $p-1$ of the
    algorithm, $u_j \in Q_{p-1}$, since $u_j \in Q_j \subseteq Q_{p-1}$.
    Because $u_j \prefleq u_p$ and $u_j \resleq{f}{k} u_p$,
    \cref{residual:algo} must take the \texttt{if} branch at step $p-1$. As a
    consequence, $u_p \not\in Q_{p}$, which is absurd.

    We have proven that the infinite branch is a \kl{bad sequence}
    for $\resleq{f}{k}$, which contradicts the assumption.
    Hence, \cref{residual:algo} must terminate.
\end{proof}

\paragraph{Validity of the algorithm.}
Let us now prove that \cref{residual:algo} is correct terminates when $f \in \NPoly[k]$, by proving that 
$\resleq{f}{k}$ is a \kl{well-quasi-ordering}, which 
is the  content of \cref{n-poly-k-implies-wqo:lemma}.

\begin{lemma}
    \label{n-poly-k-implies-wqo:lemma}
    Let $k \in \Nat$, and let $f \in \NPoly[k]$.
    Then, the relation $\resleq{f}{k}$ is a \kl{well-quasi-ordering}
    for $\Sigma^*$.
\end{lemma}
\begin{proof}
  Because $f \in \NPoly[k]$, there exists a finite monoid $M$,
  a morphism $\mu \colon \Sigma^* \to M$ and a production 
  function $\pi \colon M^{k+1} \to \NPoly[k]$
  such that $f = \pi^\dagger$. We will use the fact that the
  \kl{residuals} of a function represented in that way
  can be explicitly expressed using $\mu$ and $\pi$, hence,
  that their \kl{non-commutative derivatives} can also be computed.

  Let $u,w \in \Sigma^*$ be words, then $uw = s_1 \cdots s_{k+1}$ implies that
  there exists $\ell \leq k+1$, such that $u = u_1 \cdots u_{\ell}$, $w
  = w_{\ell} \cdots w_{k+1}$, with $s_i = u_i$ for $1 \leq i \leq \ell - 1$,
  $w_i = s_i$ for $\ell + 1 \leq i \leq k+1$, and $s_{\ell} = u_{\ell}
  w_{\ell}$.
  Therefore,
  \begin{align*}
    \app{f}{u}(w) &= f(uw) \\
                  &=
    \sum_{uw = s_1 \cdots s_{k+1}} \pi(\mu(s_1), \dots, \mu(s_{k+1})) \\
    &=
    \sum_{\ell = 1}^{k+1}
    \sum_{w = w_{\ell} \cdots w_{k+1}}
    \sum_{u = u_1 \cdots u_{\ell}}
    \pi(\mu(u_1), \dots, \mu(u_{\ell-1}), \mu(u_{\ell} w_{\ell}), \mu(w_{\ell+1}), \dots, \mu(w_{k+1})) \\
    &=
    \sum_{\ell = 1}^{k+1}
    \sum_{w = w_{\ell} \cdots w_{k+1}}
    \underbrace{
    \sum_{u = u_1 \cdots u_{\ell}}
  \pi(\mu(u_1), \dots, \mu(u_{\ell-1}), \mu(u_\ell) \mu( w_\ell), \mu(w_{\ell +1}), \dots, \mu(w_{k+1}))}_{
      \defined \theta_u^\ell(\mu(w_{\ell +1}), \dots, \mu(w_{k+1}))}
    \\
    &=
    \sum_{\ell = 1}^{k+1}
    \sum_{w = w_{\ell} \cdots w_{k+1}}
    \theta_u^\ell(\mu(w_{\ell}), \dots, \mu(w_{k+1}))
    \\
    &=
    \sum_{\ell = 1}^{k+1}
    (\theta_u^\ell)^\dagger(w)
  \end{align*}

  In particular, given two words $u,v \in \Sigma^*$, we have
  for every $w \in \Sigma^*$,
  \begin{align*}
      \Deriv{f}{v}{u}(w)
      &= f(vw) - f(uw)
      = 
      \sum_{\ell = 1}^{k+1}
      \left((\theta_v^\ell)^\dagger(w) - (\theta_u^\ell)^\dagger(w)\right)
      \\
      &=
      \sum_{\ell = 1}^{k+1}
      (\theta_{v}^{\ell} - \theta_{u}^{\ell})^\dagger(w)
  \end{align*}

  The functions $\theta_v^{\ell} - \theta_u^{\ell}$ are of the form $M^{k + 2 -
  \ell} \to \Rel$, and if one could prove that they only output non-negative
  values, we would immediately conclude that $\Deriv{f}{v}{u} \in \NPoly$.
  Furthermore, if we can prove that $\theta_v^{1} - \theta_u^1$ is constant
  equal to zero, we would conclude that $\Deriv{f}{v}{u} \in \NPoly[k-1]$.

  To that end, let us associate to a word $u \in \Sigma^*$ the following vector
  of values: first, the value of $\mu(u) \in M$, then, the tables of the
  functions $\theta_u^{\ell}$ for $1 \leq \ell \leq k+1$, that are described by
  finitely many numbers in $\Nat$. Let us write $\langle u \rangle$ for this
  vector. Note that $\langle u \rangle$ takes values in a naturally
  quasi-ordered set: $\langle u \rangle \leq \langle v \rangle$ if and only if
  $\mu(u) = \mu(v)$ and for all $1 \leq \ell \leq k+1$,
  $\theta_u^{\ell} \leq \theta_v^{\ell}$ pointwise.

  Let $u, v \in \Sigma^*$ such that $\langle u \rangle \leq \langle v \rangle$.
  Then, for all $1 \leq \ell \leq k+1$,
  $\theta_v^{\ell} - \theta_u^{\ell}$ only takes non-negative values, guaranteeing that
  $\Deriv{f}{v}{u} \in \NPoly$.
  Furthermore, for every tuple $(m_1, \dots, m_{k+1}) \in M^{k + 1}$,
  \begin{align*}
  \theta_v^1(m_1, \dots, m_{k+1})
  &= 
  \pi(\mu(v) m_1, m_2, \dots, m_{k+1}) \\ 
  &=
  \pi(\mu(u) m_1, m_2, \dots, m_{k+1}) \\
  &=
  \theta_u^1(m_1, \dots, m_{k+1})
  \end{align*}
  therefore, $\theta_v^1 - \theta_u^1$ is constant equal to zero,
  and we conclude that $\Deriv{f}{v}{u} \in \NPoly[k-1]$.

  Recall that $\Nat^p$ is a \kl{well-quasi-ordering} when endowed with the
  product ordering, and that $M$ is finite. As a consequence of Dickson's
  lemma, we conclude that the quasi-ordered set $\setof{\langle u \rangle}{u
  \in \Sigma^*}$ is a \kl{well-quasi-ordering}.

  Let $\seqof{u_i}{i \in \Nat}$ be an infinite sequence of $\Sigma^*$.
  By definition of being a \kl{well-quasi-ordering}, there exists $i < j$ such that
  $\langle u_i \rangle \leq \langle u_j \rangle$.
  We conclude that $\Deriv{f}{u_i}{u_j} \in \NPoly[k-1]$,
  i.e., that $u_i \resleq{f}{k} u_j$.
  We just showed that every infinite sequence of $\Sigma^*$
  contains two indices $i < j$ such that $u_i \resleq{f}{k} u_j$,
  hence that $(\Sigma^*, \resleq{f}{k})$ is a \kl{well-quasi-ordering}.
\end{proof}

\AP We now have all the ingredients to prove the main result of this section,
which is a characterization of $\NPoly$ in terms of the existence of
\kl{$k$-residual transducers} and the \kl{well-quasi-ordering} of the relation
$(\resleq{f}{k})$. Note however that this theorem is not effective, as it
requires the ability to decide if a function belongs to $\NPoly[k]$ which 
remains open in the non-commutative case.

\begin{theorem}
    \label{non-commutative-npoly:thm}
    Let $f \in \ZPoly$ be a non-negative function, 
    and $k \in \Nat$,
    the following are equivalent:
    \begin{enumerate}
        \item \label{n-poly-1-transd:item} $f$ is \kl{computed}
            by an \kl{$\NPoly[k-1]$-transducer};
        \item \label{n-poly-k:item} $f \in \NPoly[k]$;
        \item \label{n-poly-wqo:item} $(\Sigma^*, \resleq{f}{k})$ is a
            \kl{well-quasi-ordering};
        \item \label{n-poly-well:item} every $\prefleq$-increasing sequence
            of $\Sigma^*$  is a \kl{good sequence}
            for $\resleq{f}{k}$;
        \item \label{n-poly-residual:item} The
            \kl{$k$-residual transducer}
            of 
            $f$ exists.
    \end{enumerate}
    If $f$ is \kl{commutative}, the  
    properties are decidable, and the conversions effective.
\end{theorem}
\begin{proof}
    \cref{n-poly-1-transd:item} implies \cref{n-poly-k:item} by
    definition. Then,
    \cref{n-poly-k:item} implies \cref{n-poly-wqo:item} by
    \cref{n-poly-k-implies-wqo:lemma}.
    The implication \cref{n-poly-wqo:item} $\Rightarrow$ \cref{n-poly-well:item}
    is obvious.
    Then, \cref{wqo-implies-termination:lemma} proves
    that \cref{n-poly-well:item} implies \cref{n-poly-residual:item}.
    Furthermore, because a \kl{$k$-residual transducer} is a \kl{$\NPoly[k-1]$-transducer},
    \cref{n-poly-residual:item} implies \cref{n-poly-1-transd:item}.
    Finally, \cref{residual:algo}
    is effective as soon as $\resleq{f}{k}$ is decidable, which 
    is the case when $f$ is a \kl{commutative}
    \kl{$\Rel$-polyregular} function, as shown in
    \cref{decidable-n-poly:thm}.
\end{proof}

\subsection{Aperiodicity and Star-Free Functions}
\label{aperiodic-star-free:sec}

\AP In this section, we investigate the connection between \kl{star-free}
\kl{$\Nat$-polyregular} functions and structural properties of their
\kl{residual transducers}. This is motivated by the fact that in the case of
$\Rel$ outputs, belonging to $\ZSF[k]$ is equivalent to having a
\kl{$k$-residual transducer} without \kl{counters} and with $\ZSF[k-1]$ labels
on its transitions (see \cref{H-transducers:thm}).

\AP Unfortunately, the \kl{residual transducer} does not seem to provide any
characterization of \kl{star-free} \kl{$\Nat$-polyregular} functions. Indeed,
we can exhibit a function $f \in \NSF[1]$ such that its \kl{$1$-residual
transducer} contains a \kl{counter}, as shown in
\cref{non-aperiodic-residual-transd:ex}. Even more, we can prove
that any choice of \kl{$\NPoly[1]$-transducer} with at most $2$ states computing
our example function will contain a \kl{counter}.

\begin{example}
    \label{non-aperiodic-residual-transd:ex}
    Let us define
    $\BadExKo(\varepsilon) = 1$,
    $\BadExKo(a) = 0$,
    $\BadExKo(a^2) = 1$,
    and $\BadExKo(a^n) = n - 3$ for all $n \geq 3$.
    The \kl{$0$-residual transducer} of $\BadExKo$ has a \kl{counter} and two states,
    and is in fact a \kl{$\NSF[0]$-transducer}.
    Furthermore,
    every \kl{$\NPoly[0]$-transducer} with at most two states contains a \kl{counter}.
\end{example}
\begin{proof}
    It is quite clear that $\BadExKo$ cannot be computed by a \kl{$\NPoly[0]$-transducer}
    with one state, as $\Deriv{f}{a}{\varepsilon}$ takes negative values.
    Let us now consider a \kl{$\NPoly[0]$-transducer} $\aTransd$  with two states $q_0$ and $q_1$ without 
    \kl{counters},
    such that $q_0$ is the initial state, $\delta(q_0, a) = q_1$, 
    that computes $\BadExKo$.
    Because $\aTransd$ has no \kl{counters}, we conclude that $\delta(q_1,a) = q_1$.
    Then, because $\aTransd$ \kl{computes} $\BadExKo$, we must have
    $\aTransd(q_0, a) = \BadExKo(a) = 0 = \BadExKo(aaa) = \aTransd(q_0, aaa)$.
    As a consequence, 
    $F(q_1) = 0$, 
    $\lambda(q_1, a)(\varepsilon) = 0$,
    and
    $\lambda(q_1, a)(a) = 0$.
    However, this means 
    that $\aTransd(q_0, aa) = F(q_1) + \lambda(q_1, a)(a)  + \lambda(q_1, a)(\varepsilon) = 0$,
    which contradicts the fact that $\BadExKo(aa) = 1$.

    Let us now exhibit the \kl{$0$-residual transducer} of $\BadExKo$. It has
    two states $q_0$ (representing $\varepsilon$) and $q_1$ (representing $a$),
    with $q_0$ being the initial state. The map $F$ is defined by $F(q_0) = 1$
    and $F(q_1) = 0$. Then, $\delta(q_0, a) = q_1$, and $\delta(q_1, a) = q_0$.
    Finally, $\lambda(q_0, a)(w) = 0$ for all $w \in \Sigma^*$,
    and $\lambda(q_1, a)(w)$ is defined as $0$ if $\card{w} \leq 2$ and
    $2$ otherwise.
    This is correct because
    $\lambda(q_0, a) = \Deriv{f}{a}{a} = 0$
    and
    $\lambda(q_a, a) = \Deriv{f}{aa}{\varepsilon} = 2 \times \ind{\card{w} > 2}$.
    Notice that the defined function are \kl{star-free}, hence that
    the \kl{$0$-residual transducer} of $\BadExKo$ is an \kl{$\NSF[0]$-transducer}.
\end{proof}

In the specific case of $\NPoly[0]$, which corresponds to $\Nat$-linear
combination of indicator functions of regular languages, we can however recover
a characterization of \kl{star-free} functions in terms of the \kl{residual
transducer}.\footnote{This was remarked by Antonio Casares-Santos 
in a personal communication.}

\begin{lemma} 
    \label{aperiodic-iff-residual:lem}
    Let $f \in \NPoly[0]$. Then,
    $f \in \NSF$ if and only if 
    $f \in \ZSF$, 
    if and only if 
    the \kl{$0$-residual transducer} of $f$ is \kl{counter-free}.
\end{lemma}
\begin{proof}
    It is clear that $\NSF \subseteq \ZSF$. Furthermore, if the \kl{$0$-residual
    transducer} of $f$ is \kl{counter-free}, then $f \in \NSF$
    because $f$ can be computed as a $\Nat$-linear combination of
    indicator functions of star-free languages (computed by the automaton).

    Assume now that $f \in \ZSF$, and let us prove that the \kl{$0$-residual
    transducer} of $f$ is \kl{counter-free}. Note that because $f \in
    \NPoly[0]$, $u \resleq{f}{0} v$ if and only if $\app{f}{u} = \app{f}{v}$.
    In particular, in a \kl{$0$-residual transducer} of $f$, two states that
    represent the same \kl{residual} must be incomparable for the prefix relation.

    Let $(q,w^n)$ be a counter with $n \geq 1$. This means that $\delta(q, w^n)
    = q$ in the automaton, and implies that $q \resleq{f}{0} qw^n$, hence that
    $\app{f}{q} = \app{f}{qw^n}$. Because $f \in \ZSF$, we know that
    $\app{f}{qw^n} = \app{f}{qw^{n+1}}$, hence that $\app{f}{qw} = \app{f}{q}$.

    Let us write $t \defined \delta(q,w) = \delta(q,w^{n+1})$. We know that
    $\app{f}{q} = \app{f}{t}$. Assume by contradiction that $t$ and $q$ are
    incomparable for the prefix relation. Let us split $w = w_1 w_2$ where
    $w_1$ is the shortest prefix of $w$ such that $s_0 \defined \delta(q,w_1)$
    is an ancestor of $q$ and of $t$ for the prefix relation, it must exist
    because $\delta(q,w_1 w_2) = t$.

    Now, consider $s_1 \defined \delta(t, w_1)$. Assume by contradiction that
    $s_0$ is not comparable with $s_1$ for the prefix relation. Then, consider
    the smallest prefix $v$ of $w_1$ such that $\delta(t, v)$ is a strict
    prefix of $s_0$. It must exist, otherwise $s_0$ is always a prefix of
    $s_1$. Because $\app{f}{t} = \app{f}{q}$, we conclude that $\app{f}{tv} =
    \app{f}{qv}$. However, this contradicts the minimality of $w_1$, since
    $\delta(t,v)$ is an ancestor of $q$ and $t$.

    We have proven that $s_0$ and $s_1$ are comparable, hence they are equal,
    since $\app{f}{s_1} = \app{f}{t w_1} = \app{f}{q w_1} = \app{f}{s_0}$.
    Finally, we have proven that $\delta(q, w_1) = s_0$, $\delta(s_0, w_2) =
    t$, and $\delta(s_0, w_2) = \delta(t, w_1w_2) = q$ which is absurd.
    As a consequence $t$ and $q$ were comparable for the prefix relation,
    hence equal, and therefore $\delta(q, w) = q$.
\end{proof}

\AP In the hope of generalizing this result to higher growth rates, we
introduce the notion of \intro{aperiodic ordering} of $\Sigma^*$, designed to
mimic the notion of \kl{aperiodic monoid} in the context of regular
languages. Let us recall that a monoid $M$ is \kl(monoid){aperiodic} whenever for
all $x \in M$, there exists $n \in \Nat$ such that $x^n = x^{n+1}$. In the
specific case of finite monoids, this $n$ can be chosen uniformly for all
elements of the monoid. We therefore state that an ordered set $(\Sigma^*,
\leq)$ is \reintro[aperiodic ordering]{aperiodic} whenever for all $u, w \in
\Sigma^*$, there exists $N_0 \in \Nat$, such that the sequence $\seqof{uw^n}{n
\geq N_0}$ is non-decreasing. Finally, we introduce the \emph{star-free
variant} $(\intro*\resleqsf{f}{k})$ of $(\resleq{f}{k})$, defined by $u
\reintro*\resleqsf{f}{k} v$ whenever $\Deriv{f}{v}{u} \in \NSF[k-1]$.
With these definitions at hand, we are ready to state our main conjecture.

\begin{conjecture}
    \label{sf-no-periods-on-sequences:conj}
    For all $k \in \Nat$ and $f \colon \Sigma^* \to \Rel$,
    $f \in \NSF[k]$ if and only if $(\Sigma^*, \resleqsf{f}{k})$ is an \kl[aperiodic ordering]{aperiodic}
    \kl{well-quasi-ordering}.
\end{conjecture}

Note that \cref{sf-no-periods-on-sequences:conj} cannot rely on the current
definition of a \kl{residual transducer}, even when using its \emph{star-free}
variant (with $\NSF[k-1]$ labels), because of
\cref{non-aperiodic-residual-transd:ex}. However, we can already prove one
direction of this conjecture: if $f \in \NSF[k]$, then $(\Sigma^*,
\resleqsf{f}{k})$ is an \kl(ordering){aperiodic} \kl{well-quasi-ordering}.

\begin{lemma}
    \label{sf-no-periods-on-sequences:lemma}
    Let $k \in \Nat$, and $f \in \NPoly[k]$. If $f \in \NSF[k]$, then
    $(\Sigma^*, \resleqsf{f}{k})$ is an
    \kl[aperiodic ordering]{aperiodic} \kl{well-quasi-ordering}.
\end{lemma}
\begin{proof}
    This proof will be a refinement of the one in
    \cref{n-poly-k-implies-wqo:lemma}.
    Let $f \in \NSF[k]$. By definition, there exists 
    an \kl{aperiodic monoid} $M$, a morphism $\mu \colon \Sigma^* \to M$,
    and a production function $\pi \colon M^{k+1} \to \Nat$
    such that $f = \pi^\dagger$.

    We can do the exact same computations as in
    \cref{n-poly-k-implies-wqo:lemma} to express the \kl{non-commutative
    derivatives} of $f$ using \emph{the same monoid $M$} but changing the
    production function $\pi$. In particular, reading the proof again, one
    concludes not only that the derivatives are in $\NPoly[k-1]$, but even in
    $\NSF[k-1]$, since they are computed using an \kl{aperiodic monoid}.
    This shows that $\resleqsf{f}{k}$ is a \kl{well-quasi-ordering}
    relation.

    Let us now prove that $(\Sigma^*, \resleqsf{f}{k})$ is \kl[aperiodic
    ordering]{aperiodic}. First, let us notice that for us to conclude, it
    suffices to prove that for some $n \in \Nat$, the inequality $\langle
    uw^{n+k} \rangle \leq \langle uw^{n+k+1} \rangle$ holds for every $k \in
    \Nat$, where $\langle u \rangle$ is defined as in the proof of
    \cref{n-poly-k-implies-wqo:lemma}: $\langle u \rangle$ is the vector
    containing $\mu(u)$ and the tables of the functions $\theta_u^{\ell}$ for
    $1 \leq \ell \leq k+1$. It is clear that there exists $N_0$ such that for
    all $n \geq N_0$, $\mu(uw^n) = \mu(uw^{n+1})$ because $M$ is
    \kl(monoid){aperiodic}.

    For the other components of the vector $\langle uw^n \rangle$, it suffices
    to notice that they can themselves be expressed as \kl{commutative}
    \kl{star-free} \kl{$\Nat$-polyregular functions} taking $n$ as input! Let
    us define $g \colon \Nat \to \Nat$ as $g(X) =
    \theta_{uw^X}^{\ell}(m_{\ell}, \dots, m_{k+1})$ for some $1 \leq \ell \leq
    k+1$ and $(m_{\ell}, \dots, m_{k+1}) \in M^{k + 2 - \ell}$. It is clear 
    that $g$ is a \kl{commutative} \kl{star-free} \kl{$\Nat$-polyregular function},
    hence
    one can leverage 
    \cref{zsf-npoly-nsf:thm,decide-rat-poly-npoly:cor}
    and conclude that 
    \begin{equation*}
      \exists N_0,  \exists P \in \Nat[X],
      g(x) = P(x) \text{ for all } x \geq N_0 \quad .
    \end{equation*}
    Therefore,
    $g(x + N_0 +1) - g(x + N_0) \in \Nat$, for all $x \geq 0$.
    Because there are finitely many choices of $\ell$ and $(m_{\ell}, \dots,
    m_{k+1})$, we can take the maximum of the $N_0$'s obtained for each of
    those, and conclude that 
    $\langle uw^n \rangle \leq \langle uw^{n+1} \rangle$ for all $n \geq N_0$,
\end{proof}

Let us now prove \cref{sf-no-periods-on-sequences:conj} in the specific case of
\kl{commutative} functions, where we can leverage the decidability results of
\cref{decidable-n-poly:thm,zsf-npoly-nsf:thm}.

\begin{theorem}
  \label{sf-no-periods-commutative:thm}
  Let $k \in \Nat$, and let $f \in \ZPoly[k]$ be a \kl{commutative}
  function.
  Then, the following are equivalent:
  \begin{enumerate}
    \item \label{sf-no-periods-commutative:item}
      $f \in \NSF[k]$,
    \item \label{sf-no-periods-commutative-wqo:item}
      $(\Sigma^*, \resleqsf{f}{k})$ is an
        \kl[aperiodic ordering]{aperiodic} \kl{well-quasi-ordering},
    \item \label{sf-no-periods-commutative-ultpoly:item}
      $f$ is \kl{ultimately polynomial}
      and $f$ is \kl{combinatorial}.
  \end{enumerate}
\end{theorem}
\begin{proof}
  We already proved that \cref{sf-no-periods-commutative:item} implies
  \cref{sf-no-periods-commutative-wqo:item} in
  \cref{sf-no-periods-on-sequences:lemma}.
  We also know that \cref{sf-no-periods-commutative-ultpoly:item}
  implies \cref{sf-no-periods-commutative:item}
  from \cref{decidable-n-poly:thm,zsf-npoly-nsf:thm}.
  It remains to prove that \cref{sf-no-periods-commutative-wqo:item}
  implies \cref{sf-no-periods-commutative-ultpoly:item}.

  Note that from \cref{non-commutative-npoly:thm}, we know that $f \in
  \NPoly[k]$, hence that $f$ is \kl{combinatorial}. It remains to show that $f$
  is \kl{ultimately polynomial}. Let $q \colon \Nat^k \to \Sigma^*$ be a
  \kl{pumping pattern}, because we already know that $f \in \NPoly[k]$, it
  suffices to consider patterns of the form $q(X) = u v^X w$ for some $u, v, w
  \in \Sigma^*$ \cite[Remark V.19]{CDTL23}.
  Because $(\Sigma^*, \resleqsf{f}{k})$ is an \kl[aperiodic
  ordering]{aperiodic} \kl{well-quasi-ordering}, there exists $N_0 \in \Nat$
  such that for all $x_1 \geq N_0$, $q(x_1) \resleqsf{f}{k} q(x_1 + 1)$.

  As a
  consequence, one can build a \kl{counter-free}  \kl{$\NSF$-transducer} for $f \circ
  q$ as follows: the states are numbers up to $N_0$, and the transitions are of
  the form $x \rightarrow x+1$ for all $x < N_0$, and $N_0 \rightarrow N_0$. The
  labels are defined as $0$ for all transitions $x \rightarrow x+1$, and
  $\Deriv{f}{q(N_0 + 1)}{q(N_0)}$ for the loop on $N_0$. Finally, the final
  output associated to state $x$ is $\app{f}{q(x)}$ for all $x \leq N_0$. This
  transducer \kl{computes} $f \circ q$, and has no \kl{counters}, hence $f \circ q \in
  \NSF$, and in particular, \kl{ultimately polynomial}. Using the identity 
  map $p \colon n \mapsto n$, we conclude that $f \circ q \circ p$
  is ultimately a polynomial, i.e., that $f$ is \kl{ultimately polynomial}.
\end{proof}

\subsection{The special case of linear growth}
\label{growth-rate-one:sec}

\AP In this section, we will upgrade the results of
\cref{aperiodic-star-free:sec} to a full characterization of \kl{star-free}
\kl{$\Nat$-polyregular} functions with linear growth rate, that correspond to
\kl{regular functions} ($\Regular$) with unary output alphabet. First, let us
recall that a function $f \colon \Sigma^* \to \Nat$ can be in $\ZPoly \setminus
\NPoly$, as \cref{non-neg-not-nrat:ex} shows. However, this specific example,
based on the polynomial $(X - Y)^2$, has a quadratic \kl{growth rate}. We will
prove in \cref{linear-npoly-zpoly:lem} that one recovers the \kl{Fatou
property} in the case of linear growth rate: any function $f \in \ZPoly[1]$
with non-negative values is in $\NPoly[1]$.

\begin{lemma}
    \label{linear-npoly-zpoly:lem}
    Let $f \in \ZPoly[1]$ be a function with non-negative values.
    Then, $f \in \NPoly[1]$.
\end{lemma}
\begin{proof}
  Let us first remark that the claim is obvious when $f \in \ZPoly[0]$:
  as $f$ is a finite linear combination of indicator functions of regular languages,
  the pre-image of a given constant $c$ is a regular language, hence
  $f$ can be re-computed as follows: 
  \begin{equation*}
    f(w) = \sum_{c \in \Nat} c \times \ind{f^{-1}(\set{c})}(w) \quad ,
  \end{equation*}
  which is a finite linear combination of indicator functions of regular languages
  where all coefficients are non-negative (because $f$ has non-negative values).
  If we knew that $f$ was in $\ZSF[0]$, we could even conclude that $f \in \NSF[0]$,
  since all the languages $f^{-1}(\set{c})$ would be star-free.

  Let us now turn to the difficult case, and consider a function $f \in
  \ZPoly[1]$ with non-negative values. We want to prove that $f \in \NPoly[1]$,
  and to do that, it suffices to prove that $(\Sigma^*, \resleq{f}{1})$ is a
  \kl{well-quasi-ordering}. To that end, let us consider an infinite sequence
  $\seqof{u_i}{i \in \Nat}$ of $\Sigma^*$. Without loss of generality, because
  $f \in \ZPoly[1]$, we can assume that $u_i \resequiv{f}{1} u_{i+1}$ for all
  $i \in \Nat$. To prove that $u_i \resleq{f}{1} u_j$ for some $i < j$, it
  suffices to prove that $\Deriv{f}{u_j}{u_i} \in \NPoly[0]$, and we already
  know that $\Deriv{f}{u_j}{u_i} \in \ZPoly[0]$. Leveraging the case of
  constant growth rate treated above, it suffices to prove that
  $\Deriv{f}{u_j}{u_i}$ has non-negative values for some $i < j$.

  Because $f \in \ZPoly[1]$, there exists a finite monoid $M$, a surjective\footnote{One can always assume surjectivity by considering a submonoid of $M$.} morphism $\mu
  \colon \Sigma^* \to M$ and a production function $\pi \colon M^{2} \to \Rel$
  such that $f = \pi^\dagger$. To a word $u \in \Sigma^*$, we associate the
  following vector of numbers in $\Rel^{\card{M}}$: $\llbracket u \rrbracket(m)
  \defined \sum_{u = s_1 s_2} \pi(\mu(s_1), \mu(s_2) m)$.
  If one could argue that there exists $i < j$ such that $\llbracket
  u_i \rrbracket \leq \llbracket u_j \rrbracket$ pointwise, and $\mu(u_i) =
  \mu(u_j)$,
  then we would
  immediately conclude, indeed,
  \begin{align*}
    \Deriv{f}{u_j}{u_i}(w)
    &= f(u_j w) - f(u_i w) \\
    &= \sum_{u_j = s_1 s_2} \pi(\mu(s_1), \mu(s_2) \mu(w))
     + \sum_{w = t_1 t_2} \pi(\mu(u_j), \mu(t_1) \mu(t_2)) \\
    &- \sum_{u_i = s_1 s_2} \pi(\mu(s_1), \mu(s_2) \mu(w))
     - \sum_{w = t_1 t_2} \pi(\mu(u_i), \mu(t_1) \mu(t_2)) \\
    &= \llbracket u_j \rrbracket(\mu(w)) - \llbracket u_i \rrbracket(\mu(w)) \geq 0 \quad .
  \end{align*}

  We will first prove that $\llbracket u_i \rrbracket$ cannot contain
  arbitrarily large negative values, that is, there exists a finite $B \in
  \Nat$ such that for all $i \in \Nat$, $\min_{m \in M} \llbracket u_i
  \rrbracket(m) \geq -B$. Assume towards a contradiction that this is not the
  case, then there exists an infinite subsequence $\seqof{u_{\alpha(i)}}{i \in
  \Nat}$ such that $\min_{m \in M} \llbracket u_{\alpha(i)} \rrbracket(m)$
  tends to $-\infty$ as $i$ tends to infinity. Without loss of generality
  (because $M$ is finite), a single component $m_1 \in M$ tends to infinity,
  and all the elements $u_{\alpha(i)}$ evaluate to some $m_0 \in M$ under
  $\mu$. Since $\mu$ is surjective, there is a word $v$ such that $\mu(v) =
  m_1$. Therefore, $f(u_{\alpha(i)} v) = \llbracket u_{\alpha(i)}
  \rrbracket(m_1) + \sum_{v = t_1 t_2} \pi(m_0\mu(t_1), \mu(t_2))$, and since
  the second term is constant (independent of $i$), we conclude that
  $f(u_{\alpha(i)} v)$ tends to $-\infty$ as $i$ tends to infinity,
  contradicting the fact that $f$ has non-negative values.

  Therefore, up to adding a constant $B$ to all the functions $\llbracket u_i
  \rrbracket$, we can assume that they all take non-negative values, and apply
  Dickson's lemma to conclude that there exists $i < j$ such that $\llbracket
  u_i \rrbracket \leq \llbracket u_j \rrbracket$ pointwise.
  This concludes our proof for $\NPoly[1]$.
\end{proof}

If $f \in \ZSF[1]$ and is non-negative, one cannot rely on a similar argument
to conclude that $f \in \NSF[1]$, because we do not know whether
\cref{sf-no-periods-on-sequences:conj} holds or not. However, it was noticed in
\cref{aperiodic-intersection:remark} that in the specific case of
\emph{rational functions} $\Rational$, an analogue of \kl{ultimately
polynomial} (adapted to non-unary outputs) guarantees belonging to $\ARational$
(see \cref{eq:aperiodic-sequential-pumping}). The following lemma restates one
of the main results of \cite[Section 4]{RESCH95} to the specific case of unary
outputs.

\begin{lemma}
    \label{rational-ultpoly-aperiodic:lem}
    Let $f \in \NPoly[1]$ be \kl{ultimately polynomial}.
    Then, $f \in \NSF[1]$.
\end{lemma}
\begin{proof}
  As $f \in \NPoly[1]$, one can see $f$ as a \kl{rational function}
  with unary output alphabet. Because $f$ is \kl{ultimately polynomial},
  it satisfies \cref{eq:aperiodic-sequential-pumping}, hence 
  it is computable by a function in
  $\ARational$, that is an aperiodic rational function 
  \cite[Section 4, page 236]{RESCH95}.
  Therefore, $f \in \NSF[1]$.
\end{proof}

Combining \cref{linear-npoly-zpoly:lem,rational-ultpoly-aperiodic:lem}, we
obtain the following characterization of \kl{star-free} \kl{$\Nat$-polyregular}
functions with linear growth rate: we have proven that for such functions,
$\ZSF[1] \cap \NPoly[1] = \NSF[1]$, and that $\ZPoly[1] \cap \set{
\text{non-negative functions} } = \NPoly[1]$. From these observations, we
deduce interesting corollaries.

\begin{corollary}
  \label{linear-commutative-nsf:cor}
  Let $f \in \ZPoly[1]$.
  \begin{itemize}
    \item One can decide whether $f$ is in $\NPoly[1]$,
    \item One can decide whether $f$ is non-negative,
    \item One can decide whether $f$ is in $\NSF[1]$,
    \item If $f$ is non-negative, the preimage of any $n \in \Nat$ is a
      regular language, and one can compute a finite automaton recognizing it,
    \item If $f$ is non-negative and in $\ZSF[1]$, then the preimages of finite sets
      are even
      star-free languages.
  \end{itemize}
\end{corollary}
\section{Outlook}
\label{sec:ccl}

Let us end on a more general discussion regarding the status of commutative
input functions in the study of unary output polyregular functions. A
\emph{quantitative pumping argument} for polyregular function $f \colon
\Sigma^* \to \Rel$ states that $f$ has property $X$ if and only if for all
\kl{pumping pattern} $q \colon \Nat^k \to \Sigma^*$, $f \circ q$ has property
$X$. Let us formalize such a statement for \kl{growth rate} and
\kl(polyregular){aperiodicity} respectively in \cref{pre-compose-growth-commut:lemma,pre-compose-sf-commut:lemma}. Note that we generalized \kl{pumping
patterns} to \kl{commutative} \kl{star-free polyregular functions} to simplify
the statements.

\begin{lemma}[restate=pre-compose-growth-commut:lemma,label=pre-compose-growth-commut:lemma]
    \proofref{pre-compose-growth-commut:lemma}
    Let $f \in \ZRat$, and $d \in \Nat$. Then,
    $f \in \ZPoly[d]$ if and only if 
    for every \kl{commutative} \kl{star-free polyregular function} $h$
            of \kl{growth rate} $l \in \Nat$,
            $(f \circ h) \in \ZPoly[d \times l]$.
\end{lemma}
\begin{proof}
    If $f \in \ZPoly[k]$, then 
    for every \kl{polyregular function} $g \in \Poly[\ell]$,
    $f \circ g \in \ZPoly[k \times \ell]$
    \cite{CDTL23}.

    Conversely, it was proven in \cite[Theorem III.3]{CDTL23}
    that $f \in \ZPoly[k]$ if and only if
    for all $\alpha_0, \cdots, \alpha_k$,
    for all $w_1, \dots, w_k$,
    we have 
    \begin{equation}
        \label{k-pumpable:eqn}
        f\left(
            \alpha_0 \prod_{i = 1}^k w_i^{X_i} \alpha_i
        \right)
        = \bigO( (X_1 + \cdots X_k)^{k} )
        \quad .
    \end{equation}
    Now, the \kl{commutative} \kl{star-free polyregular function} 
    $h \colon \set{1, \dots, k}^* \to \Sigma^*$ that maps
    a word $u$
    to $\alpha \prod_{i = 1}^k w_i^{\card[i]{u}}$
    has \kl{growth rate} $1$.
    Hence, 
    $f \circ h \in \ZPoly[k]$, 
    and we indeed conclude that 
    \cref{k-pumpable:eqn}
    holds, i.e. that $f \in \ZPoly[k]$.
\end{proof}

\begin{lemma}[restate=pre-compose-sf-commut:lemma,label=pre-compose-sf-commut:lemma]
    \proofref{pre-compose-sf-commut:lemma}
    Let $f \in \ZPoly$. Then, $f \in \ZSF$,
    if and only if for every \kl{commutative} \kl{star-free polyregular function} $h$,
            $(f \circ h) \in \ZSF$.
\end{lemma}
\begin{proof}
    If $f \in \ZSF$, then for all $h \in \SF$,
    $f \circ h \in \ZSF$, and \emph{a fortiori}
    for \kl{commutative} functions $h$.

    Conversely, assume that $f \circ h \in \ZSF$
    for all \kl{commutative} functions $h \in \SF$.
    Using \cite[Theorem V.13]{CDTL23},
    to conclude that $f \in \ZSF$,
    it suffices to prove that
    for all $\alpha_0, \cdots, \alpha_k \in \Sigma^*$,
    for all $w_1, \dots, w_k \in \Sigma^*$,
    there exists a polynomial $P \in \Rel[X_1, \dots,X_k]$
    and a constant $N_0 \in \Nat$,
    such that if $X_1, \dots, X_k \geq N_0$:
    \begin{equation}
        \label{ultimate-polynomial:eqn}
        f\left(
            \alpha_0 \prod_{i = 1}^k w_i^{X_i} \alpha_i
        \right)
        = P(X_1, \dots, X_k)
        \quad .
    \end{equation}
    Let us consider
    the \kl{commutative} \kl{star-free polyregular function}
    $h \colon \set{1, \dots, k}^* \to \Sigma^*$ that maps
    a word $u$
    to $\alpha \prod_{i = 1}^k w_i^{\card[i]{u}}$.
    We know that
    $f \circ h \in \ZSF$, hence, that 
    there exists $Q \in \Rel[X_1, \dots, X_n]$
    and $M_0 \in \Nat$,
    such that for all $X_1, \dots, X_n \geq N_0$,
    $f \circ h( \prod_{i = 1}^k \underline{i}^{X_i}) = Q(X_1, \dots, X_n)$.
    In particular,
    we can take $N_0 = M_0$, and $P = Q$ to conclude that
    \cref{ultimate-polynomial:eqn} holds, hence, that
    $f \in \ZSF$.
\end{proof}

Remark that if \cref{pre-compose-sf-commut:lemma} were to hold for
\kl{$\Nat$-polyregular functions}, then the decidability of $\NPoly$ inside
$\ZPoly$, and the decidability of $\NSF$ inside $\NPoly$ would immediately
follow. On the one hand, one can guess a candidate function and check for
equivalence, on the other hand, one can guess a \kl{commutative} \kl{star-free
polyregular function} and check membership (which is decidable thanks to this
paper). This is restated in our concluding conjecture.

\begin{conjecture}
    \label{npoly-zpoly:conjecture}
    Let $f \in \ZPoly$. Then, $f \in \NPoly$ if and only if for every
    \kl{commutative} \kl{star-free polyregular function} $h$,
    $(f \circ h) \in \NPoly$.
\end{conjecture}
 
\clearpage

\bibliographystyle{plainurl}

\begin{thebibliography}{10}

\bibitem{BGMP15}
Félix Baschenis, Olivier Gauwin, Anca Muscholl, and Gabriele Puppis.
\newblock {One-way Definability of Sweeping Transducer}.
\newblock In Prahladh Harsha and G.~Ramalingam, editors, {\em 35th IARCS Annual Conference on Foundations of Software Technology and Theoretical Computer Science (FSTTCS 2015)}, volume~45 of {\em Leibniz International Proceedings in Informatics (LIPIcs)}, pages 178--191, Dagstuhl, Germany, 2015. Schloss Dagstuhl -- Leibniz-Zentrum f{\"u}r Informatik.
\newblock URL: \url{https://drops-dev.dagstuhl.de/entities/document/10.4230/LIPIcs.FSTTCS.2015.178}, \href {https://arxiv.org/abs/1706.01668v3} {\path{arXiv:1706.01668v3}}, \href {https://doi.org/10.4230/LIPIcs.FSTTCS.2015.178} {\path{doi:10.4230/LIPIcs.FSTTCS.2015.178}}.

\bibitem{BERE10}
Jean Berstel and Christophe Reutenauer.
\newblock {\em Noncommutative rational series with applications}, volume 137 of {\em Encyclopedia of Mathematics and its Applications}.
\newblock Cambridge University Press, 2010.
\newblock \href {https://doi.org/10.1017/CBO9780511760860} {\path{doi:10.1017/CBO9780511760860}}.

\bibitem{BOJA14}
Mikołaj Bojańczyk.
\newblock {Transducers with Origin Information}.
\newblock In Javier Esparza, Pierre Fraigniaud, Thore Husfeldt, and Elias Koutsoupias, editors, {\em Automata, Languages, and Programming}, pages 26--37. Springer Berlin Heidelberg, 2014.
\newblock \href {https://arxiv.org/abs/1309.6124v1} {\path{arXiv:1309.6124v1}}, \href {https://doi.org/10.1007/978-3-662-43951-7_3} {\path{doi:10.1007/978-3-662-43951-7_3}}.

\bibitem{BOJA18}
Mikołaj Bojańczyk.
\newblock Polyregular functions, 2018.
\newblock URL: \url{https://arxiv.org/abs/1810.08760}.

\bibitem{BDK18}
Mikołaj Bojańczyk, Laure Daviaud, and Shankara~Narayanan Krishna.
\newblock Regular and first-order list functions.
\newblock In {\em Proceedings of the 33rd Annual ACM/IEEE Symposium on Logic in Computer Science}, LICS '18, page 125–134, New York, NY, USA, 2018. Association for Computing Machinery.
\newblock URL: \url{https://www.mimuw.edu.pl/~bojan/upload/conflicsBojanczykDK18.pdf}, \href {https://arxiv.org/abs/1803.06168} {\path{arXiv:1803.06168}}, \href {https://doi.org/10.1145/3209108.3209163} {\path{doi:10.1145/3209108.3209163}}.

\bibitem{BOKL19}
Mikołaj Bojańczyk, Sandra Kiefer, and Nathan Lhote.
\newblock {String-to-String Interpretations With Polynomial-Size Output}.
\newblock In Christel Baier, Ioannis Chatzigiannakis, Paola Flocchini, and Stefano Leonardi, editors, {\em 46th International Colloquium on Automata, Languages, and Programming (ICALP 2019)}, volume 132 of {\em Leibniz International Proceedings in Informatics (LIPIcs)}, pages 106:1--106:14, Dagstuhl, Germany, 2019. Schloss Dagstuhl -- Leibniz-Zentrum f{\"u}r Informatik.
\newblock URL: \url{https://drops-dev.dagstuhl.de/entities/document/10.4230/ LIPIcs.ICALP.2019.106}, \href {https://doi.org/10.4230/LIPIcs.ICALP.2019.106} {\path{doi:10.4230/LIPIcs.ICALP.2019.106}}.

\bibitem{CACHA1996}
Paul-Jean Cahen and Jean-Luc Chabert.
\newblock {\em Integer-Valued Polynomials}.
\newblock American Mathematical Society, December 1996.
\newblock URL: \url{https://people.math.rochester.edu/faculty/doug/otherpapers/ Cahen-Chabert.pdf}, \href {https://doi.org/10.1090/surv/048} {\path{doi:10.1090/surv/048}}.

\bibitem{CADA15}
Olivier Carton and Luc Dartois.
\newblock {Aperiodic Two-way Transducers and FO-Transductions}.
\newblock In Stephan Kreutzer, editor, {\em 24th EACSL Annual Conference on Computer Science Logic (CSL 2015)}, volume~41 of {\em Leibniz International Proceedings in Informatics (LIPIcs)}, pages 160--174, Dagstuhl, Germany, 2015. Schloss Dagstuhl -- Leibniz-Zentrum f{\"u}r Informatik.
\newblock URL: \url{https://drops-dev.dagstuhl.de/entities/document/10.4230/LIPIcs.CSL.2015.160}, \href {https://doi.org/10.4230/LIPIcs.CSL.2015.160} {\path{doi:10.4230/LIPIcs.CSL.2015.160}}.

\bibitem{CHOF03}
Christian Choffrut.
\newblock Minimizing subsequential transducers: a survey.
\newblock {\em Theoretical Computer Science}, 292(1):131--143, 2003.
\newblock Selected Papers in honor of Jean Berstel.
\newblock URL: \url{https://www.sciencedirect.com/science/article/pii/S0304397501002195}, \href {https://doi.org/10.1016/S0304-3975(01)00219-5} {\path{doi:10.1016/S0304-3975(01)00219-5}}.

\bibitem{CDTL23}
Thomas Colcombet, Gaëtan Douéneau-Tabot, and Aliaume Lopez.
\newblock {Z}-polyregular functions.
\newblock In {\em 2023 38th Annual ACM/IEEE Symposium on Logic in Computer Science (LICS)}, pages 1--13, Los Alamitos, CA, USA, jun 2023. IEEE Computer Society.
\newblock URL: \url{https://doi.ieeecomputersociety.org/10.1109/ LICS56636.2023.10175685}, \href {https://doi.org/10.1109/LICS56636.2023.10175685} {\path{doi:10.1109/LICS56636.2023.10175685}}.

\bibitem{DGK21}
Luc Dartois, Paul Gastin, and Shankara~Narayanan Krishna.
\newblock Sd-regular transducer expressions for aperiodic transformations.
\newblock In {\em 2021 36th Annual ACM/IEEE Symposium on Logic in Computer Science (LICS)}, pages 1--13, 2021.
\newblock \href {https://doi.org/10.1109/LICS52264.2021.9470738} {\path{doi:10.1109/LICS52264.2021.9470738}}.

\bibitem{DJR16}
Luc Dartois, Ismaël Jecker, and Pierre-Alain Reynier.
\newblock Aperiodic string transducers.
\newblock In Srecko Brlek and Christophe Reutenauer, editors, {\em Developments in Language Theory - 20th International Conference, {DLT} 2016, Montr{\'{e}}al, Canada, July 25-28, 2016, Proceedings}, volume 9840 of {\em Lecture Notes in Computer Science}, pages 125--137. Springer, 2016.
\newblock \href {https://doi.org/10.1007/978-3-662-53132-7\_11} {\path{doi:10.1007/978-3-662-53132-7\_11}}.

\bibitem{DAVIS1973}
Martin Davis.
\newblock Hilbert’s tenth problem is unsolvable.
\newblock {\em The American Mathematical Monthly}, 80(3):233--269, March 1973.
\newblock URL: \url{http://dx.doi.org/10.1080/00029890.1973.11993265}, \href {https://doi.org/10.1080/00029890.1973.11993265} {\path{doi:10.1080/00029890.1973.11993265}}.

\bibitem{DOUE21}
Gaëtan Douéneau-Tabot.
\newblock {Pebble Transducers with Unary Output}.
\newblock In Filippo Bonchi and Simon~J. Puglisi, editors, {\em 46th International Symposium on Mathematical Foundations of Computer Science (MFCS 2021)}, volume 202 of {\em Leibniz International Proceedings in Informatics (LIPIcs)}, pages 40:1--40:17, Dagstuhl, Germany, 2021. Schloss Dagstuhl -- Leibniz-Zentrum f{\"u}r Informatik.
\newblock URL: \url{https://drops-dev.dagstuhl.de/entities/document/10.4230/ LIPIcs.MFCS.2021.40}, \href {https://doi.org/10.4230/LIPIcs.MFCS.2021.40} {\path{doi:10.4230/LIPIcs.MFCS.2021.40}}.

\bibitem{DOUE22}
Gaëtan Douéneau-Tabot.
\newblock {Hiding Pebbles When the Output Alphabet Is Unary}.
\newblock In Mikołaj Bojańczyk, Emanuela Merelli, and David~P. Woodruff, editors, {\em 49th International Colloquium on Automata, Languages, and Programming (ICALP 2022)}, volume 229 of {\em Leibniz International Proceedings in Informatics (LIPIcs)}, pages 120:1--120:17, Dagstuhl, Germany, 2022. Schloss Dagstuhl -- Leibniz-Zentrum f{\"u}r Informatik.
\newblock URL: \url{https://drops-dev.dagstuhl.de/entities/document/10.4230/ LIPIcs.ICALP.2022.120}, \href {https://doi.org/10.4230/LIPIcs.ICALP.2022.120} {\path{doi:10.4230/LIPIcs.ICALP.2022.120}}.

\bibitem{DOUE23}
Gaëtan Douéneau-Tabot.
\newblock {\em Optimization of string transducers}.
\newblock PhD thesis, Université Paris Cité, 2023.
\newblock URL: \url{https://gdoueneau.github.io/pages/DOUENEAU- TABOT_Optimization-transducers_v2.pdf}.

\bibitem{DRGA19}
Manfred Droste and Paul Gastin.
\newblock {Aperiodic Weighted Automata and Weighted First-Order Logic}.
\newblock In Peter Rossmanith, Pinar Heggernes, and Joost-Pieter Katoen, editors, {\em 44th International Symposium on Mathematical Foundations of Computer Science (MFCS 2019)}, volume 138 of {\em Leibniz International Proceedings in Informatics (LIPIcs)}, pages 76:1--76:15, Dagstuhl, Germany, 2019. Schloss Dagstuhl -- Leibniz-Zentrum f{\"u}r Informatik.
\newblock URL: \url{https://drops-dev.dagstuhl.de/entities/document/10.4230/LIPIcs.MFCS.2019.76}, \href {https://arxiv.org/abs/1902.08149v3} {\path{arXiv:1902.08149v3}}, \href {https://doi.org/10.4230/LIPIcs.MFCS.2019.76} {\path{doi:10.4230/LIPIcs.MFCS.2019.76}}.

\bibitem{EILE74}
Samuel Eilenberg.
\newblock {\em Automata, Languages, and Machines}, volume~A of {\em Automata, Languages, and Machines}.
\newblock Academic Press, 1974.
\newblock URL: \url{https://books.google.pl/books?id=vVh0xgEACAAJ}, \href {https://doi.org/10.5555/540244} {\path{doi:10.5555/540244}}.

\bibitem{ENMA02}
Joost Engelfriet and Sebastian Maneth.
\newblock Two way finite state transducers with nested pebbles.
\newblock In Krzysztof Diks and Wojciech Rytter, editors, {\em Mathematical Foundations of Computer Science 2002}, volume 2420, pages 234--244, Berlin, Heidelberg, 2002. Springer Berlin Heidelberg.
\newblock \href {https://doi.org/10.1007/3-540-45687-2_19} {\path{doi:10.1007/3-540-45687-2_19}}.

\bibitem{FGL16}
Emmanuel Filiot, Olivier Gauwin, and Nathan Lhote.
\newblock {Aperiodicity of Rational Functions Is PSPACE-Complete}.
\newblock In Akash Lal, S.~Akshay, Saket Saurabh, and Sandeep Sen, editors, {\em 36th IARCS Annual Conference on Foundations of Software Technology and Theoretical Computer Science (FSTTCS 2016)}, volume~65 of {\em Leibniz International Proceedings in Informatics (LIPIcs)}, pages 13:1--13:15, Dagstuhl, Germany, 2016. Schloss Dagstuhl -- Leibniz-Zentrum f{\"u}r Informatik.
\newblock URL: \url{https://drops-dev.dagstuhl.de/entities/document/10.4230/LIPIcs.FSTTCS.2016.13}, \href {https://doi.org/10.4230/LIPIcs.FSTTCS.2016.13} {\path{doi:10.4230/LIPIcs.FSTTCS.2016.13}}.

\bibitem{FGLM18}
Emmanuel Filiot, Olivier Gauwin, Nathan Lhote, and Anca Muscholl.
\newblock On canonical models for rational functions over infinite words.
\newblock In Sumit Ganguly and Paritosh Pandya, editors, {\em 38th IARCS Annual Conference on Foundations of Software Technology and Theoretical Computer Science (FSTTCS 2018)}, volume 122 of {\em Leibniz International Proceedings in Informatics (LIPIcs)}, pages 30:1--30:17, Dagstuhl, Germany, 2018. Schloss Dagstuhl -- Leibniz-Zentrum f{\"u}r Informatik.
\newblock URL: \url{https://drops-dev.dagstuhl.de/entities/document/10.4230/LIPIcs.FSTTCS.2018.30}, \href {https://doi.org/10.4230/LIPIcs.FSTTCS.2018.30} {\path{doi:10.4230/LIPIcs.FSTTCS.2018.30}}.

\bibitem{FGRS13}
Emmanuel Filiot, Olivier Gauwin, Pierre-Alain Reynier, and Frederic Servais.
\newblock From two-way to one-way finite state transducers.
\newblock In {\em Proceedings of the 2013 28th Annual ACM/IEEE Symposium on Logic in Computer Science}, LICS '13, page 468–477. IEEE Computer Society, 2013.
\newblock \href {https://arxiv.org/abs/1301.5197v2} {\path{arXiv:1301.5197v2}}.

\bibitem{FKT14}
Emmanuel Filiot, Shankara~Narayanan Krishna, and Ashutosh Trivedi.
\newblock {First-order Definable String Transformations}.
\newblock In Venkatesh Raman and S.~P. Suresh, editors, {\em 34th International Conference on Foundation of Software Technology and Theoretical Computer Science (FSTTCS 2014)}, volume~29 of {\em Leibniz International Proceedings in Informatics (LIPIcs)}, pages 147--159, Dagstuhl, Germany, 2014. Schloss Dagstuhl -- Leibniz-Zentrum f{\"u}r Informatik.
\newblock URL: \url{https://drops-dev.dagstuhl.de/entities/document/10.4230/LIPIcs.FSTTCS.2014.147}, \href {https://doi.org/10.4230/LIPIcs.FSTTCS.2014.147} {\path{doi:10.4230/LIPIcs.FSTTCS.2014.147}}.

\bibitem{HILB1902}
David Hilbert.
\newblock Mathematical problems.
\newblock {\em Bulletin of the American Mathematical Society}, 8(10):437--479, 1902.
\newblock URL: \url{https://www.ams.org/journals/bull/1902-08-10/S0002-9904- 1902-00923-3/}, \href {https://doi.org/10.1090/s0002-9904-1902-00923-3} {\path{doi:10.1090/s0002-9904-1902-00923-3}}.

\bibitem{KARH77}
Juhani Karhumäki.
\newblock Remarks on commutative {N}-rational series.
\newblock {\em Theoretical Computer Science}, 5(2):211--217, 1977.
\newblock URL: \url{https://www.sciencedirect.com/science/article/pii/ 0304397577900081}, \href {https://doi.org/10.1016/0304-3975(77)90008-1} {\path{doi:10.1016/0304-3975(77)90008-1}}.

\bibitem{lopez:LIPIcs.STACS.2025.67}
Aliaume Lopez.
\newblock {Commutative $\mathbb{N}$-Rational Series of Polynomial Growth}.
\newblock In Olaf Beyersdorff, Micha{\l} Pilipczuk, Elaine Pimentel, and Nguy\~{ê}n~Kim Thắng, editors, {\em 42nd International Symposium on Theoretical Aspects of Computer Science (STACS 2025)}, volume 327 of {\em Leibniz International Proceedings in Informatics (LIPIcs)}, pages 67:1--67:16, Dagstuhl, Germany, 2025. Schloss Dagstuhl -- Leibniz-Zentrum f{\"u}r Informatik.
\newblock URL: \url{https://drops.dagstuhl.de/entities/document/10.4230/LIPIcs.STACS.2025.67}, \href {https://doi.org/10.4230/LIPIcs.STACS.2025.67} {\path{doi:10.4230/LIPIcs.STACS.2025.67}}.

\bibitem{MATI1970}
Yuri~Vladimirovich Matiyasevich.
\newblock The diophantineness of enumerable sets.
\newblock {\em Doklady Akademii Nauk SSSR}, 191:279--282, 1970.
\newblock in Russian.

\bibitem{MNPA71}
Robert McNaughton and Seymour~A. Papert.
\newblock {\em Counter-Free Automata}.
\newblock The MIT Press, 1971.
\newblock \href {https://doi.org/10.5555/1097043} {\path{doi:10.5555/1097043}}.

\bibitem{MEAL55}
George~H. Mealy.
\newblock A method for synthesizing sequential circuits.
\newblock {\em The Bell System Technical Journal}, 34(5):1045--1079, 1955.
\newblock \href {https://doi.org/10.1002/j.1538-7305.1955.tb03788.x} {\path{doi:10.1002/j.1538-7305.1955.tb03788.x}}.

\bibitem{LENP21}
Lê Thành Dũng~(Tito) Nguyễn, Camille Noûs, and Cécilia Pradic.
\newblock Comparison-free polyregular functions.
\newblock In Nikhil Bansal, Emanuela Merelli, and James Worrell, editors, {\em 48th International Colloquium on Automata, Languages, and Programming (ICALP 2021)}, volume 198 of {\em Leibniz International Proceedings in Informatics (LIPIcs)}, pages 139:1--139:20, Dagstuhl, Germany, 2021. Schloss Dagstuhl -- Leibniz-Zentrum für Informatik.
\newblock URL: \url{https://drops-dev.dagstuhl.de/entities/document/10.4230/LIPIcs.ICALP.2021.139}, \href {https://arxiv.org/abs/2105.08358v2} {\path{arXiv:2105.08358v2}}, \href {https://doi.org/10.4230/LIPIcs.ICALP.2021.139} {\path{doi:10.4230/LIPIcs.ICALP.2021.139}}.

\bibitem{POLYA1915}
G.~Pólya.
\newblock Über ganzwertige ganze {Funktionen}.
\newblock {\em Rend. Circ. Mat. Palermo}, 40:1--16, 1915.
\newblock URL: \url{https://zbmath.org/?format=complete&q=an:45.0655.02}, \href {https://doi.org/10.1007/BF03014836} {\path{doi:10.1007/BF03014836}}.

\bibitem{REUT80}
Christophe Reutenauer.
\newblock Séries formelles et algèbres syntactiques.
\newblock {\em Journal of Algebra}, 66(2):448--483, 1980.
\newblock URL: \url{https://www.sciencedirect.com/science/article/pii/0021869380900976}, \href {https://doi.org/10.1016/0021-8693(80)90097-6} {\path{doi:10.1016/0021-8693(80)90097-6}}.

\bibitem{RESCH95}
Christophe Reutenauer and Marcel~Paul Schützenberger.
\newblock Variétés et fonctions rationnelles.
\newblock {\em Theoretical Computer Science}, 145(1–2):229–240, July 1995.
\newblock URL: \url{http://dx.doi.org/10.1016/0304-3975(94)00180-Q}, \href {https://doi.org/10.1016/0304-3975(94)00180-q} {\path{doi:10.1016/0304-3975(94)00180-q}}.

\bibitem{SCHU62}
Marcel~P. Schützenberger.
\newblock Finite counting automata.
\newblock {\em Information and control}, 5(2):91--107, 1962.
\newblock URL: \url{https://www.sciencedirect.com/science/article/pii/ S0019995862902449}, \href {https://doi.org/10.1016/S0019-9958(62)90244-9} {\path{doi:10.1016/S0019-9958(62)90244-9}}.

\bibitem{SCHU65}
Marcel~P. Schützenberger.
\newblock On finite monoids having only trivial subgroups.
\newblock {\em Information and Control}, 8(2):190--194, 1965.
\newblock \href {https://doi.org/10.1016/S0019-9958(65)90108-7} {\path{doi:10.1016/S0019-9958(65)90108-7}}.

\bibitem{SCHU77}
Marcel~P. Schützenberger.
\newblock Sur une variante des fonctions sequentielles.
\newblock {\em Theoretical Computer Science}, 4(1):47--57, 1977.
\newblock URL: \url{https://www.sciencedirect.com/science/article/pii/030439757790055X}, \href {https://doi.org/10.1016/0304-3975(77)90055-X} {\path{doi:10.1016/0304-3975(77)90055-X}}.

\bibitem{THOM97}
Wolfgang Thomas.
\newblock Languages, automata, and logic.
\newblock In Grzegorz Rozenberg and Arto Salomaa, editors, {\em Handbook of formal languages}, pages 389--455. Springer, 1997.
\newblock \href {https://doi.org/10.1007/978-3-642-59136-5} {\path{doi:10.1007/978-3-642-59136-5}}.

\end{thebibliography}
\appendix

\section{Proofs of section \ref{polynomials:sec}}

\begin{proofof}[n-poly-combinatorics:lem]
	Let us fix a finite monoid $M$, a morphism $\mu \colon \Sigma^* \to M$, and
	let $\omega$ be an idempotent power for the finite monoid $M$,
	i.e., a number such that for all $x \in M$,
	$(x^{\omega})^2 = x^{\omega}$.
	We will prove by induction on $k \in \Nat_{\geq 1}$
	that for all
	$\pi \colon M^k \to \Nat$,
	the lemma holds for the function
	$f = \pi^\dagger$, i.e., such that
	$f(w)$ is the sum over all factorizations of $w$
	into $k$ words $(w_1, \dots, w_k) \in \Sigma^*$
	of the value $\pi(\mu(w_1), \dots, \mu(w_k))$.
	Since every \kl{$\Nat$-polyregular function}
	is obtained via some choice of $M, \mu, \pi$ (\cref{nat-rel-poly:def})
	we derive the desired result.

	In all cases, we will use a \kl{pumping pattern} $q \colon \Nat^p \to
		\Sigma^*$ defined by $q(X_1, \dots, X_p) \defined \alpha_0 \prod_{i = 1}^p
		(u_i^{\omega \times X_i} \alpha_i)$ using some words $\alpha_i$ (for $0 \leq i
		\leq p$) and $u_i$ (for $1 \leq i \leq p)$ in $\Sigma^*$. That is, we will
	implicitly multiply all variables by $\omega$ in order to make the
	equations more readable. Note that when using an induction hypothesis, we
	will therefore have to check that the polynomial $q$ will have this
	specific form.

	\textbf{Base case: $k = 1$.}
	Let us show that $f \circ q$ is
	actually a constant (positive) function (and in particular a \kl{natural binomial function}), as for
	all $n_1, \dots, n_p \geq 1$,
	\begin{align*}
		f(q(n_1, \dots, n_p))
		 & \eqdef \pi\left(\mu(q(n_1, \dots, n_p))\right)                                      \\
		 & \eqdef \pi\left(\mu(\alpha_0 \prod_{i = 1}^p u_i^{\omega n_i} \alpha_i)\right)
		\\
		 & = \pi\left(\mu(\alpha_0) \prod_{i = 1}^p \mu(u_i)^{\omega n_i} \mu(\alpha_i)\right)
		 & \text{ morphism }
		\\
		 & = \pi\left(\mu(\alpha_0) \prod_{i = 1}^p \mu(u_i) \mu(\alpha_i)\right)
		 & \text{ idempotent and } n_i \geq 1                                                  \\
		 & = c \geq 0
	\end{align*}

	\textbf{Induction hypothesis.}
	The proof will essentially follow the same schema as the base case,
	with extra care needed to properly handle the summation of polynomials
	that arises when $k \geq 2$.

	We will use the fact that given $n_1, \dots, n_p \geq 0$,
	there exists a bijection between
	partitions of $q(\omega n_1, \dots, \omega n_p)$
	into $k+1$ words, and
	triples composed of
	a position  $1 \leq \ell \leq p$,
	a ratio     $0 \leq b \leq n_\ell$,
	a remainder $0 \leq r < \omega \times \card{u_\ell}$,
	and a partition of
	$(u_\ell^\omega)_{[-r:]} u_\ell^{\omega \times (n_\ell - b)}
		\alpha_{\ell+1} \cdots \alpha_{p-1} u_p^{n_p} \alpha_{p}$
	into $k$ words,
	where we have used the notation
	$(v)_{[-r:]}$ to denote the word obtained from $v$
	by keeping only the last $r$ letters, in a sort of Pythonic syntax.
	That is, we look at where the first ``cut" is made.

	Now, given $(\ell, b, r)$ respecting the inequalities
	mentioned above,
	one can define the \kl{pumping patterns}
	$\CutPref{\ell,r,b} \colon \Nat^{\ell - 1} \to \Sigma^*$
	and
	$\CutSuff{\ell,r} \colon \Nat^{p - \ell + 1} \to \Sigma^*$
	so that
	$q(X_1, \dots, X_p) = \CutPref{\ell,r,b}(X_1, \dots, X_{\ell-1})
		q_{\ell, r}(X_{\ell} - b, \dots, X_p)$
	under the assumption that $X_\ell \geq b$:
	\begin{align*}
        \CutPref{\ell,r,b}(X_1, \dots, X_{\ell - 1})
        &\defined
        \alpha_0 u_1^{\omega X_1} \alpha_2 \dots \alpha_{\ell - 1}
        u_\ell^{\omega b} (u_{\ell}^\omega)_{[1:r]}
        \quad .
        \\
		\CutSuff{\ell,r}(X_{1}, \dots, X_{p - \ell + 1})
        &\defined
        (u_\ell^\omega)_{[-r:]} u_\ell^{\omega X_1}
		\alpha_{\ell+1} \cdots \alpha_{p-1} u_p^{X_{p - \ell + 1}} \alpha_{p}
		\quad .
	\end{align*}
	Remark that $\CutSuff{\ell,r}$ has all of its indeterminates
	multiplied by $\omega$, so that the induction hypothesis
	can be used.

	The key remark is that $\mu(\CutPref{\ell,r,b}(X_1, \dots, X_{\ell - 1}))$ is a
	constant function from $\Nat^{\ell-1}$ to $M$, and that its value
	only depends on whether $b = 0$ or $b > 0$.
	This holds with
	the exact same proof as the base case of this induction. Let us call
	$m_{\ell,r,1}$ and $m_{\ell,r,0}$ the two elements of the monoid that are
	obtained by $\mu(\CutPref{\ell,r,b}(X_1, \dots, X_{\ell-1}))$ respectively when
	$b = 0$ and when $b > 0$. Finally, let us define $\pi_{\ell,r,0} \colon M^k
		\to \Nat$ via $\pi_{\ell,r,0}(m_2, \dots, m_{k+1}) \defined
		\pi(m_{\ell,r,0}, m_1, \dots, m_{k+1})$.

	With all of this preliminary work done, we are now ready to conclude by
	simply unrolling the definition of
	a function
	$f$ defined using a \kl{production function}
	$\pi \colon M^{k+1} \to \Nat$.
	In the following equations, we omit the precise
	range for the sum over $(\ell, r, b)$ to save space.
	\begin{align*}
		 & f(q(n_1, \dots, n_p))           \\
		 & \eqdef
		\sum_{v_1 \dots v_{k+1} = q(n_1, \dots, n_p)}
		\pi(\mu(v_1), \dots, \mu(v_{k+1})) \\
		 & = \sum_{(\ell, b, r)}
		\sum_{v_2 \dots v_{k+1} = \CutSuff{\ell,r}(n_\ell - b, \dots, n_p)}
		\pi(\mu(\CutPref{\ell,r,b}(n_1, \dots, n_{\ell-1})), \mu(v_2), \dots, \mu(v_{k+1}))
		\\
		 & =
		\sum_{(\ell, 0, r)}
		\sum_{v_2 \dots v_{k+1} = \CutSuff{\ell,b,0}(n_\ell - b, \dots, n_p)}
		\pi(m_{\ell,r,0}, \mu(v_2), \dots, \mu(v_{k+1}))
		\\
		 & +
		\sum_{(\ell, b > 0, r)}
		\sum_{v_2 \dots v_{k+1} = \CutSuff{\ell,r}(n_\ell - b, \dots, n_p)}
		\pi(m_{\ell,r,1}, \mu(v_2), \dots, \mu(v_{k+1}))
		\\
		 & =
		\sum_{(\ell, 0, r)}
		\sum_{v_2 \dots v_{k+1} = \CutSuff{\ell,b,0}(n_\ell - b, \dots, n_p)}
		\pi_{\ell,r,0}(\mu(v_2), \dots, \mu(v_{k+1}))
		\\
		 & +
		\sum_{(\ell, b > 0, r)}
		\sum_{v_2 \dots v_{k+1} = \CutSuff{\ell,r}(n_\ell-b, \dots, n_p)}
		\pi_{\ell,r,1}(\mu(v_2), \dots, \mu(v_{k+1}))
		\\
		 & =
		\sum_{1 \leq \ell \leq p}
		\sum_{0 \leq r < \omega \times \card{u_\ell}}
		(\pi_{\ell,r,0})^\dagger (\CutSuff{\ell, 0, r}(n_\ell-b, \dots, n_p))
		\\
		 & +
		\sum_{1 \leq \ell \leq p}
		\sum_{0 \leq r < \omega \times \card{u_\ell}}
		\sum_{0 < b < n_\ell}
		(\pi_{\ell,r,1})^{\dagger}(\CutSuff{\ell,r}(n_\ell - b, \dots, n_p))
	\end{align*}
	Now, using the induction hypothesis,
	we have \kl{natural binomial functions} $F_{\ell,r,0}$ and $F_{\ell,r,1}$
	satisfying
	\begin{equation*}
		(\pi_{\ell,r,1})^{\dagger}(\CutSuff{\ell,r}(n_\ell - b, \dots, n_p))
		= F_{\ell,r,1}(\omega (n_\ell - b), \dots, \omega n_p)
		\quad
		\forall n_\ell - b, \dots, n_p \geq 1
		\quad .
	\end{equation*}
	And similarly
	\begin{equation*}
		(\pi_{\ell,r,0})^{\dagger}(\CutSuff{\ell,r}(n_1, \dots, n_{\ell-1})
		= F_{\ell,r,0}(\omega n_1, \dots, \omega n_{\ell-1})
		\quad
		\forall n_1, \dots, n_{\ell - 1} \geq 1
		\quad .
	\end{equation*}

	It is a well-known equality that for all $s,n \in \Nat$,
	$\sum_{0 < i < n} \binom{s}{i} = \binom{s+1}{n}$.
	As a consequence, \kl{natural binomial functions} are
	stable under the following summation operation:
	$\Sigma_Y \colon Q(x_1,\dots,x_n,y) \mapsto \sum_{0 < b < y} Q(x_1, \dots, x_n, b)$.
	This allows us to conclude by defining:
	\begin{equation*}
		F(x_1, \dots, x_p) \defined
		\sum_{1 \leq \ell \leq p} \sum_{0 \leq r < \omega \card{u_\ell}}
		F_{l,r,0}(x_{\ell}, \dots, \omega x_p) + \Sigma_Y F_{l,r,1}(Y,
		x_{\ell+1}, \dots, x_p)
		\quad .
	\end{equation*}
	The latter being a \kl{natural binomial function} satisfying
	the desired equality.
\end{proofof}

\begin{proofof}[n-sf-combinatorics:lem]
	We redo the proof of \cref{n-poly-combinatorics:lem}, with the extra
	assumption that the monoid $M$ is \kl(monoid){aperiodic}, hence that there
	exists $s \in \Nat$ such that $x^{s+1} = x^s$ for all $x \in M$. As a
	consequence, we can define $\omega = 1$, and
	translate inputs by $s$, instead of multiplying them by $\omega$.
\end{proofof}

\begin{proofof}[derivation-translation:lem]

    We will first prove the equivalence between \cref{d-t-transl:item} and
    \cref{d-t-correct:item} without keeping track of the constant $K$, and then
    notice that this $K$ is actually computable. This will simplify the reading
    process.

    We will first prove that \cref{d-t-transl:item} $\implies$
    \cref{d-t-correct:item}, for any choice of $K \in \Nat$. Indeed, let us
    consider some partial valuation $\nu \colon \vec{X} \topartial \Nat$. By
    assumption, $\translate{K}(\restr{P}{\nu}) \in \Nat[\vec{X}]$, and in
    particular conclude that \kl{maximal monomials} of
    $\translate{K}(\restr{P}{\nu})$ are exactly the \kl{maximal monomials} of
    $\restr{P}{\nu}$, and are therefore \kl{non-negative}.

    Conversely, the implication \cref{d-t-correct:item} $\implies$
    \cref{d-t-transl:item} is proven by induction on the number of
    indeterminates of $P$. If $P$ is a constant polynomial, then $P = n$ for
    some $n \in \Nat$, and we conclude that $\translate{0}(P) \in
    \Nat[\vec{X}]$. Otherwise, $P = P_1 + P_2$ where $P_1$ is the sum of the
    \kl{maximal monomials} of $P$. Using
    \cref{derivation-stabilises-correct:lem}, there exists a computable $K \in
    \Nat$ such that $Q \defined \Diff{K}{P_1} + \translate{K}(P_2)$ belongs to
    $\Nat[\vec{X}]$. Furthermore, for all partial functions $\nu \colon \vec{X}
    \topartial \set{0, \dots, K}$ fixing at least one indeterminate of $P$, the
    induction hypothesis yields a number $L_\nu \in \Nat$ such that
    $\translate{L_\nu}(\restr{P}{\nu}) \in \Nat[\vec{X}]$. Let us define $L_m$
    to be the maximum of $L_\nu$ where $\nu$ ranges over partial
    functions described above.

    Let us prove that for all partial functions $\nu \colon \vec{X} \topartial
    \Nat$, the translation $\translate{L_m + K}(\restr{P}{\nu})$ belongs to
    $\Nat[\vec{X}]$. Without loss of generality, $\nu$ fixes only
    indeterminates that appear in $P$. If $\nu$ fixes at least one
    indeterminate $X_i$ to a value $\nu(X_i) = k \leq K$, then we can write
    $\nu = \mu [X_i = k]$, where $\mu$ does not fix the value of $X_i$. By
    construction, we know that $\translate{L_i}(\restr{\restr{P}{X_i =
    k}}{\mu})$ belongs to $\Nat[\vec{X}]$, and therefore that $\translate{K +
    L_m}(\restr{P}{\nu})$ does so too. The only case left is when all the
    variables appearing in $\nu$ are greater than $K$.
    In that case, $\nu = \translate{K}(\mu)$ for some partial 
    valuation $\mu$.
    In this case,
    $\translate{K}(\restr{P}{\translate{K}(\mu)})
    = \restr{\translate{K}(P)}{\mu}$.
    Remark that by definition,
    $\translate{K}(P) = P_1 + Q$,
    hence that
    $\translate{K + L_m}(P) = \translate{L_m}(P_1) + \translate{L_m}(Q)$
    belongs to $\Nat[\vec{X}]$.
    Finally,
    we conclude that $\restr{\translate{K+L_m}(P)}{\mu}$
    belongs to $\Nat[\vec{X}]$
    because
    the latter is closed under partial applications.

    Now, remark that in the proof, we have computed $K$ recursively from $P$,
    and only applied partial functions using values below $K$. This not only
    shows that $K$ is computable from $P$, but also that
    \cref{d-t-transl-fin:item} and \cref{d-t-transl:item} are equivalent.

    For the effective decision procedure,
    given $P \in \Rel[\vec{X}]$,
    one applies \cref{derivation-translation:lem}
    to obtain a bound $K$, and then 
    computes $\translate{K}(\restr{P}{\nu})$ for all 
    partial valuations $\nu \colon \vec{X} \topartial \set{0, \dots, K}$.
    Then $P \in \CorrectPoly$ if and only if all those polynomials
    belong to $\Nat[\vec{X}]$, which is decidable.
\end{proofof}

\begin{proofof}[integer-binomial-polynomial:lem]
	Let us first notice that $\pbinom{X}{k}$ is \kl{integer-valued}
	when $k \in \Nat$. One way to notice this is that when $X \geq 0$,
	$\pbinom{X}{k} = \binom{X}{k}$ is the number of ways to choose $k$ elements among $X$,
	which is an integer. When
	$X < 0$, then $\pbinom{X}{k} = (-1)^k \pbinom{k - X + 1}{k} = (-1)^k \binom{k - X + 1}{k}$
	is also an integer. Because \kl{integer-valued} polynomials are closed under products,
	$\Rel$-linear combination, and 
    shifting of domain by an integer,
    we conclude that all \kl{integer binomial polynomials} are
	\kl{integer-valued}.

	For the converse direction, we follow the original proof of Pólya \cite{POLYA1915,CACHA1996}.
	The only difference is that we leverage differential operators on multiple indeterminates.
	To that end, let us consider the operator $\Delta_i$ that maps
	a polynomial $P$ to the polynomial $\Delta_i(P) \defined P(X_1, \dots, X_i + 1, \dots, X_k) - P$.
	That is, the discrete partial derivative with respect to the $i$-th indeterminate.
	Let us notice that $\Delta_i \pbinom{X_i}{\alpha} = \pbinom{X_i}{\alpha - 1}$ when $\alpha > 0$,
	indeed:
	\begin{align*}
		\Delta_i \pbinom{X_i}{\alpha} & = \pbinom{X_i + 1}{\alpha} - \pbinom{X_i}{\alpha}                                                         \\
		                              & = \frac{(X_i + 1) \cdots (X_i + 1 - \alpha + 1)}{\alpha!} - \frac{X_i \cdots (X_i - \alpha + 1)}{\alpha!} \\
		                              & = \frac{X_i \cdots (X_i - \alpha + 2)}{\alpha!} \alpha                                                    \\
		                              & = \pbinom{X_i}{\alpha - 1}
		\quad .
	\end{align*}
	Furthermore, when $\alpha = 0$, $\Delta_i \pbinom{X_i}{0} = 0$.

	Let us now remark that a simple expansion proves $\Delta_i \Delta_j P = \Delta_j \Delta_i P$ for all
	$1 \leq i,j \leq k$.
	This justifies the soundness of the notation
	$\Delta_\alpha$ where $\alpha \colon \set{1, \dots, k} \to \Nat$ is a
	multi-index, where $\Delta_{\alpha} \defined \Delta_1^{\alpha_1} \cdots
		\Delta_k^{\alpha_k}$.
	Let us write $\card{\alpha} \defined \sum_{i = 1}^k \alpha_i$ for the
	size of a multi-index.
	We claim that for all $P \in \Rat[X_1, \dots, X_k]$ of degree at most $d$,
	the following \emph{discrete Taylor expansion} holds:
	\begin{equation}
		\label{discrete-taylor:eq}
		P =
		\sum_{\card{\alpha} \leq d}
		\Delta_{\alpha} P (0, \dots, 0)
		\times
		\prod_{i = 1}^k \pbinom{X_i}{\alpha_i}
		\quad .
	\end{equation}

	To obtain \cref{discrete-taylor:eq}, we proceed by induction on the degree $d$.
	When $d = 0$, the result is holds because $P = P(0, \dots, 0)$.
	Assume that the result holds for all polynomials of degree at most $d$.
	Let $P$ of degree at most $d+1$.
	Then, there exists coefficients $\seqof{c_{\alpha}}{\card{\alpha} \leq d+1}$
	in $\Rat$
	such that
	\begin{equation*}
		P = \sum_{\card{\alpha} \leq d+1} c_{\alpha} \prod_{i = 1}^k \pbinom{X_i}{\alpha_i}
		\quad .
	\end{equation*}
	This is because the monomials of the form $\prod_{i = 1}^k \pbinom{X_i}{\alpha_i}$
	form a basis of $\Rat[X_1, \dots, X_k]$.
	In particular, when applying one partial derivative, one can leverage the induction hypothesis to obtain:
	\begin{align*}
		\Delta_1 P & = \sum_{\card{\alpha} \leq d+1} c_{\alpha} \Delta_1 \prod_{i = 1}^k \pbinom{X_i}{\alpha_i}
		=
		\sum_{\card{\alpha} \leq d+1 \wedge \alpha_1 > 0} c_{\alpha}  \pbinom{X_1}{\alpha_1 - 1} \prod_{i = 2}^k \pbinom{X_i}{\alpha_i}
		\\
		           & = \sum_{\card{\beta} \leq d} \Delta_{\beta}(\Delta_1 P)(0,\dots,0) \times \pbinom{X_1}{\beta_1} \prod_{i = 2}^k \pbinom{X_i}{\beta_i}
		\quad .
	\end{align*}

	Using again the fact that the monomials form a basis, we conclude by
	identification that for all $\card{\alpha} \leq d+1$ such that $\alpha_1 >
		0$, we have $c_{\alpha} = \Delta_{\alpha}P (0,\dots,0)$. We can repeat the
	same process with $\Delta_i$ for $1 \leq i \leq k$ to obtain that
	$c_{\alpha} = \Delta_{\alpha} P (0,\dots, 0)$ for all $\alpha$ such that $1
		\leq \card{\alpha} \leq d+1$. Now, for the specific case of $\card{\alpha}
		= 0$, which is the constant term of $P$, we simply conclude that
	$P(0,\dots,0) = c_{\alpha}$, and notice that $\Delta_{\alpha} P = P$ in
	this case.

	Let us now conclude by noticing that if $P$ is \kl{integer-valued},
	then so is $\Delta_i P$ for all $1 \leq i \leq k$. As a consequence,
	so are $\Delta_{\alpha} P$, for all multi-indexes $\alpha$, and we conclude
	that $\Delta_{\alpha}P (0,\dots,0) \in \Rel$ for all $\alpha$.
	Combined with \cref{discrete-taylor:eq}, we conclude that $P$ is an \kl{integer binomial polynomial}.
    Remark that $\Delta_{\alpha} P (0, \dots, 0)$ only ever involves evaluation of $P$ on 
    $\Nat^k$, justifying the weakening of the hypothesis.
\end{proofof}

\begin{proofof}[polyrec-integer-strong:lem]
    Because $n P \in \CorrectPoly$,
    it is represented by a \kl{star-free $\Nat$-polyregular function} $f$ thanks to
    \cref{corrected-version:thm}. Let us now use \cref{n-sf-combinatorics:lem}
    to obtain a number $s \geq 1$ (that only depends on $P$) such
    that for all \kl{pumping patterns} $q \colon \Nat^p \to \Sigma^*$, $f
    \circ q(X_1 + s, \dots, X_p + s)$ is a \kl{natural binomial function} $F$
    over $\Nat^p$. We will use this $s$ with $p = k$, where
    $k = \card{\vec{X}}$ is the number of indeterminates of $P$,
    and $q(X_1, \dots, X_k) \defined \prod_{i = 1}^k a_i^{X_i}$,
    assuming the alphabet $\Sigma$ is $\set{a_1, \dots, a_k}$.
    In particular, $F(x_1, \dots, x_k) = f \circ q (x_1, \dots, x_k) = nP(x_1, \dots, x_k)$
    for all $\vec{x} \in (\Nat_{\geq s})^k$.
    Let us write the
    \kl{natural binomial function} $F$ as follows:
    \begin{equation*}
        F(X_1, \dots, X_k) =
        \sum_{\card{\alpha} \leq d} e_{\alpha} \prod_{i = 1}^k \binom{X -
        p_{\alpha,i}}{\alpha(i)} \quad
        \text{where } \forall 
        \card{\alpha} \leq d,
        e_{\alpha} \in \Nat
        \quad .
    \end{equation*}

    Because of the equality $\binom{X - p - 1}{k} + \binom{X - p - 1}{k -
    1} = \binom{X - p}{k}$, we conclude that we can find a $K \in \Nat$ and
    coefficients $f_{\alpha} \in \Nat$ such that $F(X_1, \dots, X_k) =
    \sum_{\card{\alpha} \leq d} f_{\alpha} \prod_{i = 1}^k \binom{X_i -
    K}{\alpha(i)}$, i.e., we can assume that $p_{\alpha,i} = K$ for all
    $\card{\alpha} \leq d$.
    Let us remark that $P$ is an \kl{integer-valued} polynomial of
    degree $d$, and therefore one can use
    \cref{integer-binomial-polynomial:lem} to rewrite it in the following form:
    \begin{equation*}
        P(X_1, \dots, X_k) = \sum_{\card{\alpha} \leq d}
        c_{\alpha} \prod_{i = 1}^k \pbinom{X_i - K}{\alpha(i)}
        \quad \text{where } \forall \card{\alpha} \leq d, 
        c_{\alpha}
        \in \Rel \quad .
    \end{equation*}

    Because $F(x_1, \dots, x_k) = P(x_1, \dots, x_k)$ for all
    $\vec{x} \in (\Nat_{\geq s})^k$, one can write
    \begin{equation*}
        n \left(\sum_{\card{\alpha} \leq d} c_{\alpha} \prod_{i = 1}^k \pbinom{x_i - K}{\alpha(i)} \right)
        = 
        \sum_{\card{\alpha} \leq d} f_{\alpha} \prod_{i = 1}^k \binom{x_i - K}{\alpha(i)}
        \quad 
        \forall \vec{x} \in (\Nat_{\geq s})^k
    \end{equation*}

    Because this equality holds over $(\Nat_{\geq s})^k$, we conclude that the
    underlying polynomials are equal in $\Rel[\vec{X}]$, and therefore that:
    \begin{equation*}
        \sum_{\card{\alpha} \leq d} n c_{\alpha} \prod_{i = 1}^k \pbinom{X_i - K}{\alpha(i)}
        = 
        \sum_{\card{\alpha} \leq d} f_{\alpha} \prod_{i = 1}^k \pbinom{X_i - K}{\alpha(i)}
    \end{equation*}

    Identifying the coefficients of the two polynomials (which is possible
    because the products of binomial coefficients translated by $K$ form a
    basis of $\Rat[\vec{X}]$), we conclude that for all $\card{\alpha} \leq d$,
    $n c_{\alpha} = f_{\alpha}$. In particular,
    for all $\card{\alpha} \leq d$, $c_{\alpha} \in \Nat$,
    hence $P$ is a \kl{natural binomial polynomial}.

    Remark that this reasoning remains true whenever we partially apply $P$ to
    some values in $\Nat$, since partial applications of $nP$ continue to be in
    $\CorrectPoly$ by definition, and remain \kl{integer-valued}. We have
    proven that $P$ is a \kl{strongly natural binomial polynomial}.
\end{proofof}

\end{document}